\documentclass[useAMS,usenatbib]{mn2e}
\usepackage{graphicx}
\usepackage{amsmath}
\usepackage{ifpdf}
\ifpdf
\def\reff@jnl#1{{\rm#1\/}}

\def\aj{\reff@jnl{AJ}}                  % Astronomical Journal
\def\araa{\reff@jnl{ARA\&A}}            % Annual Review of Astron and Astrophys
\def\apj{\reff@jnl{ApJ}}                % Astrophysical Journal
\def\apjl{\reff@jnl{ApJ}}               % Astrophysical Journal, Letters
\def\apjs{\reff@jnl{ApJS}}              % Astrophysical Journal, Supplement
\def\apss{\reff@jnl{Ap\&SS}}            % Astrophysics and Space Science
\def\aap{\reff@jnl{A\&A}}               % Astronomy and Astrophysics
\def\aapr{\reff@jnl{A\&A~Rev.}}         % Astronomy and Astrophysics Reviews
\def\aaps{\reff@jnl{A\&AS}}             % Astronomy and Astrophysics, Supplement
\def\baas{\reff@jnl{BAAS}}              % Bulletin of the AAS
\def\jrasc{\reff@jnl{JRASC}}            % Journal of the RAS of Canada
\def\memras{\reff@jnl{MmRAS}}           % Memoirs of the RAS
\def\mnras{\reff@jnl{MNRAS}}            % Monthly Notices of the RAS
\def\physrep{\reff@jnl{Phys.Rep.}}
\def\pra{\reff@jnl{Phys.Rev.A}}         % Physical Review A: General Physics
\def\prb{\reff@jnl{Phys.Rev.B}}         % Physical Review B: Solid State
\def\prc{\reff@jnl{Phys.Rev.C}}         % Physical Review C
\def\prd{\reff@jnl{Phys.Rev.D}}         % Physical Review D
\def\prl{\reff@jnl{Phys.Rev.Lett}}      % Physical Review Letters
\def\pasp{\reff@jnl{PASP}}              % Publications of the ASP
\def\pasj{\reff@jnl{PASJ}}              % Publications of the ASJ
\def\skytel{\reff@jnl{S\&T}}            % Sky and Telescope
\def\solphys{\reff@jnl{Solar~Phys.}}    % Solar Physics
\def\sovast{\reff@jnl{Soviet~Ast.}}     % Soviet Astronomy
\def\ssr{\reff@jnl{Space~Sci.Rev.}}     % Space Science Reviews
\def\nat{\reff@jnl{Nature}}             % Nature

\begin{document}

\title[The Bluedisk project]
{H{\sc I} scaling relations of galaxies in the environment of H{\sc I}-rich and control 
galaxies observed by the Bluedisk project}

\author[E. Wang et al.]
{Enci Wang$^{1}$\thanks{E-mail: ecwang@shao.ac.cn},Jing Wang$^{2,3}$,
Guinevere Kauffmann$^{2}$,Gyula I. G. J\'ozsa$^{4,5,6}$,
\newauthor Cheng Li$^{1}$ \\
$^{1}$Partner Group of the Max Planck Institute for Astrophysics
at the Shanghai Astronomical  Observatory and Key Laboratory \\
for Research in Galaxies  and Cosmology of Chinese Academy of
Sciences, Nandan Road 80, Shanghai 200030, China \\
$^{2}$Max--Planck--Institut f\"ur Astrophysik,
        Karl--Schwarzschild--Str. 1, D-85741 Garching, Germany \\
$^{3}$CSIRO Astronomy and Space Science, Australia Telescope National Facility, 
        PO Box 76, Epping, NSW 1710, Australia\\
$^{4}$SKA South Africa, Radio Astronomy Research Group, 3rd Floor, 
        The Park, Park Road, Pinelands, 7405, South Africa\\
$^{5}$Rhodes University, Department of Physics and Electronics, Rhodes Centre for Radio Astronomy Techniques 
        and Technologies,\\ PO Box 94, Grahamstown, 6140, South Africa \\
$^{6}$Argelander-Institut f\"ur Astronomie, Auf dem H\"ugel 71, 53121 Bonn, Germany\\
%$^{7}$Netherlands Institute for Radio Astronomy (ASTRON), Postbus 2, 
%        7990 AA Dwingeloo, The Netherlands \\
}

\date{Accepted ........ Received ........; in original form ........}

\pagerange{\pageref{firstpage}--\pageref{lastpage}} \pubyear{2014}
\maketitle

\label{firstpage}

\begin{abstract}
Our work is based on the ``Bluedisk'' project, a program to map the neutral gas in a sample of 25 H{\sc I}-rich  spirals
and a similar number of control galaxies with 
the Westerbork Synthesis Radio Telescope (WSRT). In this paper we focus on the
H{\sc{I}} properties of the galaxies in the environment of our targeted
galaxies. In total, we extract 65 galaxies from the WSRT cubes with stellar masses between $10^8 \rm M_{\odot}$ and
$10^{11} \rm M_{\odot}$. Most of these galaxies are located on 
the same H{\sc{I}} mass-size relation and ``H{\sc{I}}-plane'' as normal spiral galaxies. We find that companions 
around H{\sc{I}}-rich galaxies tend to be H{\sc{I}}-rich as well and to
 have larger $R_{\rm 90,H{I}}/R_{\rm 50,H{I}}$.
This suggests a scenario of ``H{\sc{I}} conformity'',
similar to the colour conformity found by \cite{Weinmann-06}: galaxies tend to 
adopt the H{\sc{I}} properties of their neighbours.
We visually inspect the outliers from the H{\sc{I}} mass-size relation and galaxies which are offset from the H{\sc{I}} 
plane and find that they show morphological and kinematical signatures of recent interactions
with their environment.  We speculate  that these outliers
have been disturbed by tidal or ram-pressure stripping processes, or in a few cases, by accretion events.

\end{abstract}


\section{Introduction}\label{sec:introduction}

In the recent years, a large number of surveys have obtained multi-wavelength
imaging and spectroscopy for large samples of galaxies at different redshifts. Thanks to these surveys, 
we have learned a lot about how galaxies form and evolve. 
However, many physical  processes, such as the regulation of star formation by gas accretion,
remain poorly understood.

The accretion of cold gas in the form of gas-rich
dwarf galaxies can  bring gas to galaxies and fuel star formation 
\citep[]{Sancisi-08,Silk-Mamon-12,Conselice-13}.
The presence of extra-planar gas \citep[]{Chaves-Irwin-01, Boomsma-05, Wakker-07},
lopsided H{\sc{I}} morphologies \citep[]{Sancisi-76, Shang-98, Thilker-07}
and gas tails \citep[]{Kregel-Sancisi-01, Oosterloo-10} may directly be
linked to ongoing cold gas accretion through mergers. 
While some portion of the gas in galaxies is being acquired through the
accretion of dwarf galaxies, the inferred gas accretion rate is not sufficient 
to sustain star formation in galaxies ($\sim1.0\,{\rm M_{\odot}}\,{\rm yr}^{-1}$)
\citep{Binney-Dehnen-Bertelli-00}. As a consequence, a large  fraction
of gas should be accreted directly from the Intergalactic Medium (IGM).
Low-density galaxies  lose their gas through tidal interactions or ram pressure
stripping, which may finally quench their star formation
\citep[]{Moore-96, Calcaneo-Roldan-00, Mayer-06, McCarthy-08, Kapferer-08, Chang-Maccio-Kang-13}.
This process happens more frequently in dense environments.  

The Bluedisk project, which was carried out  with the Westerbork Synthesis Radio Telescope(WSRT), was  designed to 
map the H{\sc{I}} distribution is a sample of 25 unusually H{\sc I}-rich  galaxies.  
Galaxies with large  H{\sc{I}} excess
usually have bluer and younger outer disks \citep{Wang-11}, and more metal-poor ionized gas \citep{Moran-12}.  
A sample of 25 control galaxies was  also observed for 
comparison. These galaxies were closely matched in stellar mass, stellar surface mass density, 
redshift and inclination, but were not unusually rich in H{\sc I}. The goal of the project was to
determine whether there was any evidence for recent gas accretion onto unusually H{\sc I}-rich galaxies by 
investigating and contrasting the  H{\sc{I}} structure and environment of the gas-rich galaxies
with that of the control sample.

In the first paper of Bluedisk project, \citet[Paper I]{Wang-13} concentrated 
on the H{\sc{I}} size and morphology of the 42 targeted galaxies. They found 
that H{\sc{I}}-rich galaxies do not differ from normal 
galaxies with respect to H{\sc{I}} asymmetry indices or optical/H{\sc{I}} 
disk position angle differences. This is inconsistent with a scenario in which 
the excess gas was  brought in by mergers. In this paper, we extend 
this work to the galaxies in the neighbourhood of the targeted galaxies that lie within
the WSRT data cubes.

Our paper is organized as follows. In section 2, we briefly recap our observations, describe the
data and the data processing that we carried out for our  environmental study, including
our procedure for  accurate
primary beam correction.
In section 3, we discuss our identification of galaxy neighbours  via cross-matching with the
spectroscopic catalogue of the Sloan Digital Sky Survey (SDSS) and how we  
build a uniform catalog of these sources.          
In section 4, we examine the H{\sc{I}} mass-size relation and the correlation between
H{\sc I} and optical properties such as mass and stellar surface density for both the
neighbours around H{\sc I}-rich and control galaxies.  
In section 5, we discuss the morphology and properties of galaxies which are found to be outliers from
the standard scaling relations.
We summarize our results in section 6. 

Throughout this paper, all the distance-dependent 
parameters are computed with $\Omega=0.3$,
$\Lambda=0.7$ and $H_0= 70$ km s$^{-1}$ Mpc$^{-1}$. 

\section{Data}

\subsection{Observation and data reduction}
The 50 targets were observed with the WSRT in 2011 and 2012. Target selection and
data reduction are described in detail in Paper I. For  detailed information
on the targeted galaxies we refer the reader to table 1 in Paper I.

%Our target galaxies had been observed before June 2012 with an on-source
% integration time as 12h per galaxy. 
The H{\sc{I}} raw data cubes were reduced using a 
pipeline produced by \cite{Serra-12}, based on the Miriad
reduction package \citep{Sault-Teuben-Wright-95}.
The data used in this paper are H{\sc{I}} cubes built with 
a Robust weighting of 0.4, which provides a suitable compromise between 
sensitivity and resolution. The pixel size is 
4 arcsec and the velocity width for each channel is about 
$13$ km s$^{-1}$. The velocity resolution is $26$ km s$^{-1}$ (FWHM). 
The typical beam has Half-Power Beamwidth (HPBW) of $16\times16/$sin($\delta$) arcsec$^2$,
where $\delta$ is the declination.
Every cube has 148 channels,  covers a redshift 
range of $\Delta z=0.006$, and  has a size of 1 degree on each side, which
corresponds to a physical scale of 1.7 Mpc at a redshift $z=0.025$.
We point out that 92.0\% of the galaxies are within a systematic velocity 
difference of 500 km s$^{-1}$ from the primary galaxies. However, some 
untargeted galaxies are not in the immediate neighborhood of the primary
 galaxies: in the radial direction, the farthest galaxy is 884 km s$^{-1}$ away.
 Hence we are investigating a relatively large-scale environmental effect
 rather than the direct interaction between galaxy pairs in this paper.

We generate two-dimensional H{\sc{I}} total-intensity maps
(moment-0 maps) for each cube. First we identify 3-d regions of emission by a smoothing and clipping
algorithm. We then add all the 
detected H{\sc{I}} emission from  all velocity channels. 
We also estimate errors for all non-zero pixels in the H{\sc{I}} intensity map.

\subsection{Physical properties of galaxies}
The  physical quantities required for this work are a spectrophotometric estimate of the stellar 
mass $M_{*}$, stellar surface mass density $\mu_{*}$, and the NUV-r colour.
Stellar masses were taken from the MPA-JHU database (http://www.mpa-garching.mpg.de/SDSS/DR7/), 
and are derived from Sloan Digital Sky Survey (SDSS) photometry. The stellar surface mass density is defined as 
$\mu_{*}=M_{*}\,(2\pi R_{\rm 50,z}^2)^{-1} $, where
$R_{\rm 50,z}$ is the physical radius which contains half the 
total light in the z-band. The NUV magnitude is available from the Galaxy Evolution Explorer (GALEX)
pipeline and the NUV-r colors are corrected for Galactic extinction. 
The H{\sc{I}} size $R1$, $R_{\rm 50,H{I}}$,
$R_{\rm 90,H{I}}$ and $rs$, and H{\sc{I}} mass $M_{\rm H{I}}$, 
are measured using  our H{\sc{I}} data cubes. $R1$ is 
the radius where radially averaged face-on H{\sc{I}} column density
reaches $1\,{\rm M_{\odot}}\,{\rm pc}^{-2}$(corresponding to $1.25\times10^{20}$atoms cm$^{-2}$). 
$R_{\rm 50,H{\sc I}}$ and $R_{\rm 90,H{\sc I}}$ are the radii enclosing 50 and 90 percent of the H{\sc{I}} flux,
respectively. Note that an inclination correction is not applied when calculating $R1$ 
for unresolved galaxies ($R_{\rm 50,H{\sc I}}$ less than 15 arcsec), because their H{\sc{I}} sizes
are less than the beam size. For these galaxies, the derived value of $R1$
should be regarded as an upper limit to the true value of $R1$.
$rs$ is the scale-length of the  outer exponential disk, 
and is measured by assuming a PSF-convolved two-component model for  the radial distribution
of H{\sc{I}} (see Wang et al. 2014 for details). 
The H{\sc{I}} mass is defined as
$M_{\rm H{\sc I}}$=$2.356\times10^5$($D_{\rm lum}\,{\rm Mpc}^{-1}$)$^2$($F_{\rm tot}\,{(\rm Jy\,km\,s^{-1}})^{-1}$),
where $D_{\rm lum}$
is the luminosity distance 
and $F_{\rm tot}$ is the integrated H{\sc{I}}-line flux density. 
The measurements of these parameters are described in more  detail
in Paper I. 

\subsection{Primary beam correction}

\begin{figure*}
\centerline{
    \includegraphics[width=0.45\textwidth]{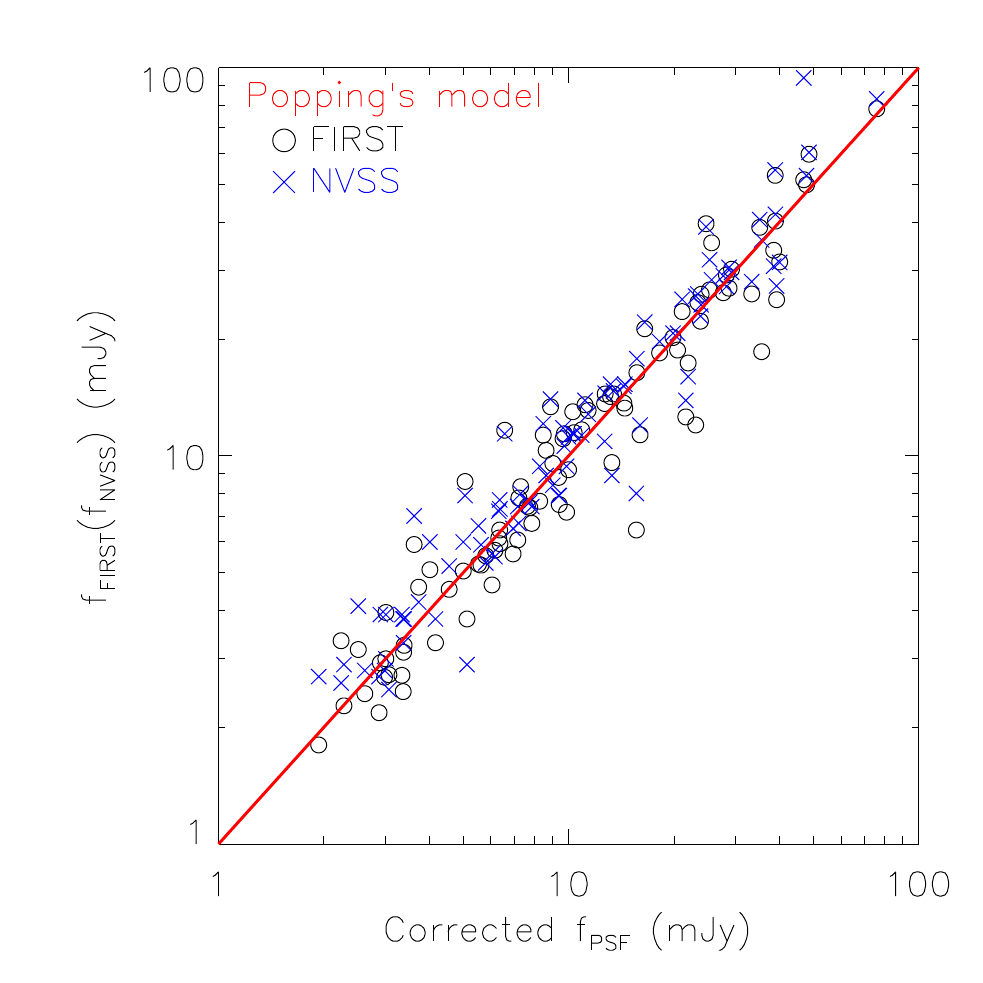}
    \includegraphics[width=0.45\textwidth]{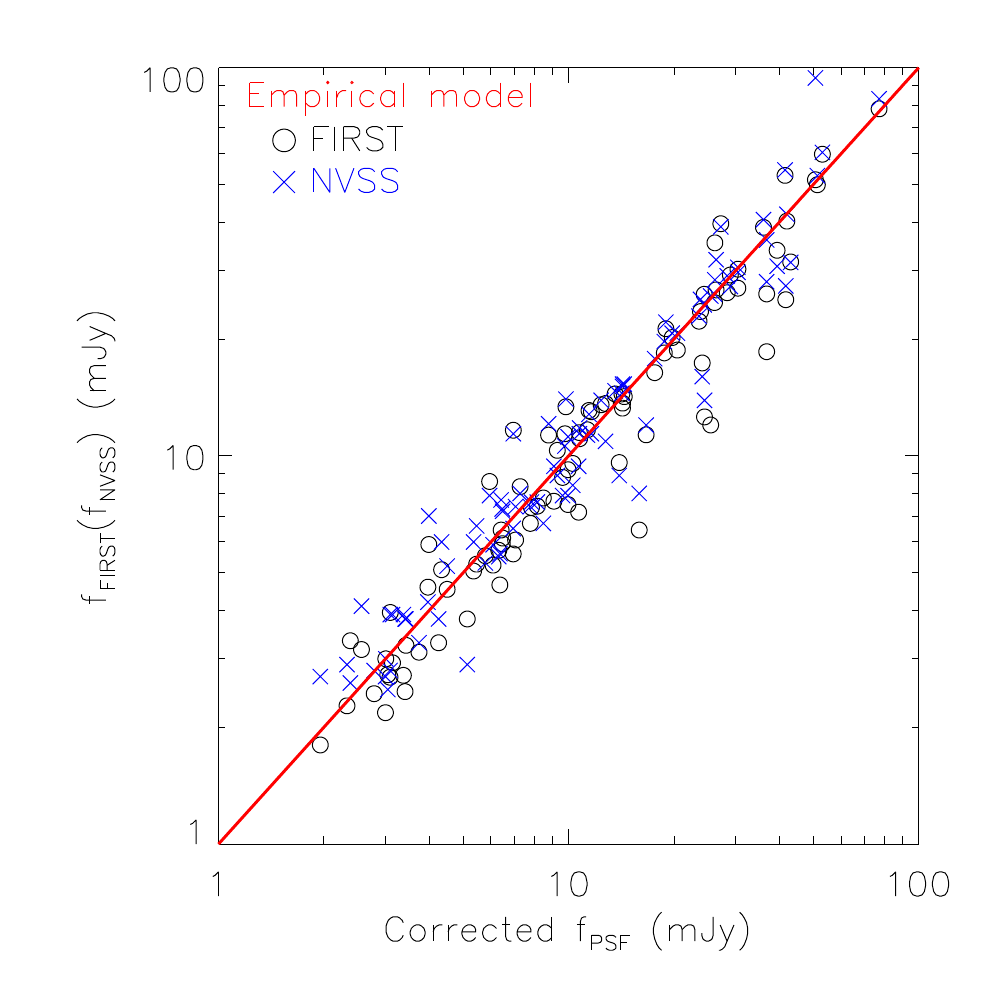}}
\caption{Test of the two methods for  primary beam correction:
1) applying the standard WSRT analytic model (with $c$=63),
and 2) a nonparametric model by Popping \& Braun (2008). 
This figure shows the relation between the primary beam
corrected flux density and the FIRST/NVSS flux density.
Black circles and blue crosses represent the FIRST data and the NVSS data,
respectively. The left and right
panels are for Popping's model and the empirical model, respectively.}
\label{fig1}
\end{figure*}

Our work depends critically on the primary beam correction, which accounts for the
attenuation by the primary beam towards large radii. We consider two methods.
One is a non-parametric model (data cube)  provided by \cite{Popping-Braun-08},
the other one is to apply a parametric correction function of the form $\cos^6(c\times \nu\times d)$
as routinely used for the WSRT, where $\nu$ is the frequency in GHz, 
and $d$ is the angular distance from the pointing centers in degrees.

To test and calibrate both approaches, we compare the un-corrected flux densities of point
sources in the radio continuum maps, which are an additional product of our data reduction 
pipeline (see Wang et al. 2013) with the flux densities listed in the NRAO FIRST
 \citep[Faint Images of the Radio Sky at Twenty-Centimeters]{Becker-White-Helfand-95} 
and NVSS \citep[NRAO VLA Sky Survey]{Condon-98} catalogues.
    
Specifically, we first extract sources from 
our continuum maps using Source Extractor (SE). Then we use the point spread function (PSF) fitting 
method to derive the flux density of each source extracted from our continuum 
maps. With a detection threshold of $5\sigma$, where $\sigma$ is the rms noise with a typical value
of $8.0\times 10^{-5}$ Jy beam$^{-1}$, we extract more than two thousand sources from all images.
This catalog is then matched to the FIRST 
and NVSS catalogs. In total, there are 4128 sources in the FIRST
catalog \citep[12Feb16 Version]{Becker-12} located in these 50 H{\sc{I}} 
observed regions, with integral flux density greater than 1 mJy.
Among them, 1322 sources are classified as point sources 
at FIRST resolution (with a major axis FWHM of 2 arcsec).
About one hundred detections in our cubes were matched with both FIRST 
and NVSS point sources. We restrict the analysis to point sources because
FIRST and NVSS flux densities are very
consistent with each other for point sources, with a scatter of 0.05 dex. 

This way, we obtain an best-fit $c$ of 63, differing distinctly from the commonly used
and recommended $c=68$ (see WSRT web pages). This difference in $c$ causes a 
0.46 dex difference in flux intensity for pixels at the farthest edge of the 
continuum map (0.71 degree from the map center at redshift $z=0.025$), 
and a largest difference of 0.17 dex for our farthest detected source. 
On average, the flux densities corrected with $c=63$ are 0.04 dex smaller 
than that of $c=68$ for our untargeted sources. 
Fig. 1 shows the comparison between the primary beam corrected 
PSF flux densities from our continuum maps and 
the FIRST/NVSS flux densities using both methods, the empirical analytic (standard) model,
and Popping's nonparametric model.
We can see a good linear correlation 
between the corrected PSF flux density and the FIRST/NVSS data. It is difficult to tell which 
model is better according to the data points, because the scatter is  
similar ($\sim$ 0.7 dex).  However, we remark that the standard
recommended analytic model with $c=68$ would have provided an insufficient primary
 beam correction in the wavelength range considered in this work. 
Since we confirm that the model of \citet{Popping-Braun-08} works well, we adopt it for our primary beam 
correction for the Bluedisk data cubes.

%%%%%%%%%%%%%%%%%%%%%%%%%%%%%%%%%%%%%%%%%%%%%%%%%%%%%%%%%%%%%%%%%%%%%%%%%%%%%%%%%%%%%%%%%%%
%%%%%%%%%%%%%%%%%%%%%%%%%%%%%%%%%%%%%%%%%%%%%%%%%%%%%%%%%%%%%%%%%%%%%%%%%%%%%%%%%%%%%%%%%%%

\section{Source identification}

We use the source finder developed by \citet{Serra-12} to detect sources
 in the cubes. The pipeline uses a smooth-and-clip algorithm: it smooths 
the cube and searches for regions with a flux intensity above 3$\sigma$ of
 the cube. The resulted catalog of 1962 sources includes both real sources
 and noise peaks. We take a few steps to filter the real galaxies with 
reliable flux measurements. The first step is to match the H{\sc I} catalog with 
the SDSS spectroscopic catalog. 163 H{\sc I} sources are matched with optical 
galaxies, with an angular distance smaller than 20 arcsec (roughly the 
beam size). 10 of them have more than one matched optical counterpart. 
We further constrain in redshift by requiring $|z_{\rm spec}-z_{\rm H{\sc I}}|<0.001$, 
which left us with 120 H{\sc I} sources, and none of them have multiple 
optical counterparts.  Although we may miss the galaxies with no optical
 spectroscopic observation, we efficiently exclude most of the unreliable
 sources from the H{\sc I} catalog. 

In the following sections, we describe how we further select the sources
 with reliable H{\sc I} flux measurements.
%To identify all H{\sc{I}} sources in the cubes, we use the source finder developed by Serra et al. (2012), 
%which is a smooth-and-clip algorithm. In total, 1962 H{\sc I} sources are detected with positive H{\sc I} flux. Note
%that many of these sources are not real sources, which are caused by rms noise.
%In this section, we describe how we build a uniform catalogue with
%high reliability for H{\sc{I}} fluxes and H{\sc{I}} morphologies. 
%We match the detected sources with  the SDSS spectroscopic catalogue. 
%163 sources are matched with the angular distance is smaller than 20
%arcsec ($\sim$ the beam size). Among them, ten sources have more than one 
%optical counterpart. However, if we give the constraint that the redshift
%difference between optical and H{\sc I} sources $|z_{\rm spec}-z_{\rm H{\sc I}}|$ is smaller than 0.001, 
%120 sources are matched and each of them has only one optical counterpart. 
%Although we would miss some H{\sc I} clouds and galaxies with no spectroscopic observation, 
%this is a very effective way to extract real galaxies from H{\sc I} sources with high reliability.
 
\subsection{Defining the outermost H{\sc{I}} contour}

\begin{figure*}
\centerline{
    \includegraphics[width=0.32\textwidth]{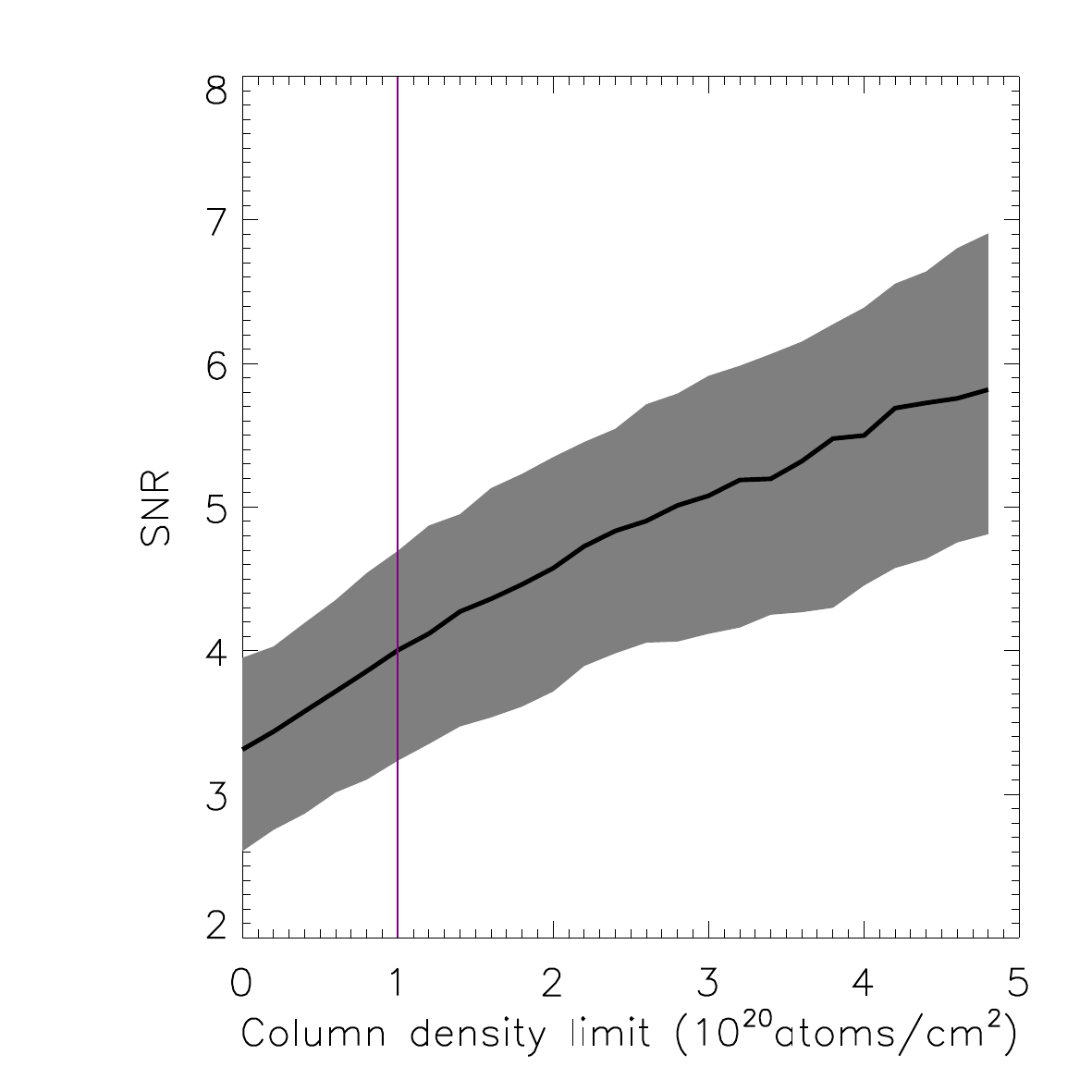}
    \includegraphics[width=0.32\textwidth]{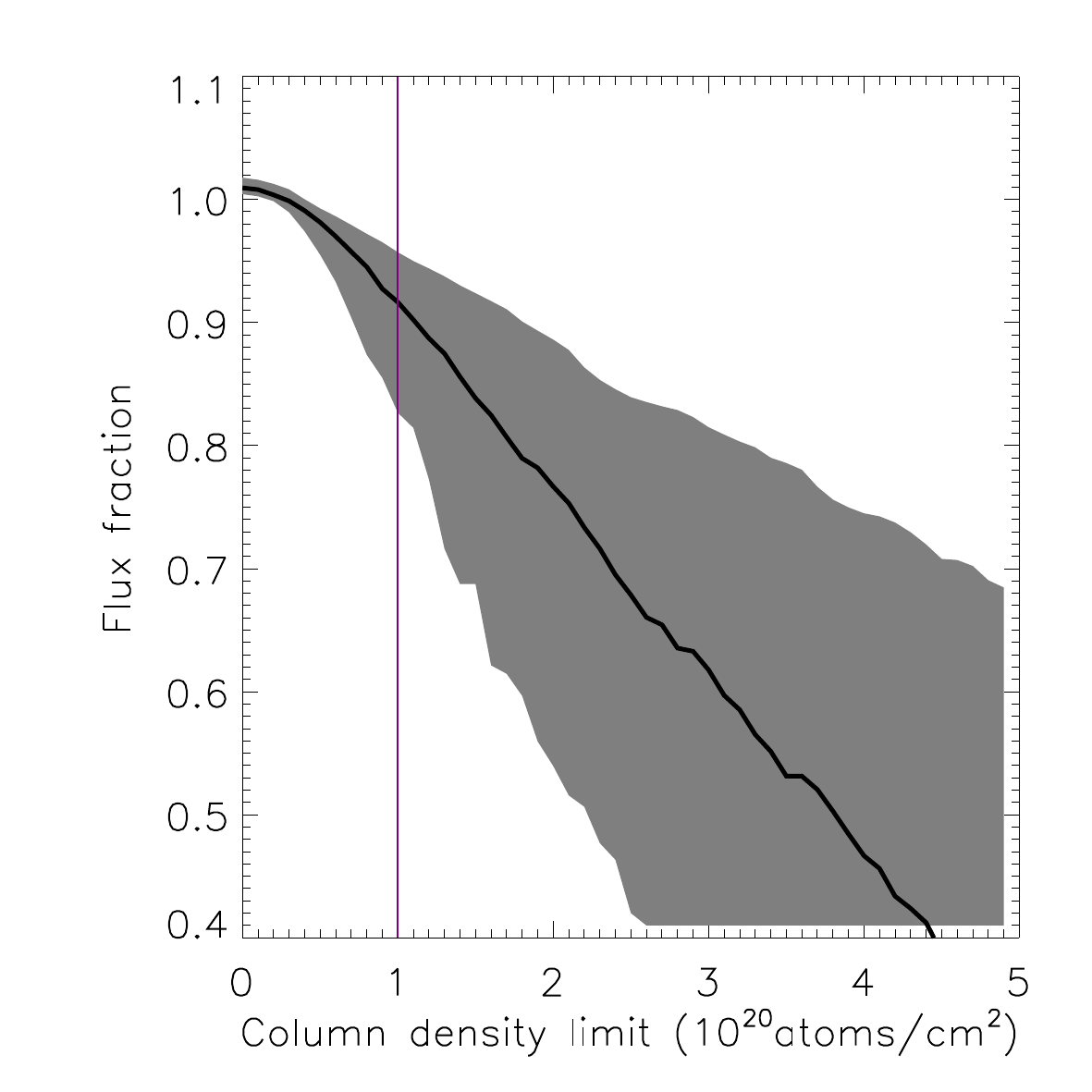}
    \includegraphics[width=0.32\textwidth]{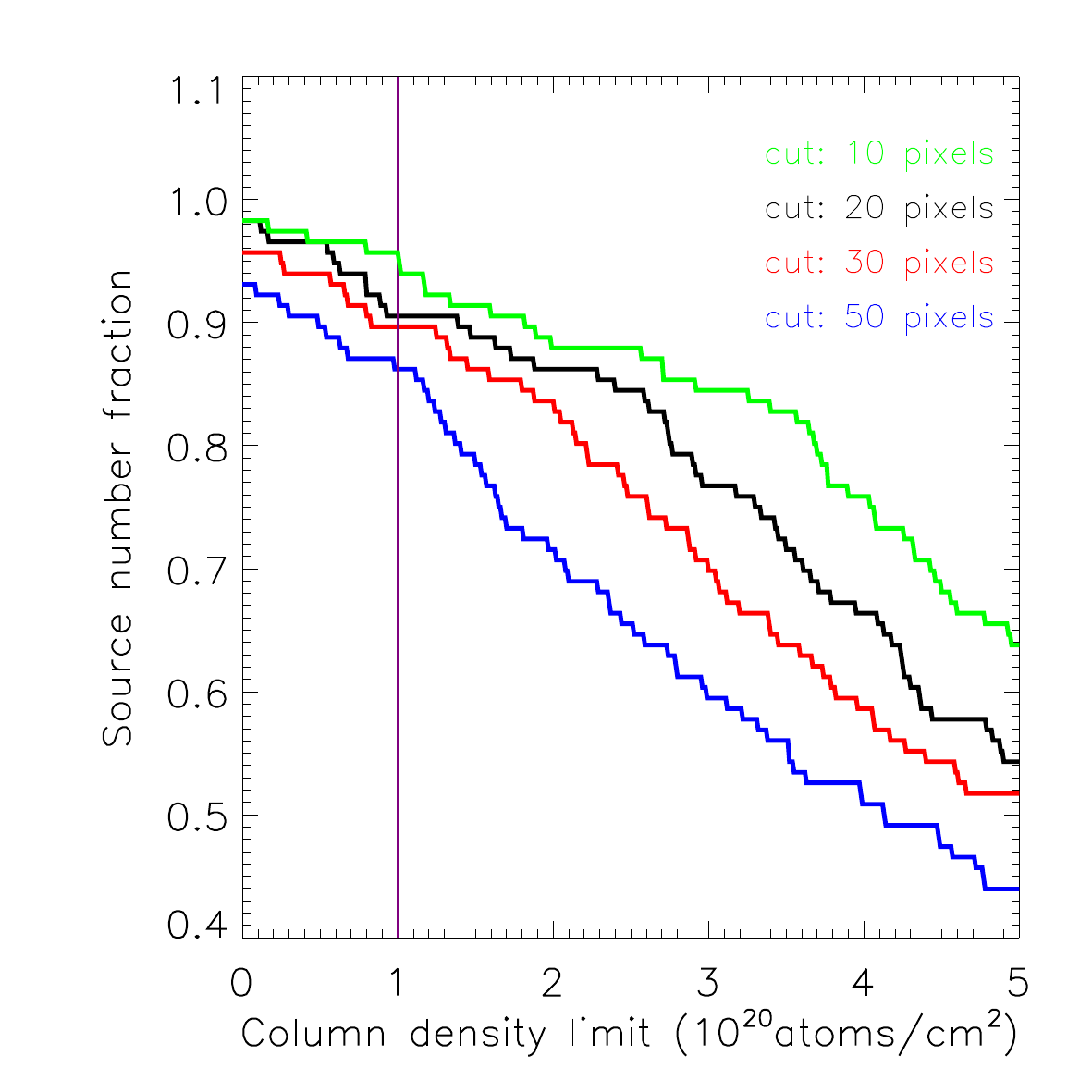}}
  \caption{  The left panel shows the SNR varies with applying different column density limits. 
The middle panel shows the flux fraction varies with increasing column density limit. The flux fraction is the ratio of
the total H{\sc{I}} flux above a specific column density limit to the H{\sc{I}} flux without
giving a column density limit.  The right panel shows the
column density limit versus the fraction of retained sources. The color
curves represent different pixel number cuts (If the effective pixel
number of the source is greater than the cut, it is treated as
a source).\label{fig2}}
\end{figure*} 

In Paper I, we describe how we build error maps for the H{\sc{I}} moment-0 images. 
We use these error maps to determine a H{\sc I} column density threshold
that reliably defines an outermost contour for  morphological analysis.
The outermost contour should also  contain most of the total flux of the
source.
The left panel of Fig. 2 shows the signal-to-noise ratio at different H{\sc I} column density levels
for all the detected sources.
The black curve shows the median signal-to-noise ratio,
 and the gray region shows the 20\% to 80\% percentile range. 
The SNR is typically above 3 when the threshold is set to be $10^{20}$ atoms cm$^{-2}$.

The middle panel of Fig. 2 shows how the fraction of the total flux of the galaxy 
varies as a function of column density limit. 
The back curve shows the median flux fraction for all sources, and 
the gray region shows the 20\% to 80\% percentile range.
\footnote {At a column density limit of zero, 
the H{\sc{I}} flux fraction is larger than 1.0 for all galaxies,
 because by clipping at that level, the noise contribution is neglected (i.e., some negative 
total flux from negative noise peaks has to be added to counterweight this effect at the edges of 
the detected sources.}
We see that most of the galaxies retain more than 90\% of their 
total flux when the limiting contour
level is $10^{20}$ atoms cm$^{-2}$.

We require sources to contain at least 20 pixels in the moment-0
maps to be included in our catalogues. In the right panel of Fig. 2, we show the fraction 
of sources that meet this criterion as a function of different density 
threshold cuts (the black curve). We also show how the curve changes if we 
change the resolving criteria to 10, 30 and 40 pixels.
We can see more than 90\% of the sources are resolved with 20 pixels at
a density detection threshold of $10^{20}$ atoms cm$^{-2}$.

Based on the analysis shown above, we adopt  $10^{20}$ atoms cm$^{-2}$
as our H{\sc I} column density threshold, because it is demonstrated to be  a good threshold for 
reliably describing the shape and size  of the H{\sc{I}} while  still retaining most of
the H{\sc{I}} mass, and it also includes most of the well-resolved H{\sc{I}} sources.

\subsection{Extracting sources with reliable H{\sc{I}} fluxes}

\begin{figure*}
  \includegraphics[width=0.45\textwidth]{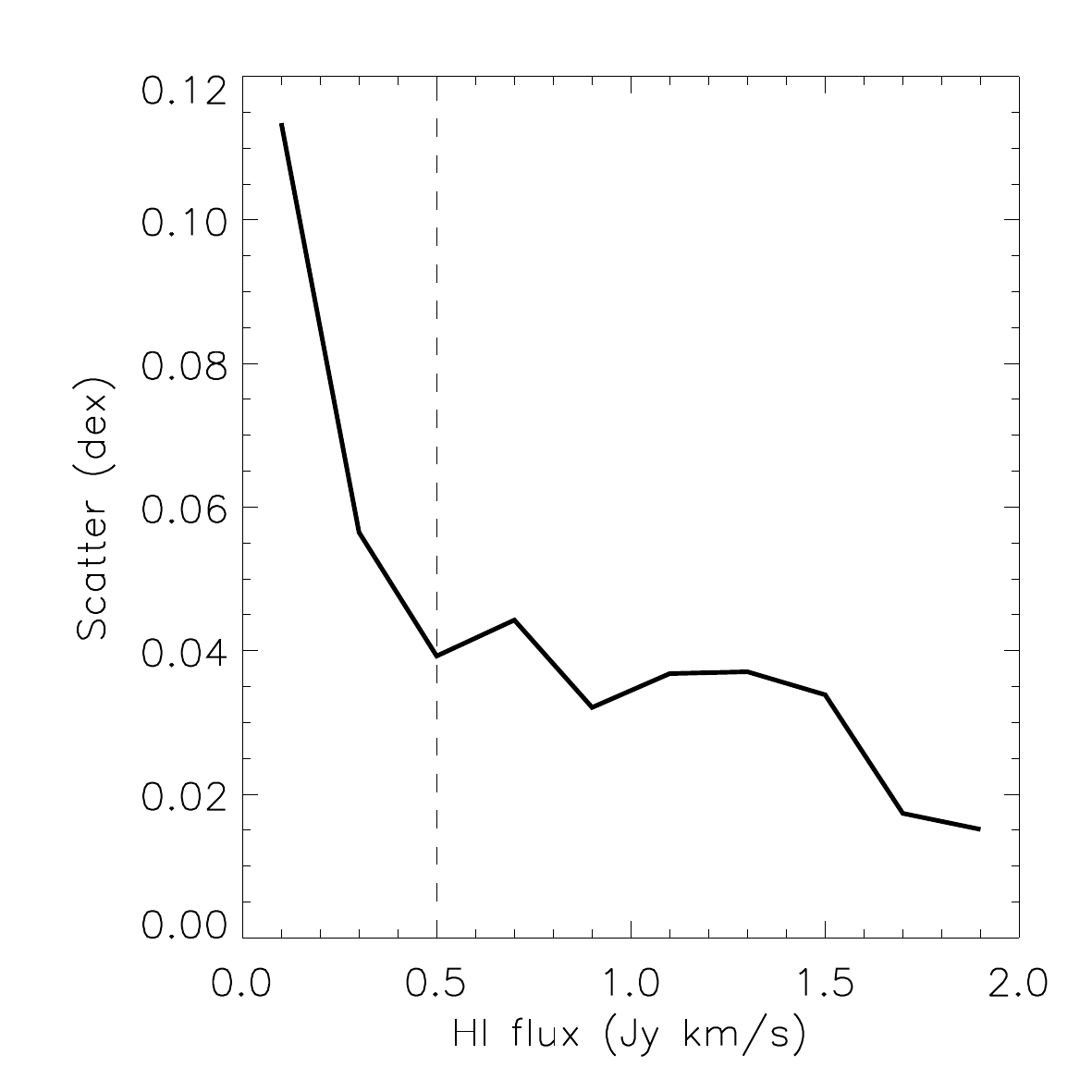}
  \includegraphics[width=0.45\textwidth]{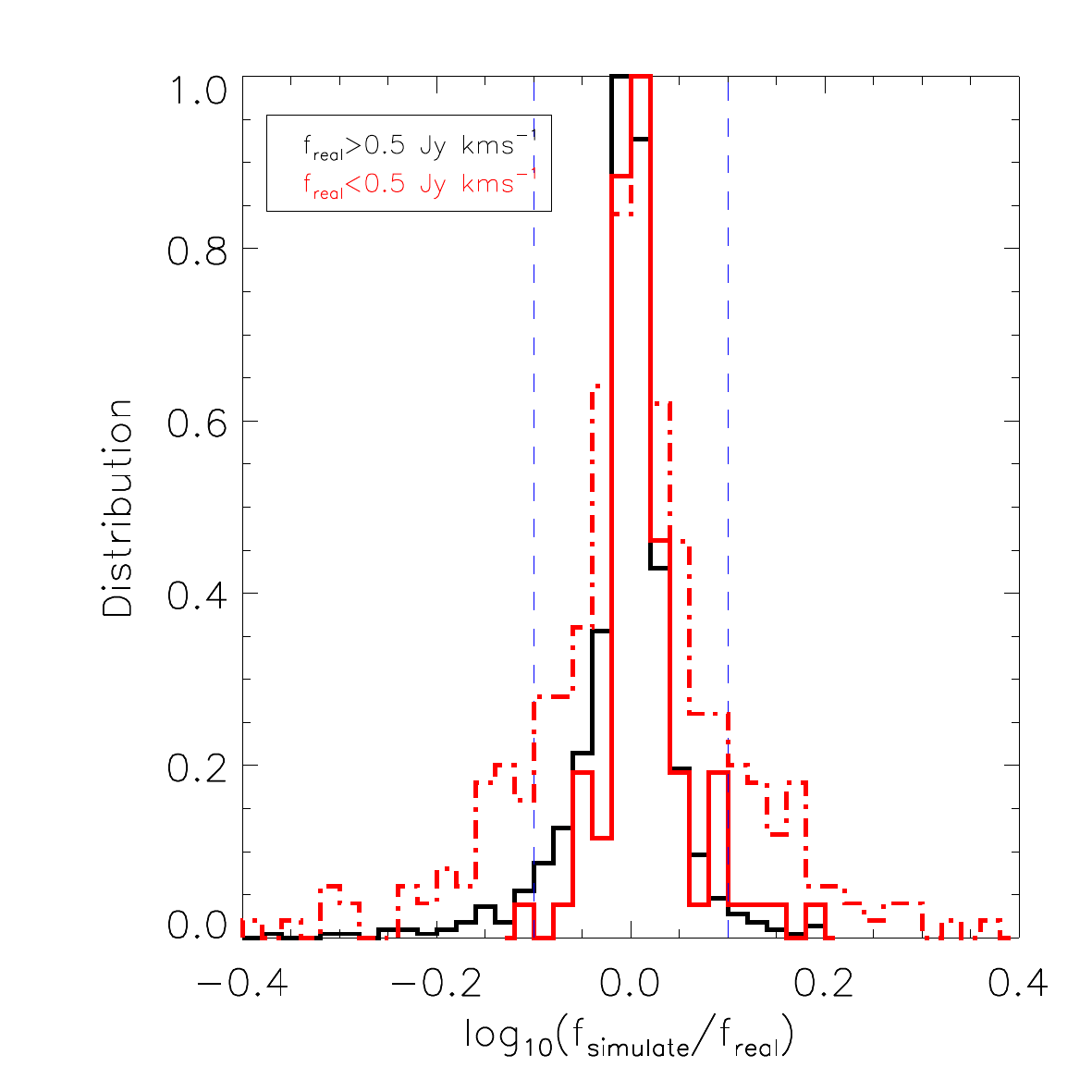}
  \caption{ The left panel shows the changing of H{\sc{I}} flux scatter with different H{\sc{I}} flux bins for
simulated sources. 
The right panel shows the distributions of H{\sc{I}} flux ratio for different H{\sc{I}} flux 
intervals for simulated sources. The black histogram represents the distribution of H{\sc{I}} flux ratio
 for H{\sc{I}} flux greater than 0.5 Jy km s$^{-1}$, and the red dot-and-dash histogram is for galaxies 
with H{\sc{I}} flux less than 0.5 Jy km s$^{-1}$. Besides, the red
histogram is also for galaxies with H{\sc{I}} flux less than 0.5 Jy km s$^{-1}$, but with SNR greater than
3. \label{fig3}}
\end{figure*}

To estimate the reliability of our determinations of the total H{\sc{I}} flux, we use a bootstrap method. We 
simulate a set of repeat observations by adding random noise with
the same characteristics as the noise in the Bluedisk cubes, and measure the
resulting variance if the  total H{\sc I} flux. 
In practice, we take the channels that have no detected sources from the data cube, and randomly repeat
them to make noise cubes that have the same size as the Bluedisk data cubes. We add the noise cubes to the original
Bluedisk cubes to make a ``perturbed'' cube. We repeat this process 10 times for each original Bluedisk
cube. 
In addition, we also generate error maps for these simulated cubes.

The left panel of Fig. 3 shows how the scatter in the derived H{\sc{I}} between the different perturbed cubes varies 
function of the original fluxes. The H{\sc{I}} flux scatter rises sharply for H{\sc{I}} fluxes 
less than 0.5 Jy km s$^{-1}$, and becomes almost flat at a value of about 0.04 dex for H{\sc{I}} fluxes greater than
0.5 Jy km s$^{-1}$.

The right panel of Fig. 3 shows the distribution of the ratios of simulated H{\sc{I}} fluxes and input H{\sc{I}} fluxes 
in two flux bins. The black histogram shows the ratio of the distribution for galaxies with H{\sc{I}} fluxes greater 
than 0.5 Jy km s$^{-1}$, and the red dot-and-dash histogram shows the distribution for galaxies with H{\sc{I}} fluxes less 
than 0.5 Jy km s$^{-1}$. The simulated H{\sc{I}} fluxes in the lower H{\sc{I}} flux bin are clearly more scattered than
in the higher H{\sc{I}} flux bin. We further select galaxies that have a SNR of at least 3 for the  
$10^{20}$ atoms cm$^{-2}$ H{\sc I} column density contour. The distribution of their flux ratios 
is shown as the red histogram in Fig. 3. 
We find it to be similar to that of galaxies with high H{\sc{I}} flux. 
In what follows,  we exclude low H{\sc{I}} flux sources (total H{\sc{I}} fluxes less than 0.5 Jy km s$^{-1}$) with
a SNR of less than 3 in the $10^{20}$ atoms cm$^{-2}$ contour. 
Our tests show that  this ensures that
the error in the   H{\sc{I}} fluxes we measured will be less than about 0.04 dex. 
 
There are 43 targeted galaxies and 65 additional galaxies 
in the final sample. Among the galaxies that were not targeted, 45 galaxies are resolved 
and 20 galaxies are unresolved. Among these additional galaxies, 15 galaxies are located in 
the ``peculiar cubes'' \citep{Wang-13}, in which the targeted galaxy is not detected 
in H{\sc{I}} or is otherwise disturbed. We exclude these galaxies 
in the analysis described below. We also divide our targeted galaxies into two parts according to
whether they lie above or below the H{\sc I}-plane defined in \cite{Catinella-10}. 
Hereafter, targeted galaxies
which lie above the H{\sc I}-plane are referred to H{\sc I}-rich galaxies (blue sample) 
and the cubes they reside in are referred as blue cubes. 
Similarly, the rest of targeted galaxies are referred as control sample and the rest of cubes are
referred as control cubes.
This definition is different to what was designed in observation, but it properly reflects
the actual H{\sc I} content of each system.
%H{\sc{I}} sources with no optical counterparts are also excluded. 
At last, 23 blue targeted galaxies, 20 control targeted galaxies, 26 untargeted galaxies in blue
cubes and 24 untargeted galaxies in control cubes are left.

The detailed properties of all the additional detected galaxies are listed in table 1.
Galaxies in ``peculiar cubes'' are also listed in this table, but not used for
 analysis in the following part of the paper.
As can be seen, they span stellar masses from $10^8-10^{11} {\rm M_{\odot}}$ and H{\sc I} masses from 
$10^{8.5}-10^{10.4} {\rm M_{\odot}}$.
Although our sample is small in size, it is unique in that it samples the environments
rare, very H{\sc I}-rich systems.
Most of the  galaxies are resolved in H{\sc{I}}, enabling us to 
study their resolved morphology and their kinematics. Compared to the Westerbork observations of neutral Hydrogen
in Irregular and SPiral galaxies (WHISP), our data are more sensitive.
\\
\begin{table*}
\begin{center}
\renewcommand{\arraystretch}{0.9}
\begin{tabular*}{1.0\textwidth}{c c c c c c c c c c c c c}
\hline \hline
 \multicolumn{1}{c}{ID} &\multicolumn{1}{c}{ra} & \multicolumn{1}{c}{dec} & \multicolumn{1}{c}{z} & \multicolumn{1}{c}{$\log M_*$} & \multicolumn{1}{c}{$\mu_*$} & \multicolumn{1}{c}{NUV-r}  & \multicolumn{1}{c}{$\log\,M_{\rm H{\sc{I}}}$} & \multicolumn{1}{c}{$Dis$}& \multicolumn{1}{c}{$R1$} & \multicolumn{1}{c}{$R_{\rm 50,H{\sc I}}$} & \multicolumn{1}{c}{$R_{\rm 90,H{\sc I}}$} & \multicolumn{1}{c}{$rs$} \\
\multicolumn{1}{c}{ } &\multicolumn{1}{c}{ } & \multicolumn{1}{c}{ } & \multicolumn{1}{c}{ } & \multicolumn{1}{c}{$\rm M_{\odot}$} & \multicolumn{1}{c}{$\rm M_{\odot} kpc^{-2}$} & \multicolumn{1}{c}{ }  & \multicolumn{1}{c}{$\rm M_{\odot}$} & \multicolumn{1}{c}{degree}& \multicolumn{1}{c}{arcsec} & \multicolumn{1}{c}{arcsec} & \multicolumn{1}{c}{arcsec} & \multicolumn{1}{c}{arcsec} \\
\hline
  14 &  112.31934 &   42.27963 &  0.02307 & 10.39 &  8.33 &  2.87 &  9.88 & 0.30 &   61 &   31 &   57 &   3.8 \\
  78 &  123.48852 &   52.64847 &  0.01820 & 11.01 &  8.99 &  3.80 & 10.40 & 0.22 &  140 &   63 &  132 &  16.4 \\
  84 &  123.03966 &   52.45524 &  0.01874 &  9.61 &  8.14 &  2.03 &  9.55 & 0.16 &   46 &   23 &   41 &     - \\
 143 &  127.30350 &   40.85448 &  0.02510 &  9.33 &  7.74 &  2.49 &  8.77 & 0.21 &   16 &   12 &   21 &  37.0 \\
 179 &  127.73492 &   55.83517 &  0.02542 &  8.91 &  7.53 &  2.56 &  8.73 & 0.39 &   15 &   11 &   21 &   6.1 \\
 221 &  129.17748 &   41.47231 &  0.02919 &  9.88 &  8.29 &  1.94 &  9.80 & 0.08 &   42 &   24 &   46 &  11.9 \\
 248 &  129.80326 &   30.92383 &  0.02569 &  8.51 &  7.10 &  2.03 &  9.19 & 0.19 &   23 &   17 &   31 &  15.4 \\
 306 &  132.05204 &   36.78074 &  0.02527 & 10.24 &  8.49 &  2.52 &  9.46 & 0.24 &   32 &   16 &   29 &   4.1 \\
 307 &  132.66660 &   36.46876 &  0.02521 &  6.98 &  6.98 &  1.51 &  9.94 & 0.35 &   57 &   25 &   55 &     - \\
 346 &  132.02870 &   41.85922 &  0.02997 & 10.11 &  7.99 &  2.64 &  9.53 & 0.26 &   32 &   21 &   36 &     - \\
 370 &  137.21930 &   44.93228 &  0.02657 & 10.27 &  9.31 &  3.62 &  8.96 & 0.13 &   14 &   14 &   43 &   7.9 \\
 375 &  137.33753 &   45.03975 &  0.02730 &  8.89 &  7.25 &  1.63 &  9.55 & 0.26 &   37 &   17 &   31 &   7.3 \\
 394 &  138.20927 &   40.49874 &  0.02759 &  9.60 &  8.03 &  2.17 &  9.73 & 0.41 &   37 &   18 &   37 &   7.4 \\
 396 &  138.39333 &   40.46574 &  0.02758 &  8.59 &  7.12 &  1.93 &  9.34 & 0.35 &   27 &   16 &   26 &   4.8 \\
 444 &  138.37436 &   51.31519 &  0.02773 &  8.80 &  7.25 &     - &  9.33 & 0.23 &   25 &   13 &   27 &  28.8 \\
 446 &  138.53751 &   51.41797 &  0.02805 &  9.51 &  8.59 &     - &  9.17 & 0.14 &   21 &   14 &   32 &   8.7 \\
 454 &  138.84323 &   51.05039 &  0.02881 &  9.68 &  7.78 &  2.41 &  9.45 & 0.32 &   31 &   19 &   28 &   4.3 \\
 482 &  139.35935 &   45.97174 &  0.02574 &  9.45 &  7.89 &  2.80 &  8.49 & 0.20 &    6 &   14 &   28 &  22.2 \\
 483 &  139.55919 &   45.65171 &  0.02690 & 10.70 &  8.47 &  2.87 & 10.16 & 0.30 &   68 &   35 &   61 &   5.0 \\
 517 &  139.90499 &   32.35320 &  0.02652 &  9.72 &  8.54 &  2.01 &  9.25 & 0.23 &   24 &   18 &   45 &  11.3 \\
 563 &  141.20293 &   49.39827 &  0.02723 &  9.73 &  8.27 &  3.30 &  8.72 & 0.24 &   15 &   10 &   19 &     - \\
 773 &  153.03446 &   46.29371 &  0.02425 & 10.34 &  8.48 &  2.63 & 10.03 & 0.40 &   47 &   34 &   66 &     - \\
 776 &  152.84976 &   45.73539 &  0.02375 &  8.69 &  7.20 &  1.60 &  8.88 & 0.23 &   19 &   12 &   21 &     - \\
 840 &  153.80716 &   56.60331 &  0.02667 &  8.82 &  7.43 &  1.24 &  9.41 & 0.07 &   28 &   17 &   33 &   8.9 \\
 889 &  154.25328 &   55.88005 &  0.02437 &  9.77 &  8.28 &  2.34 &  9.56 & 0.28 &   36 &   18 &   33 &   6.7 \\
 941 &  153.81151 &   58.69174 &  0.02295 &  8.61 &  7.73 &  1.80 &  9.12 & 0.29 &    6 &   29 &   49 &   4.5 \\
 983 &  162.50634 &   36.25677 &  0.02190 &  9.75 &  8.09 &  3.12 &  8.47 & 0.09 &    6 &   17 &   27 &     - \\
 997 &  162.75426 &   36.19258 &  0.02380 &  9.85 &  7.80 &  2.23 &  9.98 & 0.23 &   55 &   23 &   47 &     - \\
 999 &  162.49449 &   36.41499 &  0.02327 &  9.50 &  8.20 &  3.34 &  8.78 & 0.08 &   17 &   13 &   24 &   5.7 \\
1020 &  166.86889 &   35.46365 &  0.02828 &     - &     - &  3.05 &  9.87 & 0.10 &   37 &   26 &   58 &     - \\
1022 &  166.95902 &   35.40299 &  0.02840 & 10.47 &  8.81 &  3.49 &  9.56 & 0.07 &   36 &   18 &   32 &  18.0 \\
1024 &  166.95821 &   35.68483 &  0.02846 & 10.30 &  9.10 &  2.97 &  9.62 & 0.22 &   41 &   25 &   42 &   9.1 \\
1027 &  167.10516 &   35.33068 &  0.02888 &  8.96 &  7.33 &  3.81 &  8.94 & 0.16 &   18 &   13 &   25 &     - \\
1042 &  168.52972 &   34.30886 &  0.02526 &  8.70 &  6.97 &  1.91 &  9.09 & 0.16 &   25 &   18 &   31 &  17.1 \\
1104 &  177.57556 &   35.25409 &  0.02128 & 10.22 &  8.97 &  2.84 & 10.08 & 0.36 &   68 &   35 &   76 &     - \\
1143 &  185.61022 &   40.76168 &  0.02292 &  8.40 &  7.68 & -0.14 &  9.14 & 0.17 &   28 &   16 &   27 &  16.9 \\
1183 &  193.23760 &   51.82684 &  0.02762 &  8.58 &  7.78 &  1.50 &  8.90 & 0.20 &   16 &   11 &   22 &  23.2 \\
1230 &  196.76312 &   57.86514 &  0.02874 &  8.77 &  7.04 &  2.04 &  9.09 & 0.27 &   19 &   12 &   22 &   4.6 \\
1254 &  198.42330 &   47.29923 &  0.02808 &  8.80 &  7.20 &  1.22 &  9.19 & 0.20 &   23 &   15 &   25 &   6.3 \\
1259 &  198.26018 &   47.34518 &  0.02841 &  8.76 &  7.28 &  1.79 &  9.02 & 0.11 &   17 &   14 &   30 &  10.4 \\
1261 &  197.86026 &   47.44052 &  0.02861 &  9.03 &  7.29 &  2.46 &  8.92 & 0.25 &   15 &    9 &   15 &     - \\
1296 &  198.84572 &   35.17351 &  0.02309 &  8.90 &  8.12 &  1.82 &  8.98 & 0.19 &   19 &   12 &   22 &   7.1 \\
1298 &  198.67004 &   35.03638 &  0.02374 &  8.67 &  7.46 &  1.61 &  9.21 & 0.28 &   22 &   14 &   29 &  13.2 \\
1323 &  203.28755 &   40.85307 &  0.02400 &  8.88 &  7.70 &  1.24 &  8.97 & 0.33 &   10 &   22 &   33 &   3.7 \\
1407 &  212.50986 &   38.70812 &  0.02576 &  9.86 &  8.31 &  3.81 &  9.67 & 0.21 &   34 &   28 &   52 &     - \\
1410 &  212.67758 &   38.71842 &  0.02561 &  8.57 &  7.42 &  1.30 &  9.26 & 0.18 &   23 &   19 &   47 &     - \\
1411 &  212.62710 &   38.73950 &  0.02606 &  8.95 &  7.86 &  1.23 &  9.44 & 0.15 &   29 &   16 &   31 &  12.5 \\
1414 &  212.69582 &   38.75970 &  0.02570 &  8.74 &  7.46 &  1.72 &  8.81 & 0.14 &   15 &   11 &   25 &     - \\
1415 &  212.72923 &   38.78592 &  0.02580 &  8.43 &  7.14 &  1.54 &  9.13 & 0.13 &   14 &   15 &   31 &   7.2 \\
1533 &  241.43411 &   36.27508 &  0.03123 &  9.40 &  7.50 &  1.74 &  9.79 & 0.30 &   44 &   23 &   38 &     - \\
1554 &  242.24289 &   36.61088 &  0.03014 & 11.02 &  8.83 &  3.33 & 10.42 & 0.31 &   26 &   50 &  104 &  26.6 \\
1605 &  246.16476 &   41.01995 &  0.02712 &  8.79 &  7.33 &  0.66 &  9.70 & 0.10 &   33 &   15 &   31 &   6.1 \\
1632 &  246.06271 &   41.11062 &  0.02926 &  9.97 &  8.21 &  3.27 &  9.70 & 0.22 &   38 &   17 &   33 &   9.6 \\
1639 &  246.70860 &   40.91786 &  0.02915 &  8.93 &  7.68 &  1.83 &  9.21 & 0.34 &   21 &   29 &   33 &   9.0 \\
1643 &  246.13086 &   40.68445 &  0.02943 &  8.80 &  7.49 &  1.10 &  9.06 & 0.28 &   20 &   18 &   36 &     - \\
1647 &  246.08427 &   40.86501 &  0.03000 &  9.34 &  7.50 &  2.46 &  9.13 & 0.16 &   17 &   12 &   44 &     - \\
1648 &  245.99846 &   41.13817 &  0.03029 &  9.08 &  7.39 &  1.34 &  9.59 & 0.27 &   34 &   17 &   33 &  20.6 \\
1663 &  250.62296 &   42.26285 &  0.02750 &  8.78 &  7.27 &  1.27 &  9.08 & 0.14 &   20 &   16 &   32 &  15.6 \\
1669 &  250.56263 &   42.05996 &  0.02780 &  8.82 &  7.68 &  1.96 &  9.17 & 0.22 &   20 &   18 &   50 &  19.7 \\
1700 &  251.48897 &   39.98583 &  0.03020 & 10.17 &  8.22 &  3.18 &  9.69 & 0.36 &    5 &   20 &   35 &   7.2 \\
1713 &  251.87396 &   40.56570 &  0.03108 &  8.92 &  7.44 &  1.39 &  9.68 & 0.32 &   37 &   19 &   32 &     - \\
1725 &  258.80617 &   30.45146 &  0.02864 &  9.98 &  8.17 &  2.56 &  9.66 & 0.31 &   39 &   20 &   34 &     - \\
1774 &  259.68221 &   58.13513 &  0.02908 & 10.95 &  8.85 &  3.60 &  9.98 & 0.39 &   55 &   26 &   48 &  13.5 \\
1795 &  258.77994 &   58.24045 &  0.03101 & 10.20 &  8.35 &  3.08 &  9.78 & 0.26 &   40 &   18 &   38 &     - \\
1907 &  262.17343 &   57.11476 &  0.02812 &     - &     - & -0.42 &  9.03 & 0.03 &   17 &   11 &   23 &  16.7 \\
\hline \hline
\end{tabular*}
\caption{Properties of additional galaxies extracted in 
the data cubes of the  Bluedisk project. The detailed information
of the targeted galaxies is presented in Paper I.
 From left to right, columns represent galaxy ID, Right Ascension, 
Declination, redshift, base-10 logarithm of the
 stellar mass, base-10 logarithm of the stellar surface density, 
NUV--r color, base-10 logarithm of the H{\sc{I}} mass, distance to the 
central of the data cube (in degree), $R1$, $R_{\rm 50,H{\sc{I}}}$,
$R_{\rm 90,H{\sc{I}}}$ and exponentially scale length rs, 
respectively. }
\label{table}
\end{center}
\end{table*}

%%%%%%%%%%%%%%%%%%%%%%%%%%%%%%%%%%%%%%%%%%%%%%%%%%%%%%%%%%%%%%%%%%%%%%%%%%%%%%%%%%%%%%%%
%%%%%%%%%%%%%%%%%%%%%%%%%%%%%%%%%%%%%%%%%%%%%%%%%%%%%%%%%%%%%%%%%%%%%%555555%%%%%%%%%%%%

\section{Results}

\begin{figure}
  \includegraphics[width=0.45\textwidth]{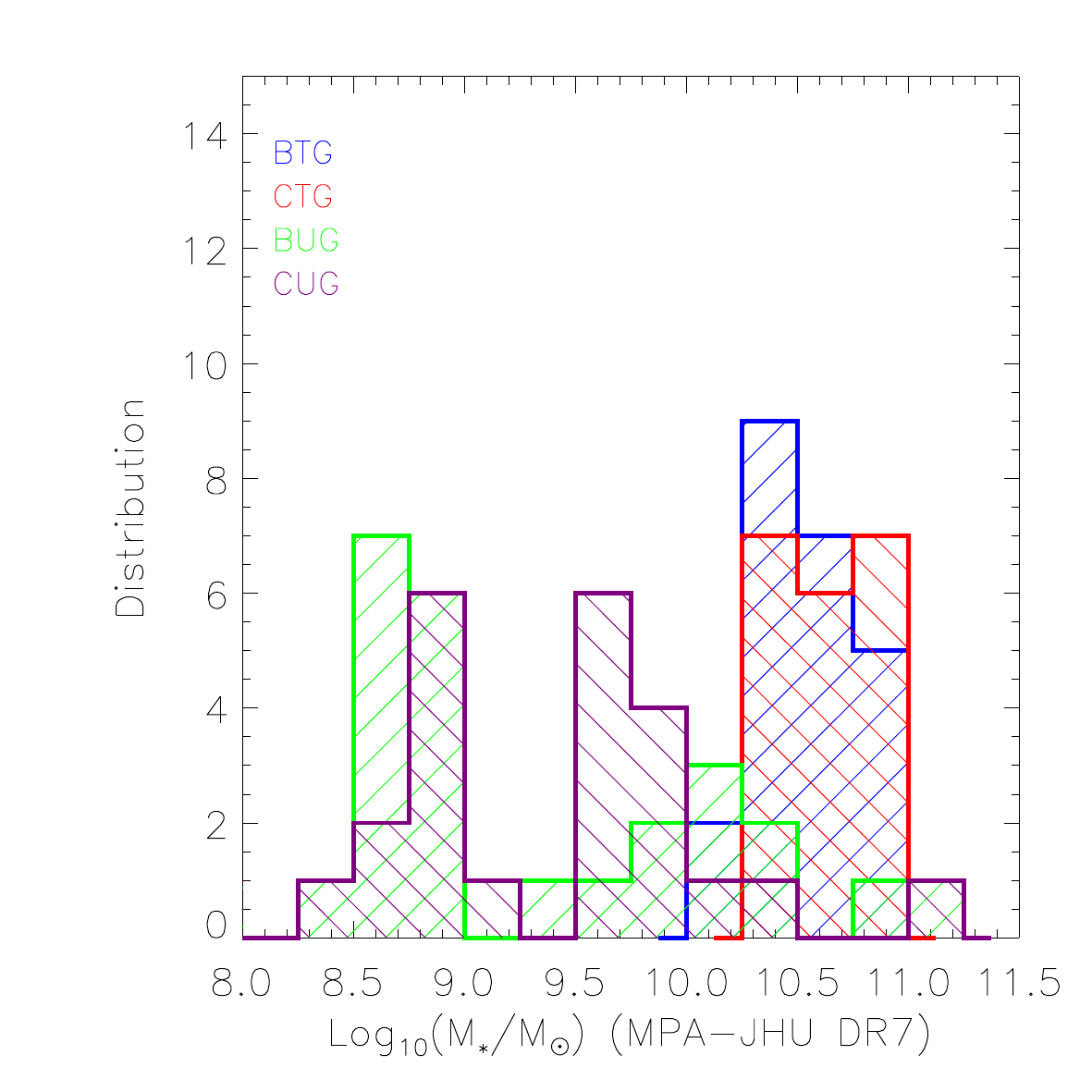}
  \caption{ Stellar mass distribution of our sample.
The stellar mass is from the MPA-JHU DR7 catalogue. Our sample is
divided into BTG (blue), CTG (red),
BUG (green) and CUG (purple).\label{fig4}}
\end{figure}

The mass distribution for our final sample is shown in Fig. 4. The 
sample is divided into four subsamples: the blue targeted galaxies (BTG)
, the control targeted galaxies (CTG), the additional 
galaxies located in blue cubes (BUG) and the additional  galaxies located in control cubes (CUG). 
By selection, all targeted galaxies are massive with $\log M_{*}/M_{\odot}>10$.
The additional galaxies have a very 
broad stellar mass distribution from $10^8 \rm M_{\odot}$ to $10^{11} \rm M_{\odot}$. 
The goal of this paper is to study their H{\sc{I}} properties. 

\subsection{H{\sc{I}} mass-size relation}
\begin{figure*}
\centerline{
\includegraphics[width=0.45\textwidth]{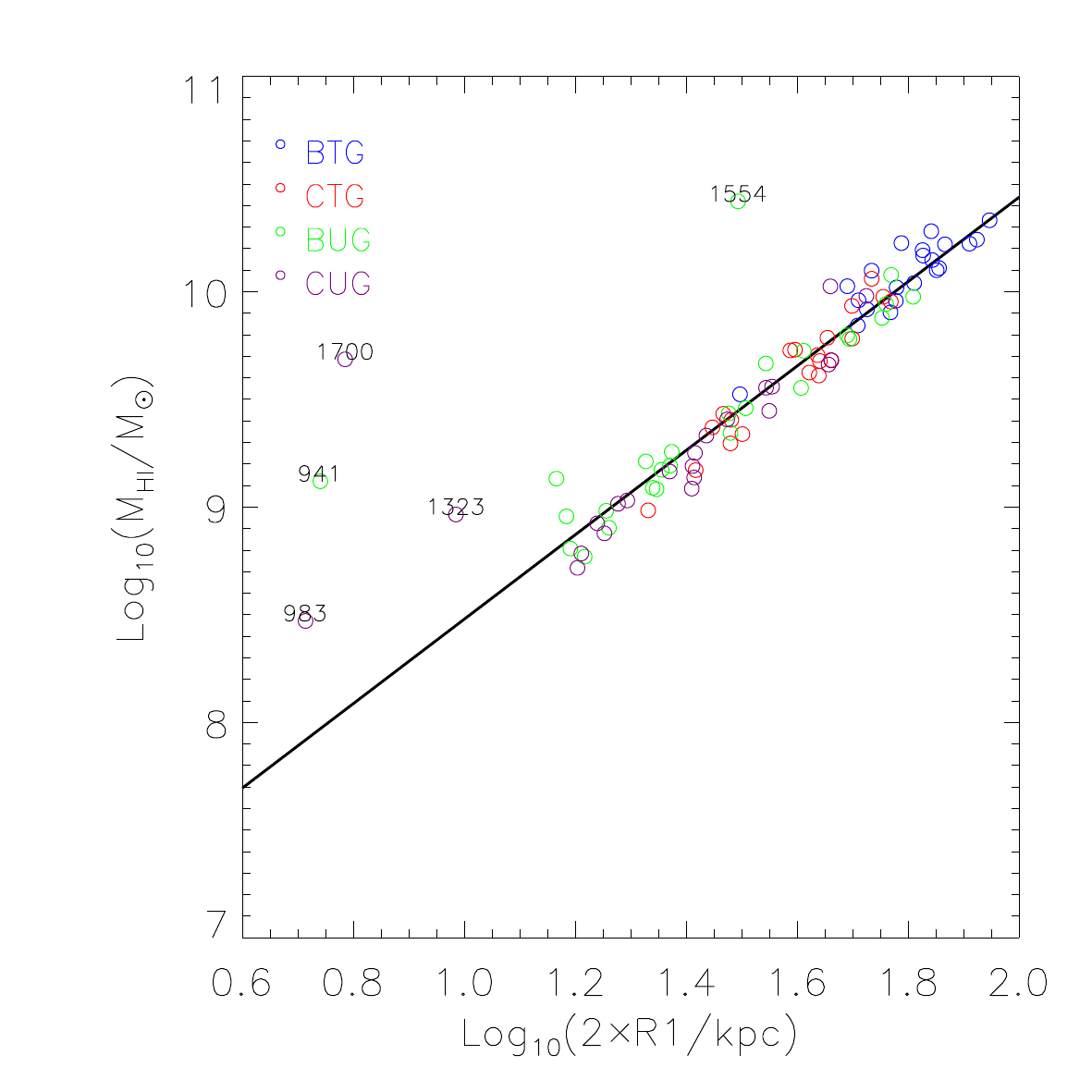}
\includegraphics[width=0.45\textwidth]{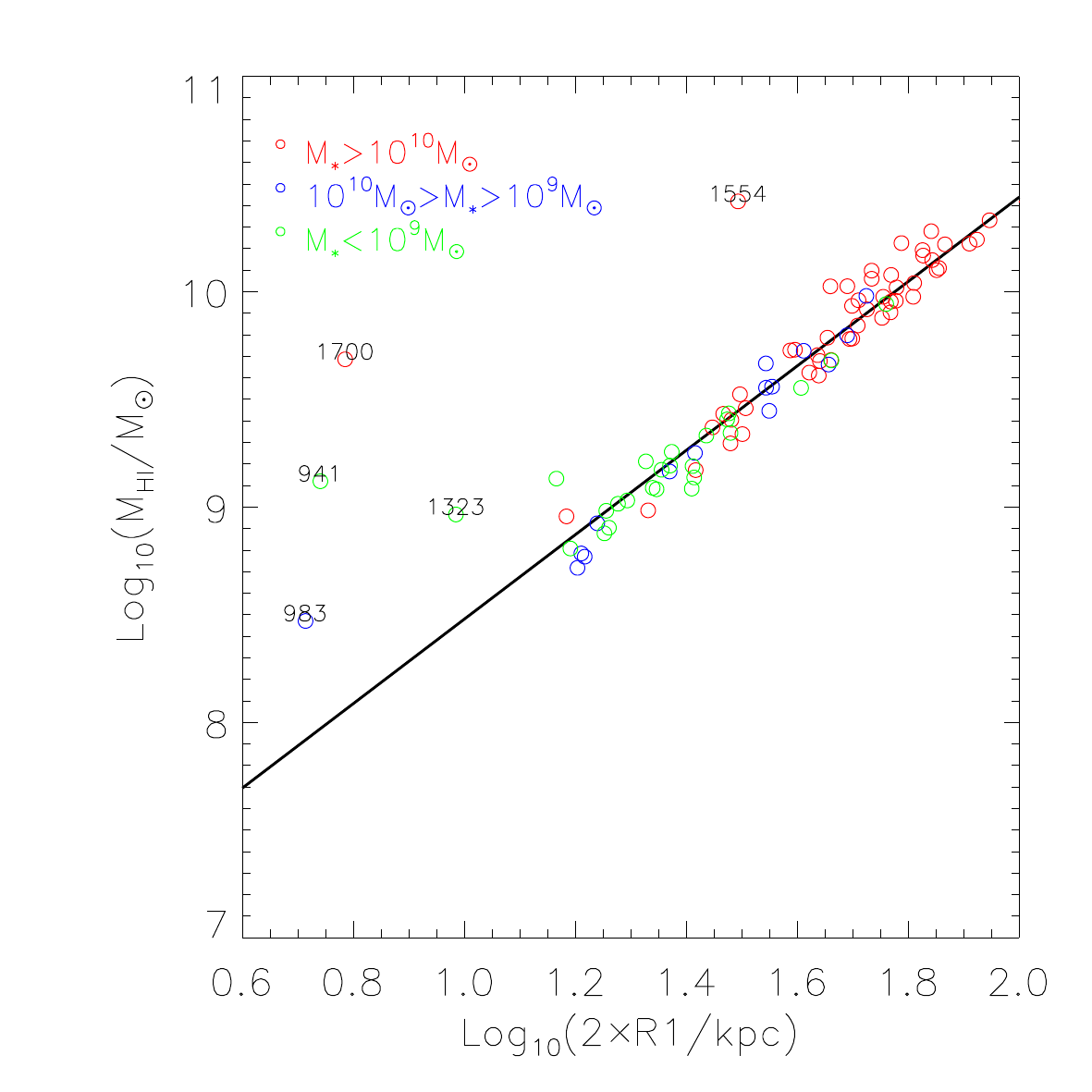} }
\caption{ The H{\sc{I}} mass-size relation for our sample.
The left panel shows the mass-size relation for BTG (blue), CTG (red), BUG (green) and CUG (purple).
The right panel is similar to left one but for different stellar
mass ranges. Some outliers in H{\sc{I}} mass-size relation are
marked by their IDs (these galaxies are discussed in detail in section 5).}
\label{fig5}
\end{figure*}

The tight relation between the diameter of the H{\sc{I}} disk and the total 
H{\sc{I}} mass in galaxies was investigated by \cite{Broeils-Rhee-97}. Later, it was found that 
this relation changes very little for different kinds 
of galaxies \citep{Swaters-02, Noordermeer-05}. Fig. 5 
shows the H{\sc{I}} mass-size relation for our sample. The left panel shows 
this relation for the four subsamples: BTG (blue), 
CTG (red), BUG (green) and CUG (purple). The black line is from 
\cite{Broeils-Rhee-97}. The right panel shows the same relation for different 
stellar mass ranges. Some outliers far from this relation are 
marked by their IDs. 

Almost all the galaxies closely
obey this relation, which is consistent with the result of Paper I.
This relation is similarly tight for galaxies with both high and low stellar masses:
the average column density of H{\sc{I}} is about same for all spiral galaxies.
\cite{Wang-14} suggest that this relation is primarily a result of atomic-to-molecular
gas conversion in the inner disk and is further enhanced by a homogeneous radial distribution
of H{\sc{I}} density in the outer disk of galaxies. We will investigate the nature
of the handful of outliers later in the paper.

\subsection{H{\sc{I}}-optical scaling relations}

\begin{figure*}
\includegraphics[width=1.0\textwidth]{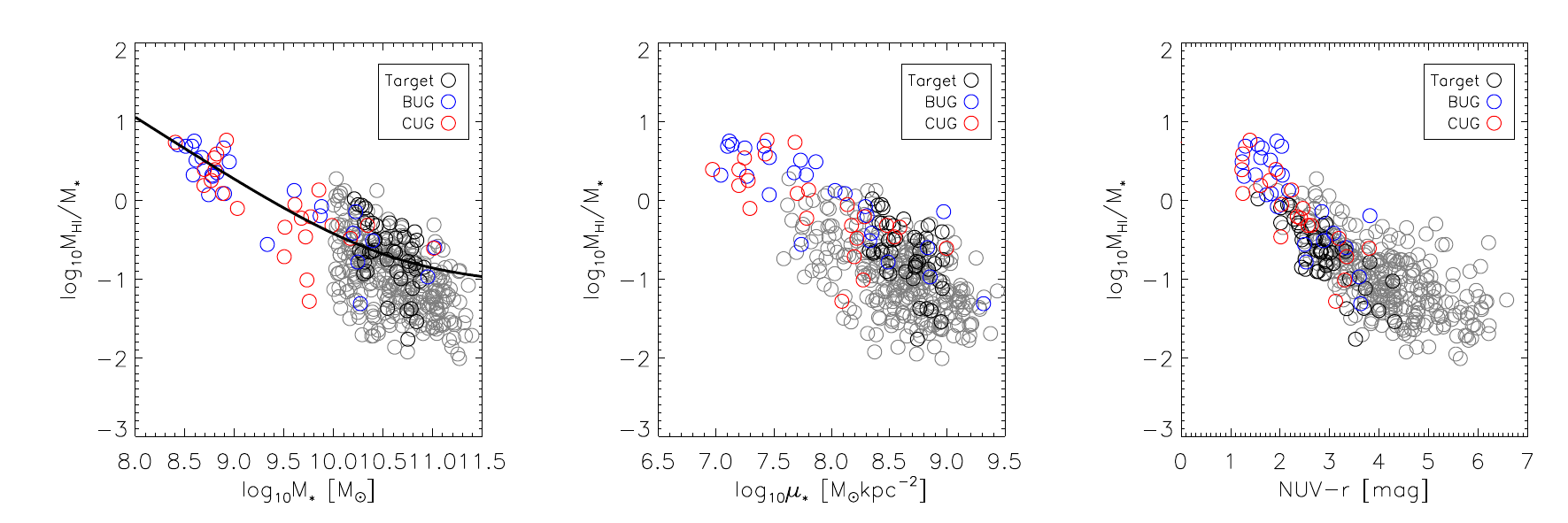}
\caption{The H{\sc{I}} mass fraction plotted as a function of stellar mass, 
stellar surface mass density and NUV-r for target
galaxies(black), other galaxies in the H{\sc I}-rich cubes (blue),
other galaxies in the  control cubes (red) and GASS sample galaxies (gray).
The black curve in the left panel shows the empirical relation of H{\sc{I}} 
mass fraction as a function of stellar mass,
which is taken from \citet{Evoli-11}.}
\label{fig6}
\end{figure*}

\begin{figure*}
\centerline{
\includegraphics[width=0.45\textwidth]{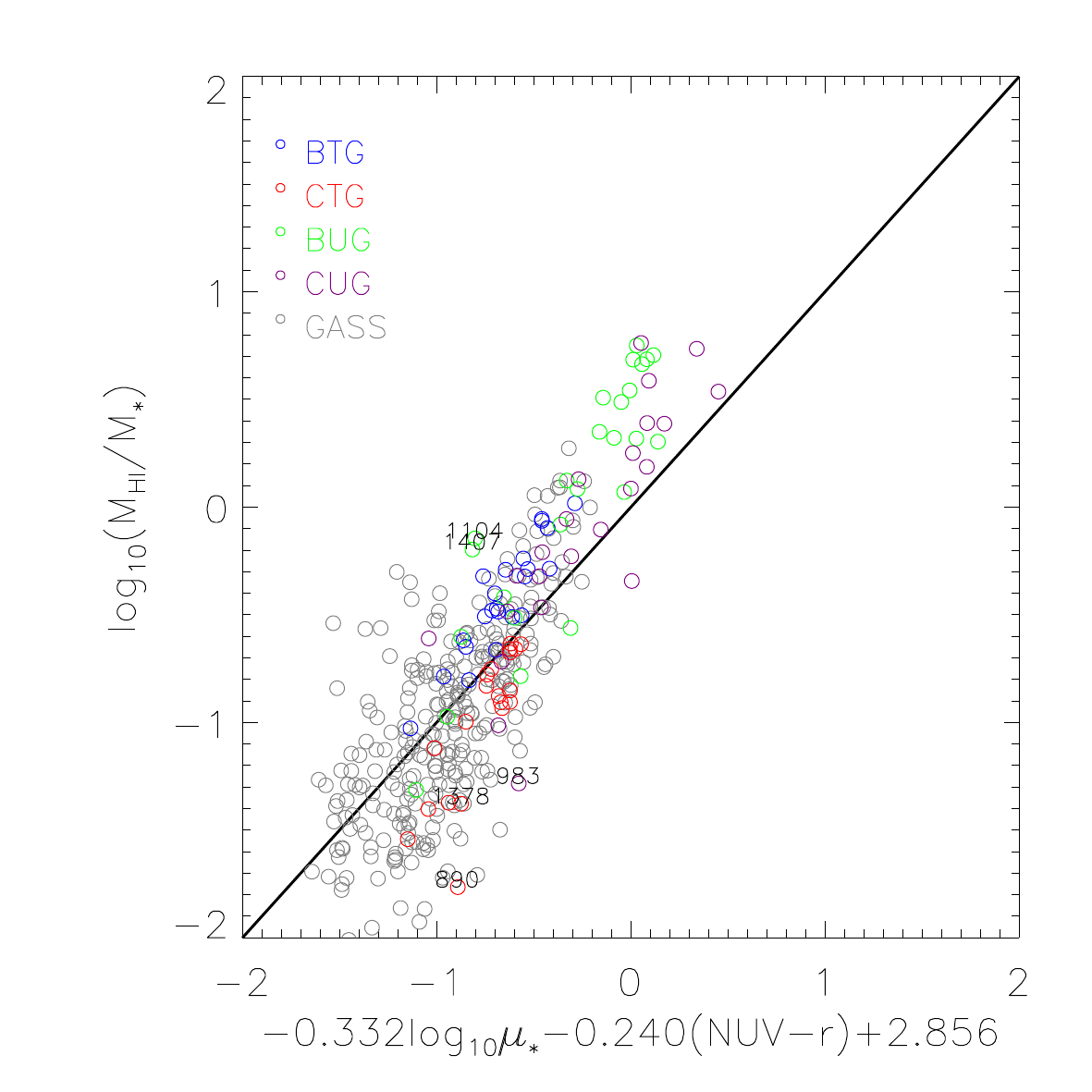}
\includegraphics[width=0.45\textwidth]{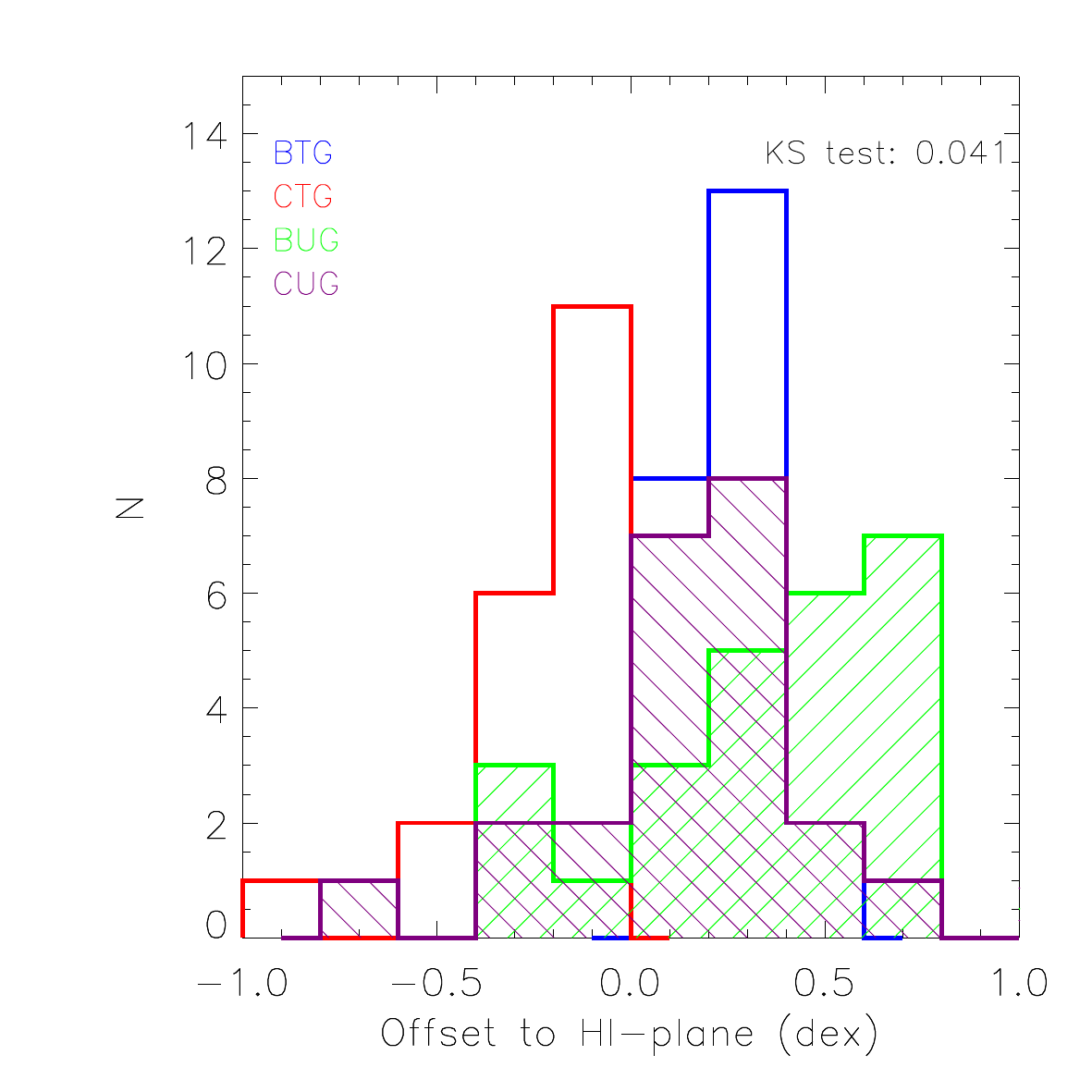} }
  \caption{
Left panel: H{\sc{I}}-plane for our sample.
This panel illustrates that the ratio of H{\sc{I}} mass to stellar
mass versus the combination of surface mass density and NUV-r.
Galaxies in the BTG (blue), CTG (red), BUT (green), CUG (purple) samples are colour-coded.
The gray data points show  the GASS sample. Right panel: The distribution of
deviations from the  H{\sc{I}}-plane (corresponding to the black line
in the left panel) for our sample. The blue, red, green and purple distributions represent
for the distributions of BTG, CTG, BUG and CUG galaxies, respectively.
The K-S test probability for the difference in deviation from the H{\sc I}-plane 
between the BUG and CUG samples is denoted at the top-right corner of the right panel.}
  \label{fig7}
\end{figure*}

\begin{figure*}
\includegraphics[width=1.0\textwidth]{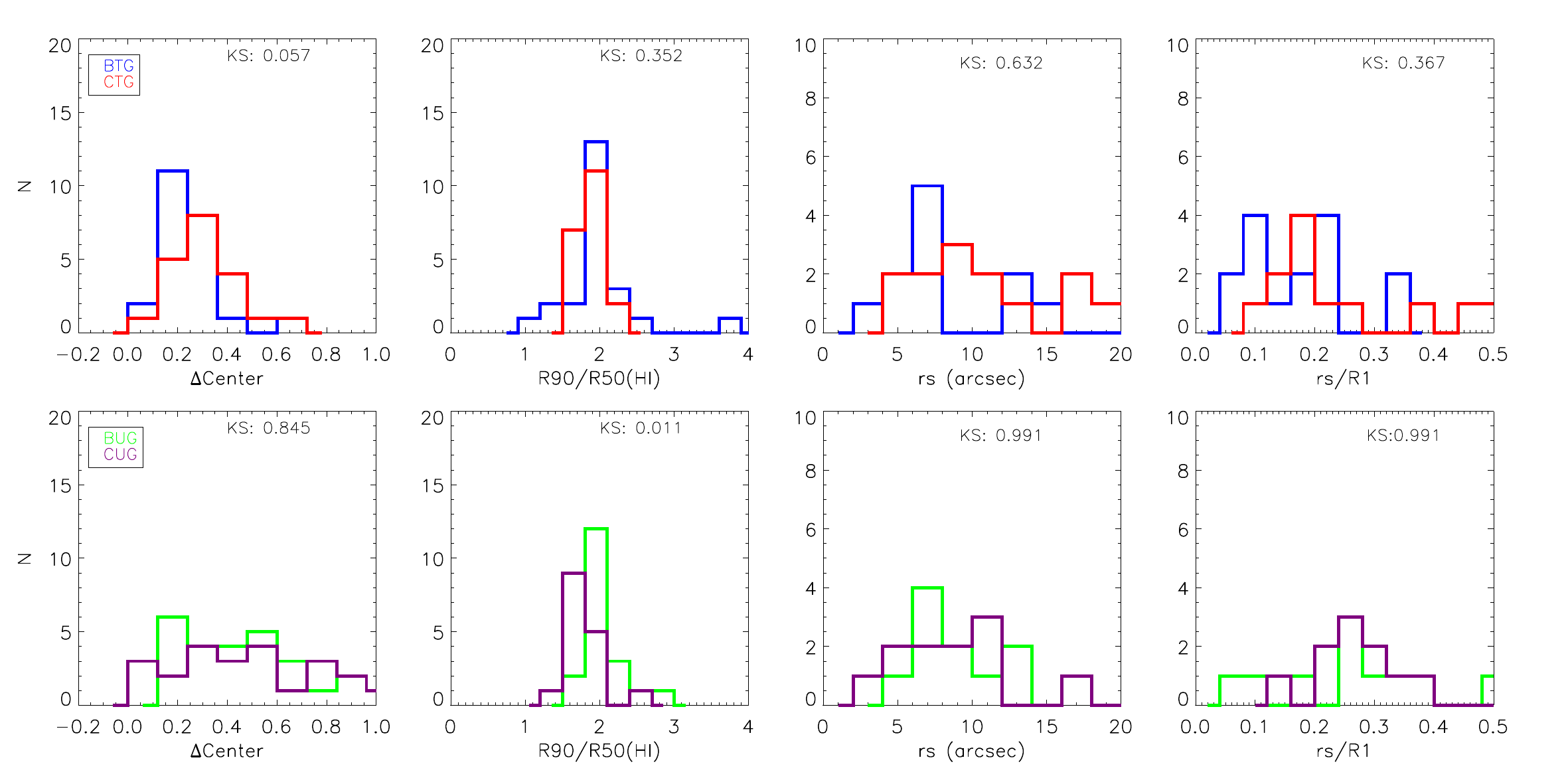}
\caption{The distributions of $\Delta$center, $R_{\rm 90,H{\sc{I}}}/R_{\rm 50,H{\sc{I}}}$, $rs$ and $rs/R1$ for BTG (blue), 
CTG (red), BUG(green) and CUG(purple). 
The K-S test probability that the H{\sc I}-rich and control cube 
distributions are drawn from the same underlying distribution
is given in the top right corner of each panel.}
\label{fig8}
\end{figure*}

%\begin{figure}
%\epsfig{figure=./plots/plothist_con.ps,width=0.45\textwidth}
%\caption{The distribution of deviations from the H{\sc I}-plane for untargeted galaxies with R90/R50(H{\sc I}) greater than
%2.0. The median value of these deviations is 0.44 dex.}
%\label{fig9}
%\end{figure}
 
In Fig. 6, we plot the H{\sc{I}}-to-stellar mass ratio as a function of stellar mass, 
stellar surface mass density and NUV-r color for our sample.
The H{\sc{I}} mass fraction correlates with stellar mass, surface mass density and NUV-r, and extends the
mean relations quantified for normal galaxies with stellar masses
greater than $10^{10} \rm M_{\odot}$ in former studies \citep[]{Catinella-10,Cortese-11,Huang-12}. 
We do not see  different H{\sc{I}} scaling relations for  galaxies in the data cubes with
a central targeted  H{\sc{I}} excess galaxy and data cubes with a central targeted control galaxy.

We plot the H{\sc{I}} plane linking  H{\sc{I}}-to-stellar mass ratio with
stellar surface mass density and NUV-r colour defined by Catinella et al. (2010) in the 
left panel of Fig. 7. The gray data points represent galaxies from  
GASS \citep{Catinella-10}. In this projection,  both
targeted and untargeted galaxies lie close to the H{\sc{I}} plane.
Differences arise  when we focus on the {\em displacement} from the plane. 
As already found in Paper I, we observe
a tilt of our observed galaxies with respect to the diagonal of the plot. The  plane underpredicts the
H{\sc{I}} content of all observed galaxies at the high-H{\sc{I}}-fraction end 
and slightly over-predicts it at the low-H{\sc{I}}-fraction end of
our sample. The underprediction at the high-H{\sc{I}}-fraction end is discussed in detail by \cite{Li-12}, who argue
that the addition of additional parameters yield a more  unbiased prediction of H{\sc I} content. 

The right panel of Fig. 7 shows the distributions
of the deviations from the H{\sc{I}}-plane (the black line in Fig. 7).  
We find a significant difference between the distributions of the galaxies in the
cubes of the H{\sc I}-rich targets and the galaxies in the control cubes.
We perform a Kolmogorov-Smirnov (K-S) test to quantify this significance, 
and obtain a probability of 0.041, suggesting a 96\% significance for the
 null hypothesis of the two samples being drawn from the same parent 
distribution to be rejected.  
It appears that on average, galaxies in the data cubes with an H{\sc{I}} excess
targeted galaxy do on average contain more H{\sc{I}} (relative to their
surface mass density and colour ) than galaxies
in the cubes containing a control galaxy.
 This suggests that galaxies associated with blue targeted galaxies are also likely to be gas-rich. 
This supports the observation pointed out by \cite{Kauffmann-Li-Heckman-10},
see also \cite{Weinmann-06}, who found that there is 
more photometrically estimated H{\sc{I}} in satellites around more star-forming primary galaxies than 
in satellites around less star-forming primary galaxies. In the following, we refer to this as ``H{\sc{I}} conformity''.
 Those authors argued that the satellites trace a large-scale
gas reservoir that is accreted onto the central galaxies.

Next, we repeat the morphological analysis of Paper I for the additional galaxies detected in the cubes. 
In Fig. 8, we plot  distributions of $\Delta$Center, $R_{\rm 90,H{\sc I}}/R_{\rm 50,H{\sc I}}$, $rs$ and $rs/R1$.
$\Delta$Center is calculated as the distance between the H{\sc{I}} center and the r-band center,
 normalized by the semi-major axis of the H{\sc{I}} ellipse. 
$R_{\rm 90,H{\sc I}}/R_{\rm 50,H{\sc I}}$ describes the concentration/extension of the H{\sc{I}} disks. 
$rs$ is the exponentially scale length of H{\sc{I}} disks.

The galaxies in the H{\sc I}-rich cubes do not differ from those in the control cubes in their distribution of 
$\Delta$Center. However, their distributions of $R_{\rm 90,H{\sc I}}/R_{\rm 50,H{\sc I}}$ do show a significant difference.  
Galaxies in the H{\sc I}-rich cubes
 tend to have larger $R_{\rm 90,H{\sc I}}/R_{\rm 50,H{\sc I}}$ than in the control cubes.  
If we investigate galaxies with $R_{\rm 90,H{\sc I}}/R_{\rm 50,H{\sc I}}$ greater than 2.0, we find almost all of them have more H{\sc{I}} gas than
the predicted by their optical properties. The distribution of the deviations from the H{\sc I}-plane for these galaxies
 is shown in Fig. 9. The median value of the deviations is 0.44 dex, which means these galaxies have 2.7 
 times H{\sc I} of the predicted values in general.
 We also find their H{\sc{I}} to extend far beyond the optical disk.
The $rs$ and $rs/R1$ distributions of the galaxies in the H{\sc I}-rich cubes do not differ significantly from 
those for galaxies in the control cubes.

To assess whether the density of the environment may play a role in these
differences, we checked the number of SDSS 
spectroscopically observed galaxies of which fall in our cubes, 
but which have no H{\sc{I}} detections. 
There are 95 galaxies with no H{\sc{I}} detections in the H{\sc I}-rich cubes and 295 galaxies with no H{\sc{I}} detections 
in the control cubes. Most of these galaxies have stellar masses in the 
range between $10^8 \rm M_{\odot}$ and $10^{10} \rm M_{\odot}$.
We also quantify the environment by using the 3-d reconstructed matter overdensity 
based on SDSS DR7 data set \citep{Jasche-10},
 which is defined as $\delta=(\rho-\bar{\rho})/\bar{\rho}$, where $\rho$ is the matter density and $\bar{\rho}$
is the mean matter density. Thus the mean overdensity of galaxies in control cubes is
10.7, while the mean over-density of galaxies in blue cubes is 4.6.
It is clear that the control cubes are located in denser regions than the H{\sc I}-rich
cubes, and the fraction of optically-identified galaxies that have detectable H{\sc{I}} masses is much higher in the blue
cubes than in the control cubes.
This may indicate that  H{\sc{I}} conformity is closely related to the environment of the galaxy.

\begin{figure}
\includegraphics[width=0.45\textwidth]{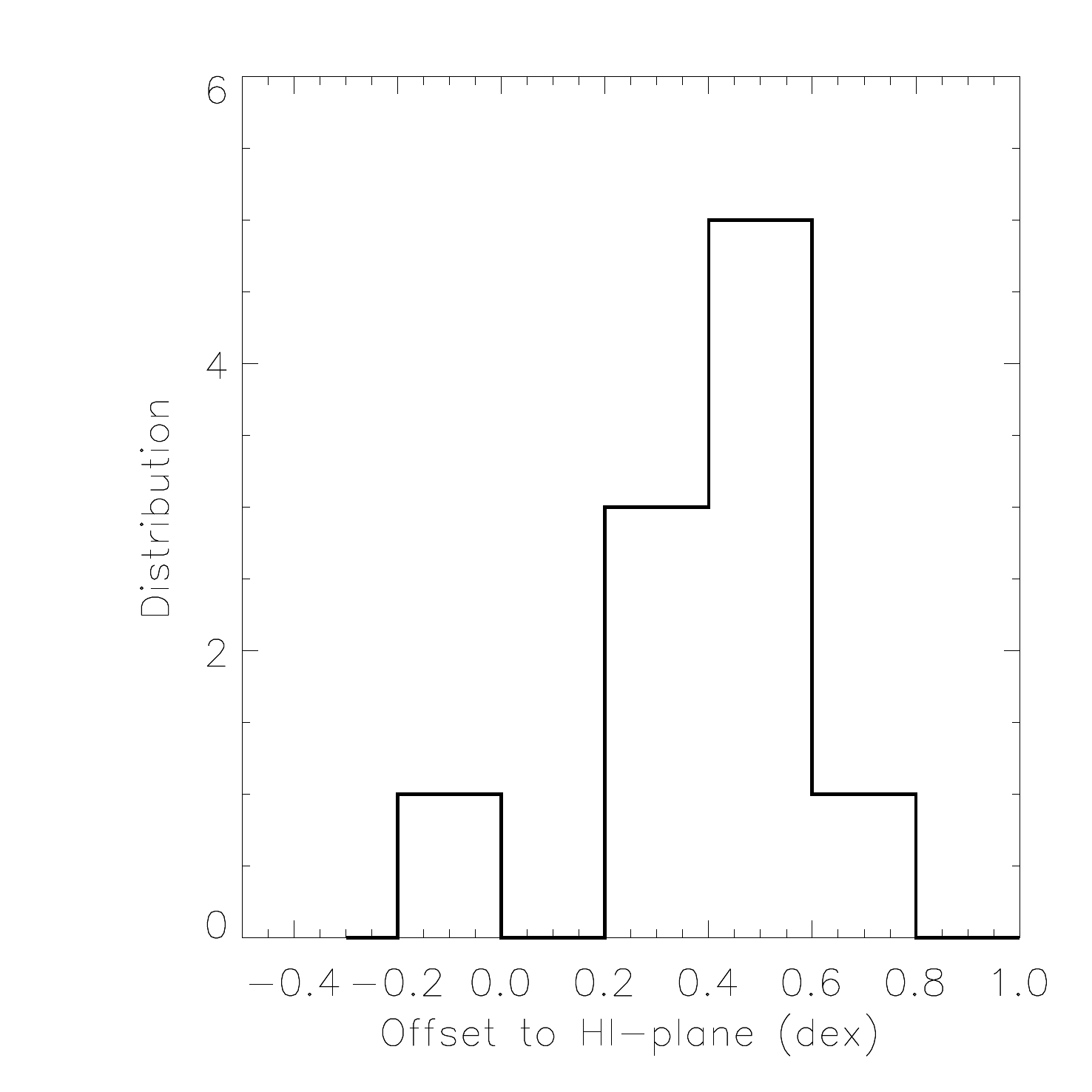}
\caption{The distribution of deviations from the H{\sc I}-plane for untargeted galaxies with $R_{\rm 90,H{\sc I}}/R_{\rm 50,H{\sc I}}$ greater than
2.0. The median value of these deviations is 0.44 dex.}
\label{fig9}
\end{figure}

\section{Morphology of outliers}

\begin{figure*}
\mbox{
\includegraphics[width=0.25\textwidth]{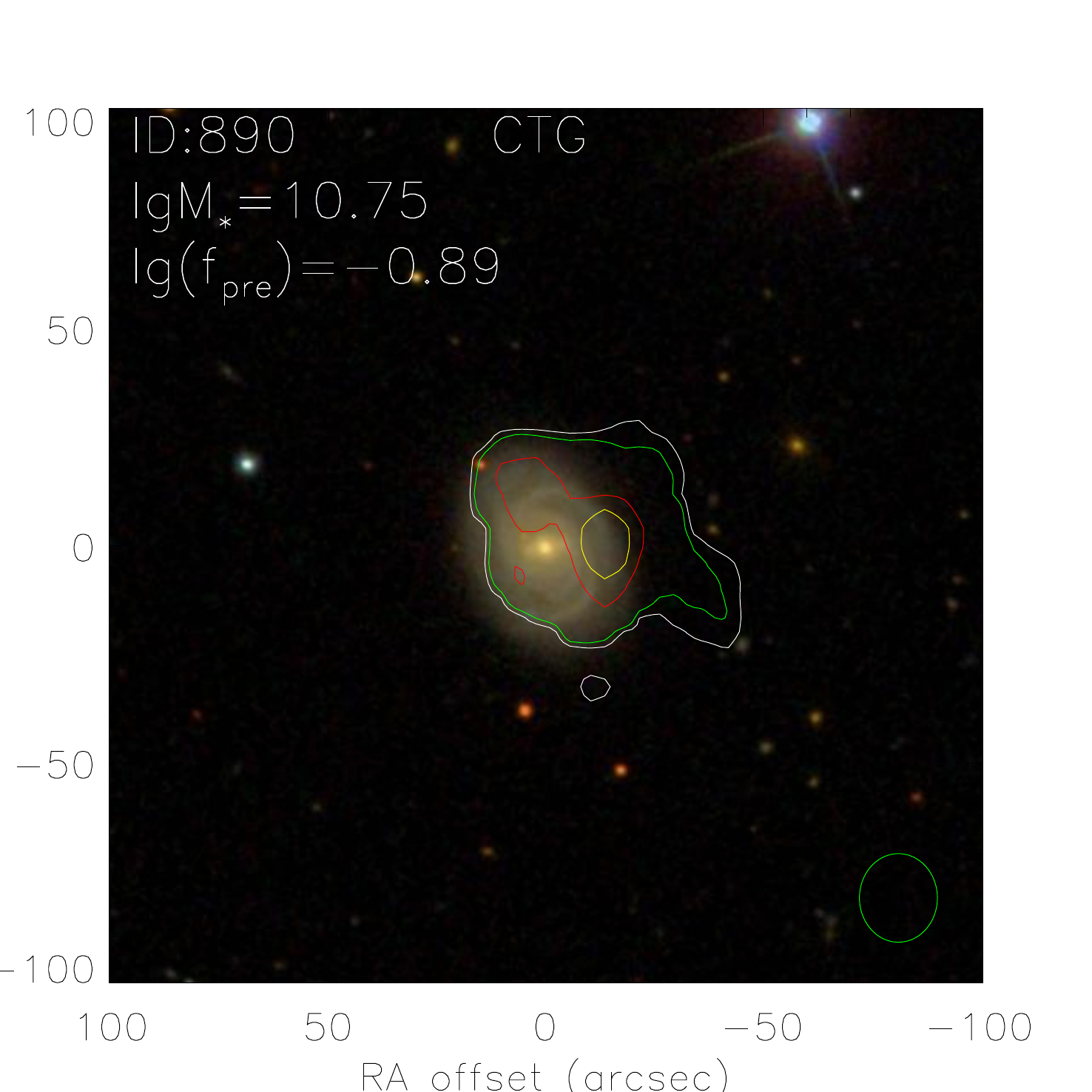}
\includegraphics[width=0.30\textwidth]{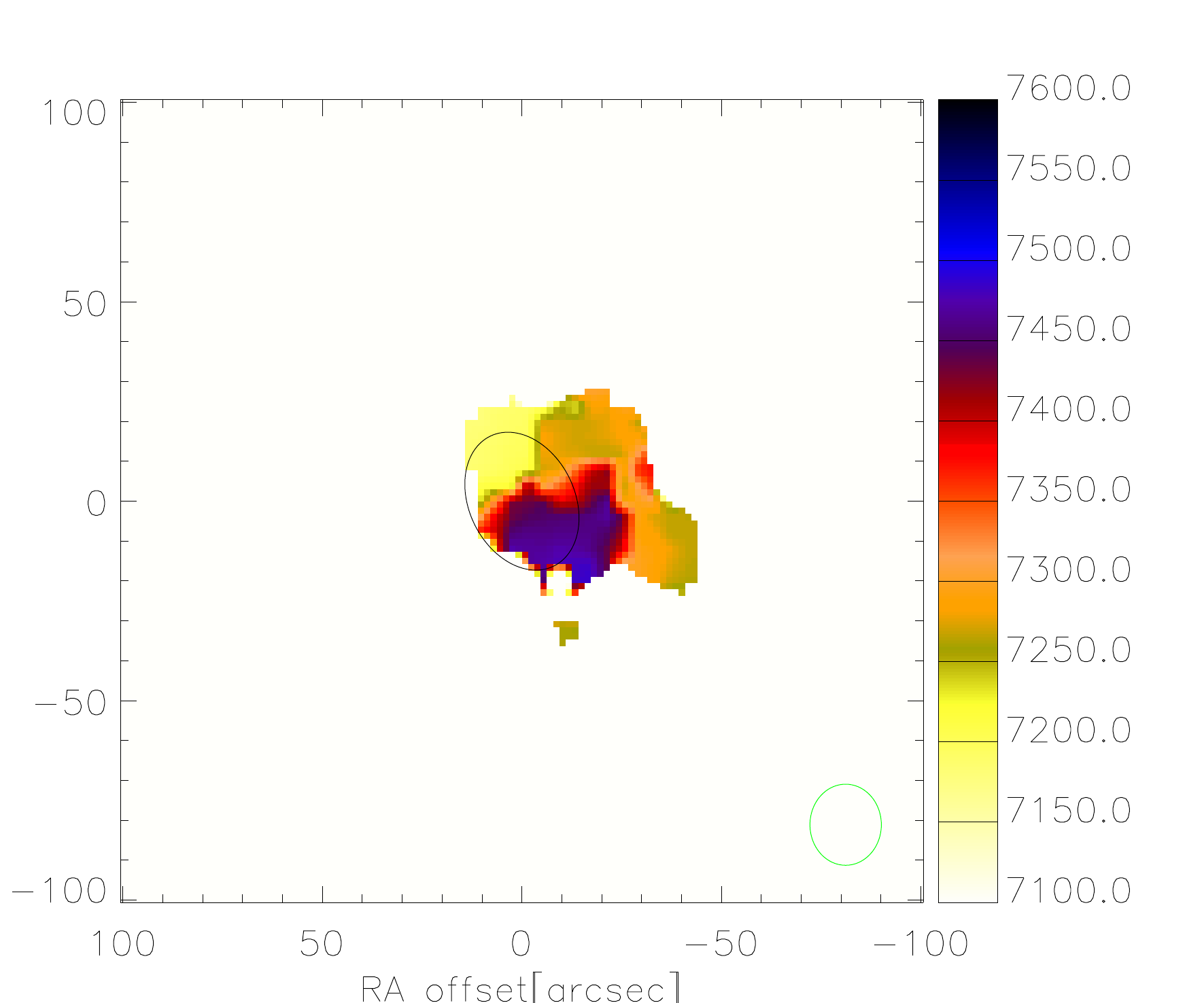}
\includegraphics[width=0.30\textwidth]{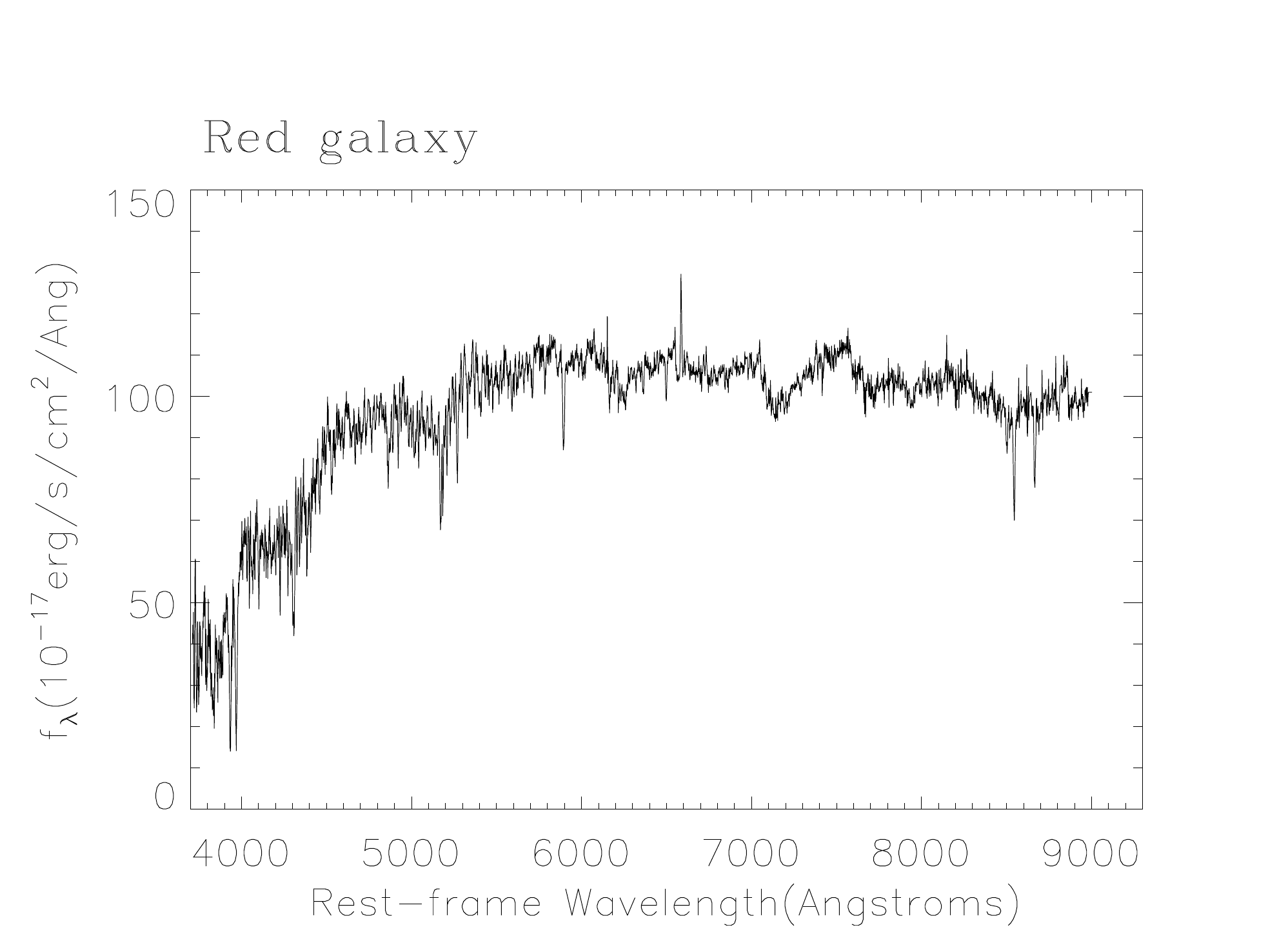}
}\par
\mbox{
\includegraphics[width=0.25\textwidth]{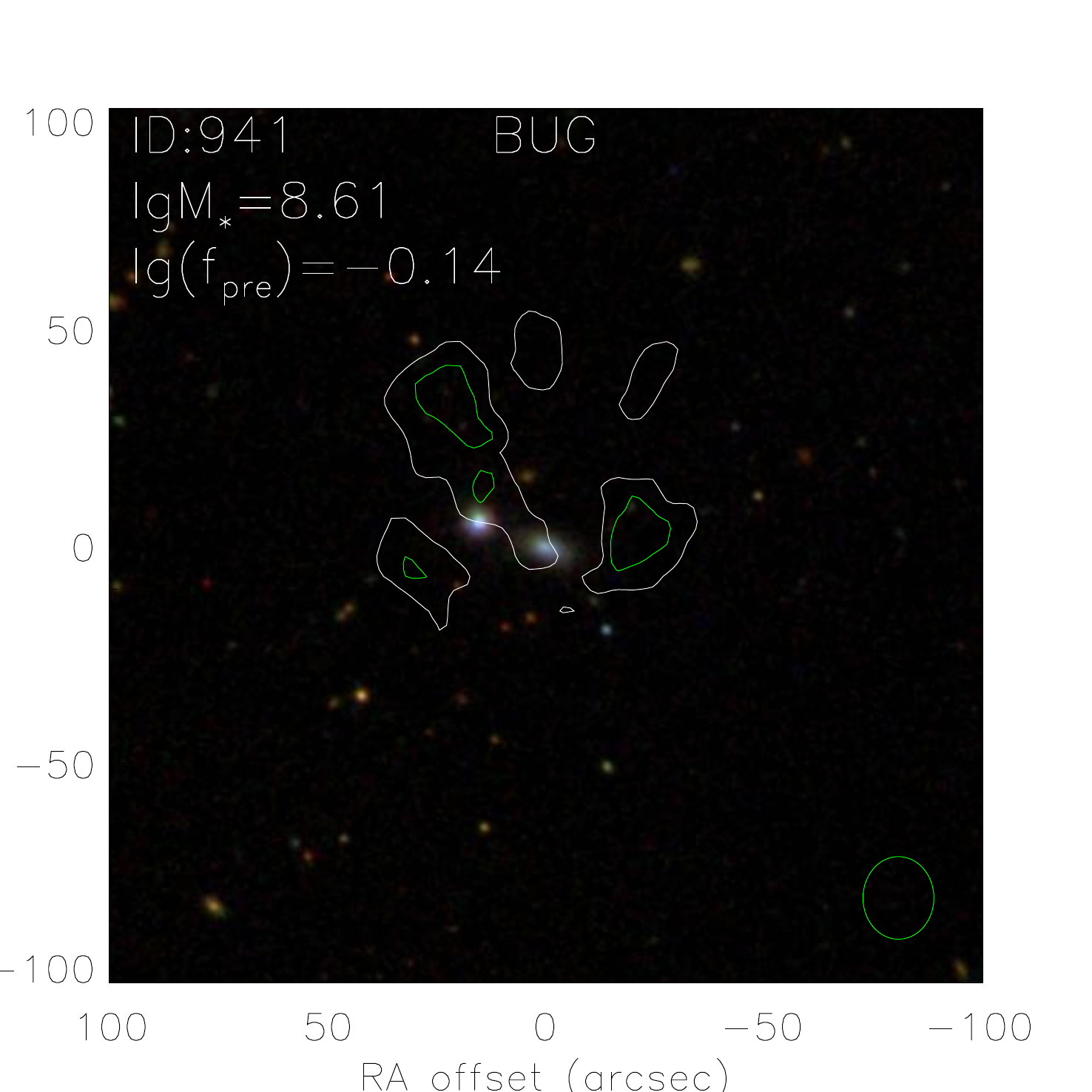}
\includegraphics[width=0.30\textwidth]{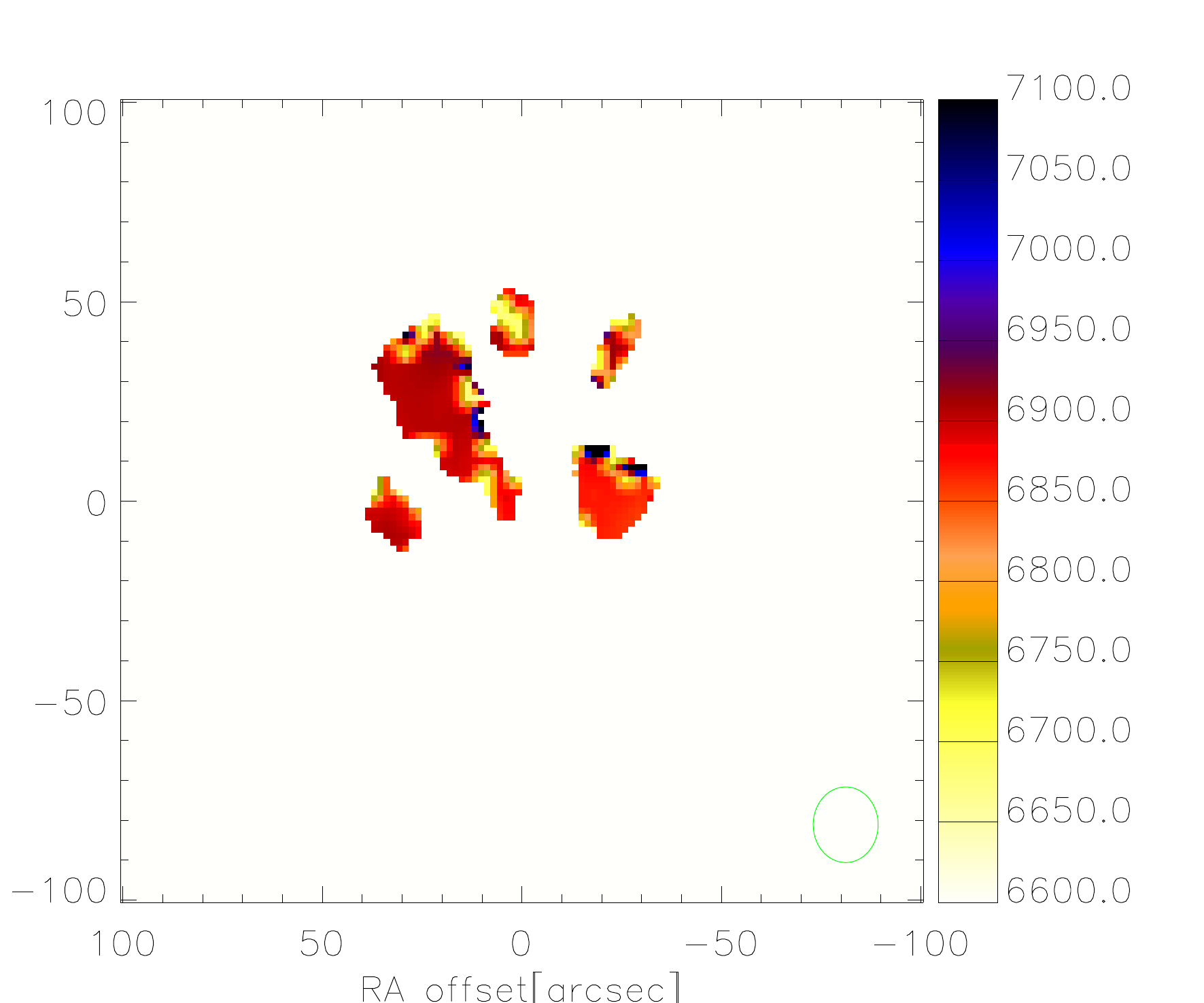}
\includegraphics[width=0.30\textwidth]{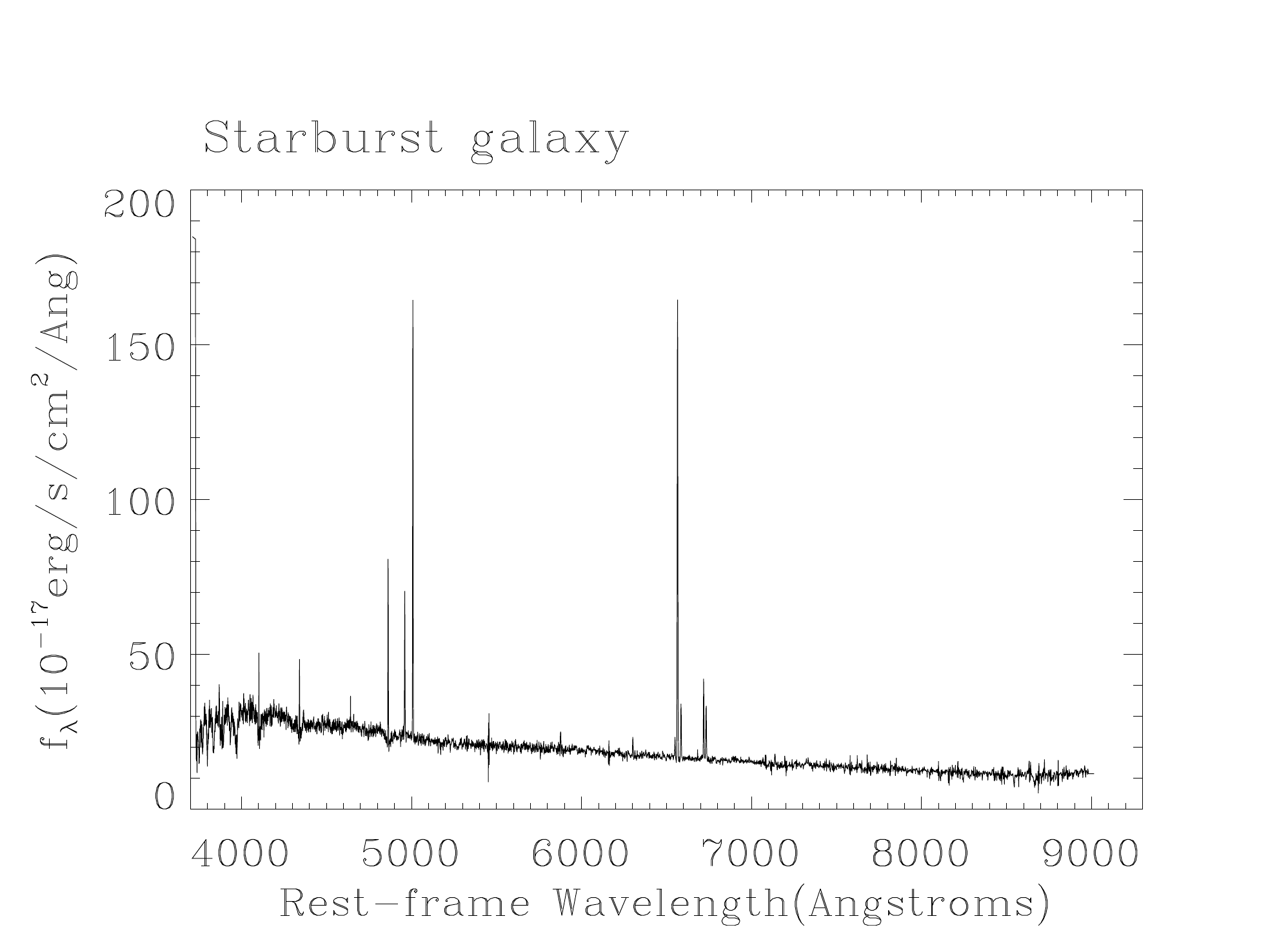}
}\par
\mbox{
\includegraphics[width=0.25\textwidth]{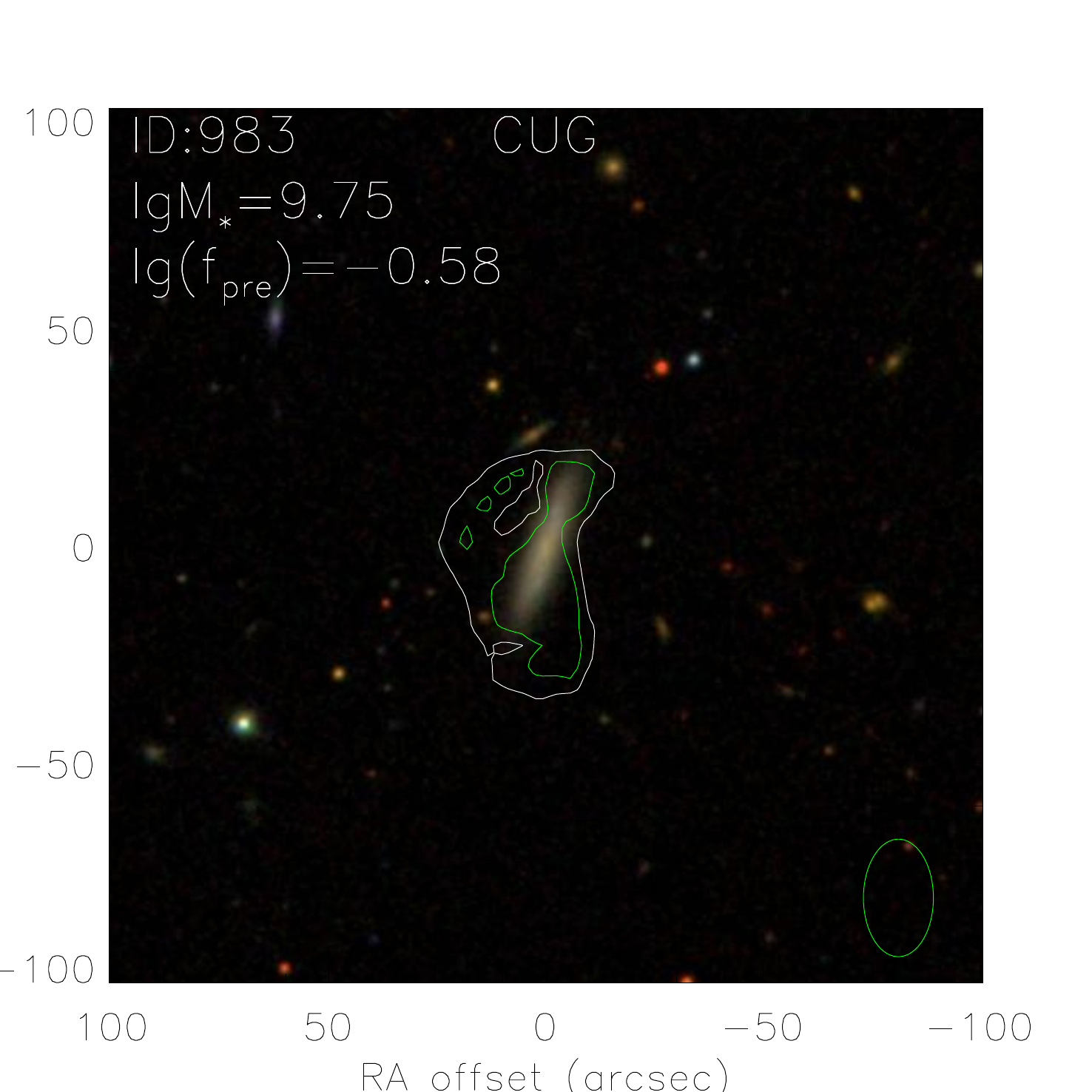}
\includegraphics[width=0.30\textwidth]{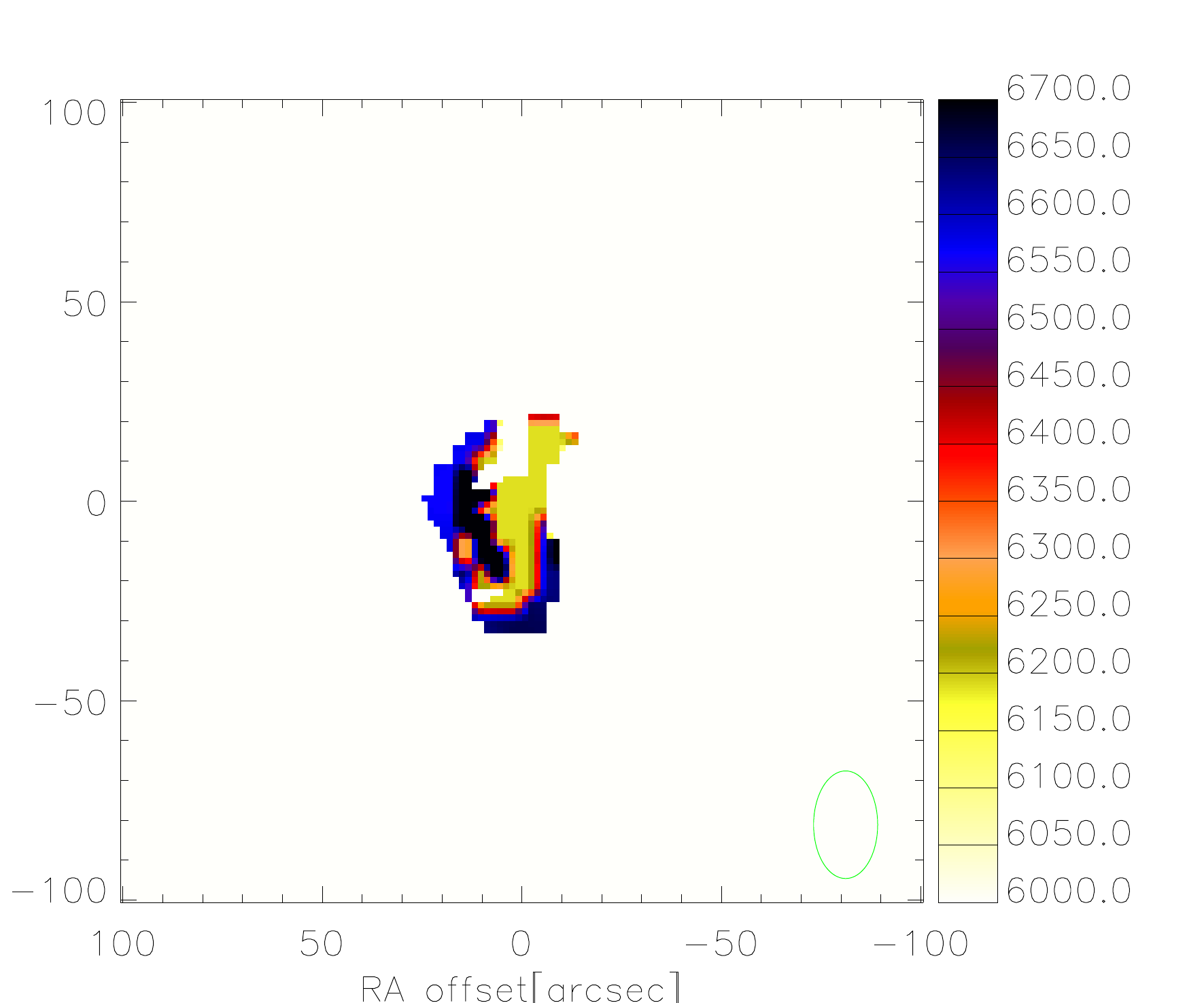}
\includegraphics[width=0.30\textwidth]{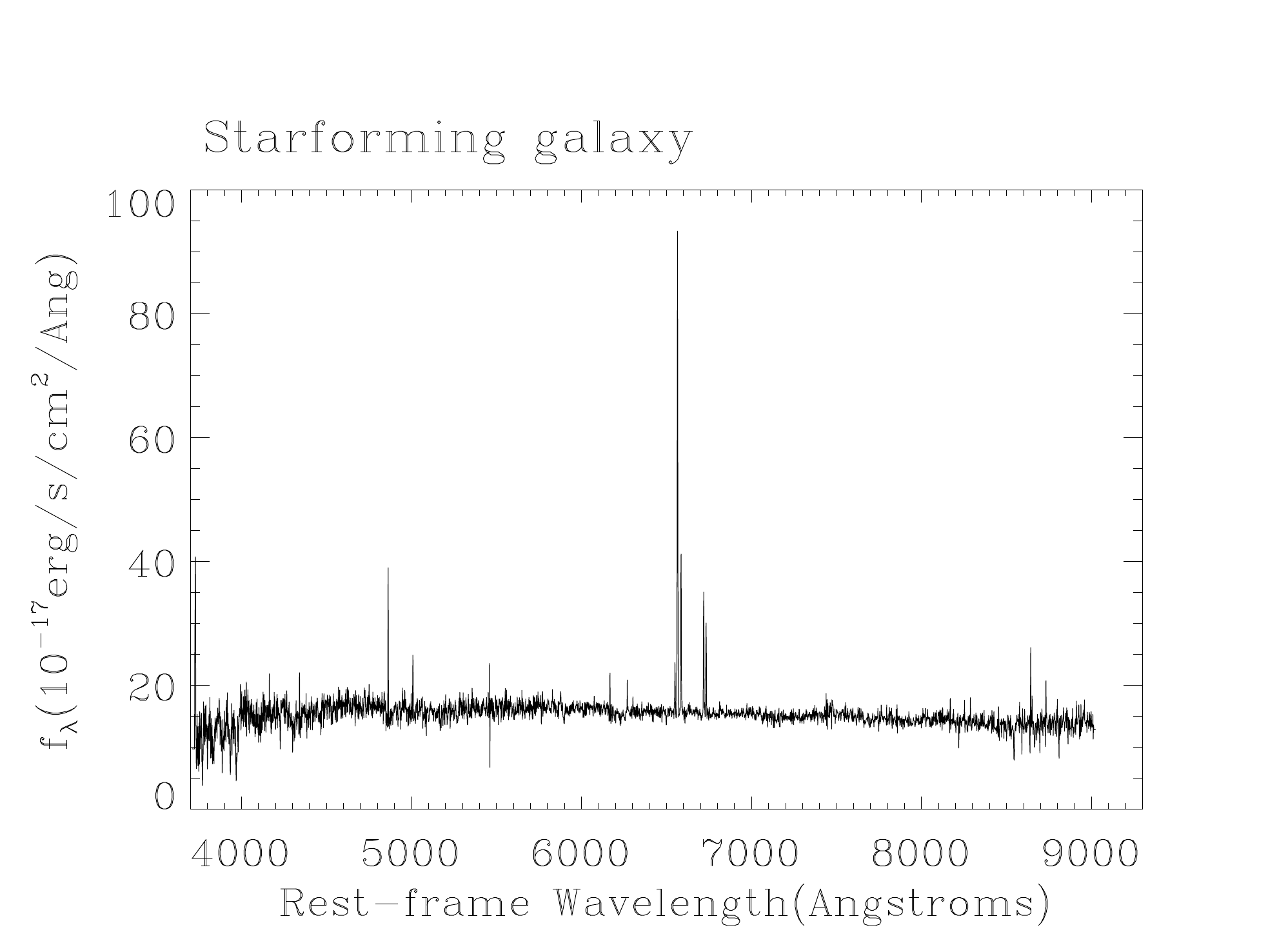}
}\par
\mbox{
\includegraphics[width=0.25\textwidth]{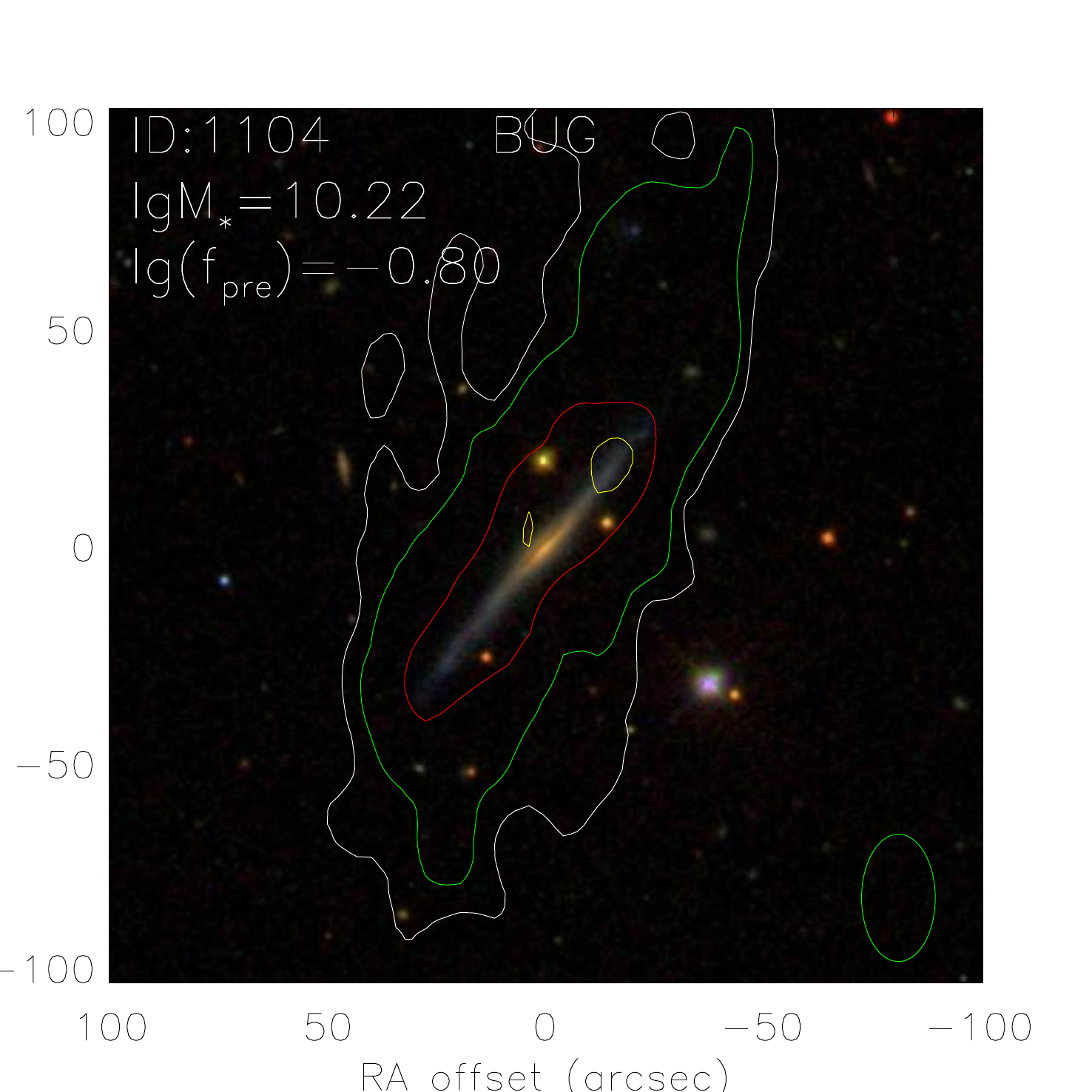}
\includegraphics[width=0.30\textwidth]{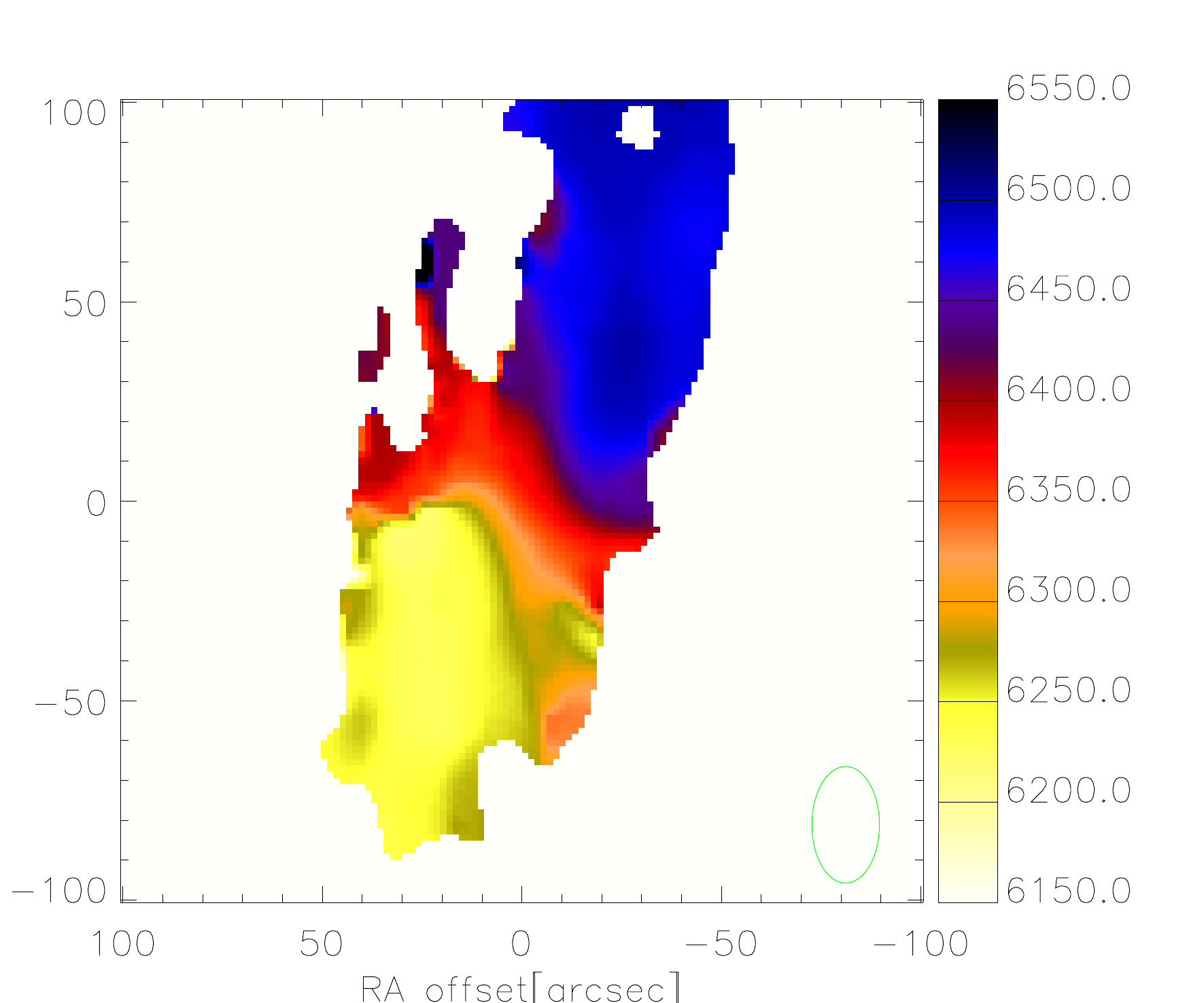}
\includegraphics[width=0.30\textwidth]{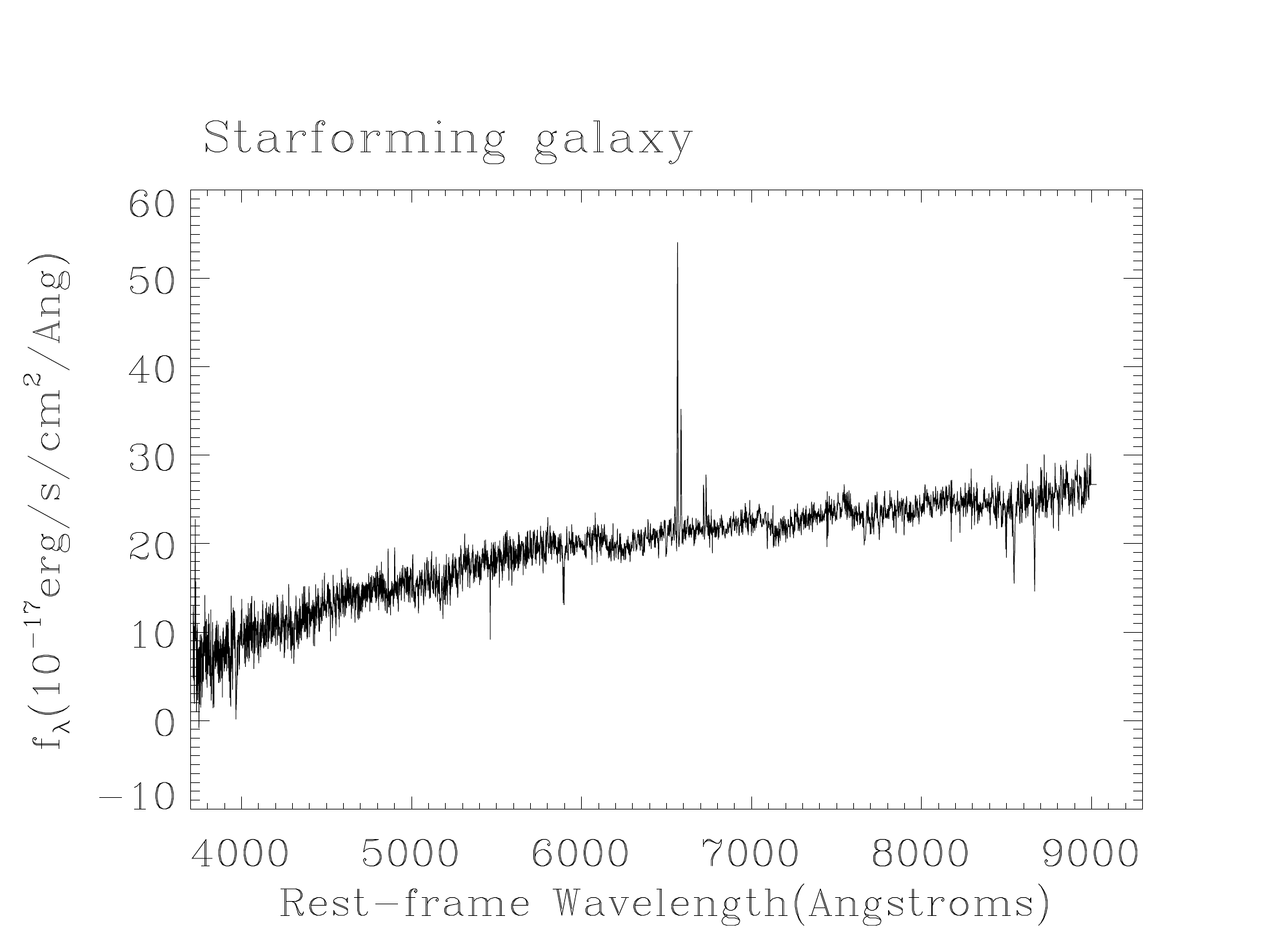}
}
\mbox{
\includegraphics[width=0.25\textwidth]{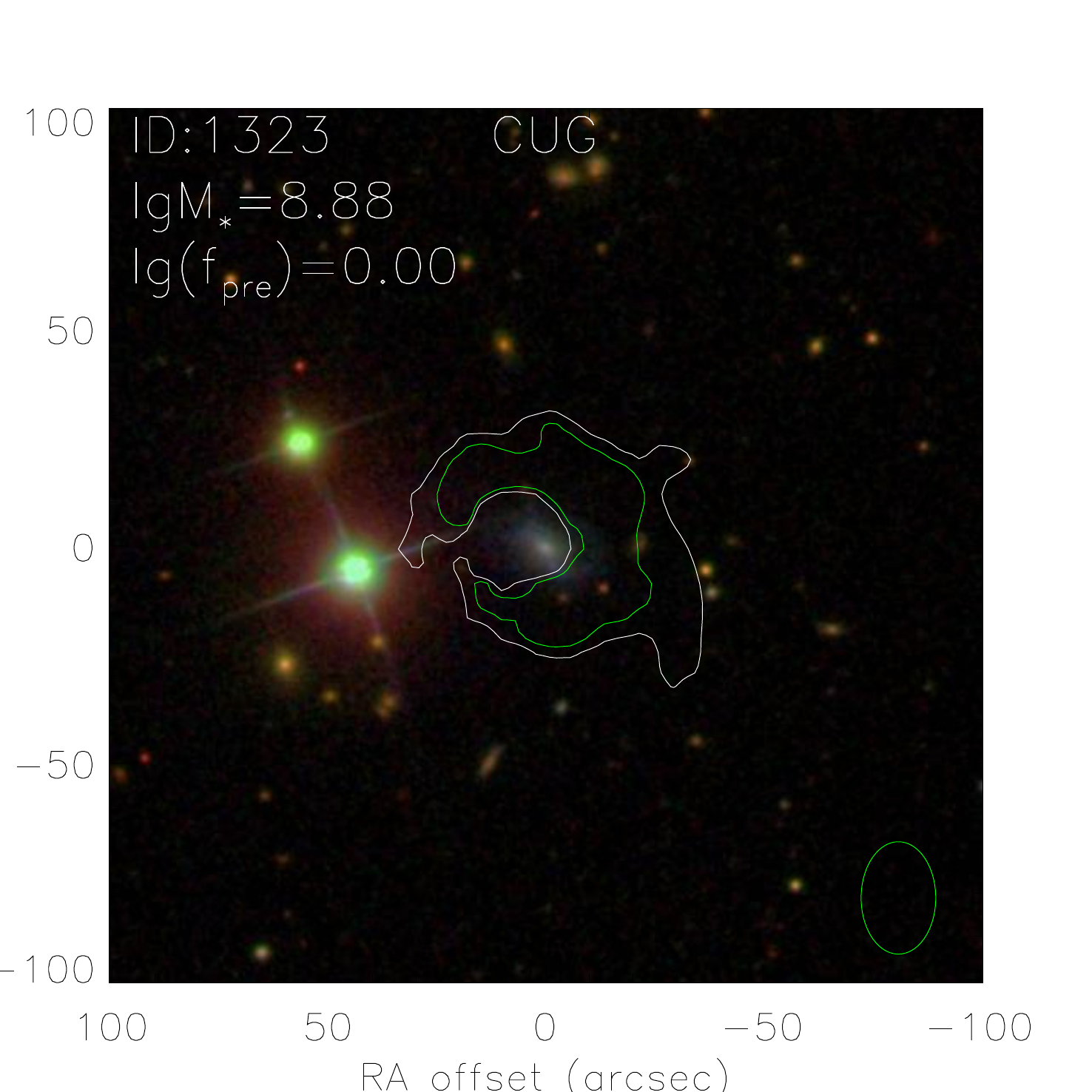}
\includegraphics[width=0.30\textwidth]{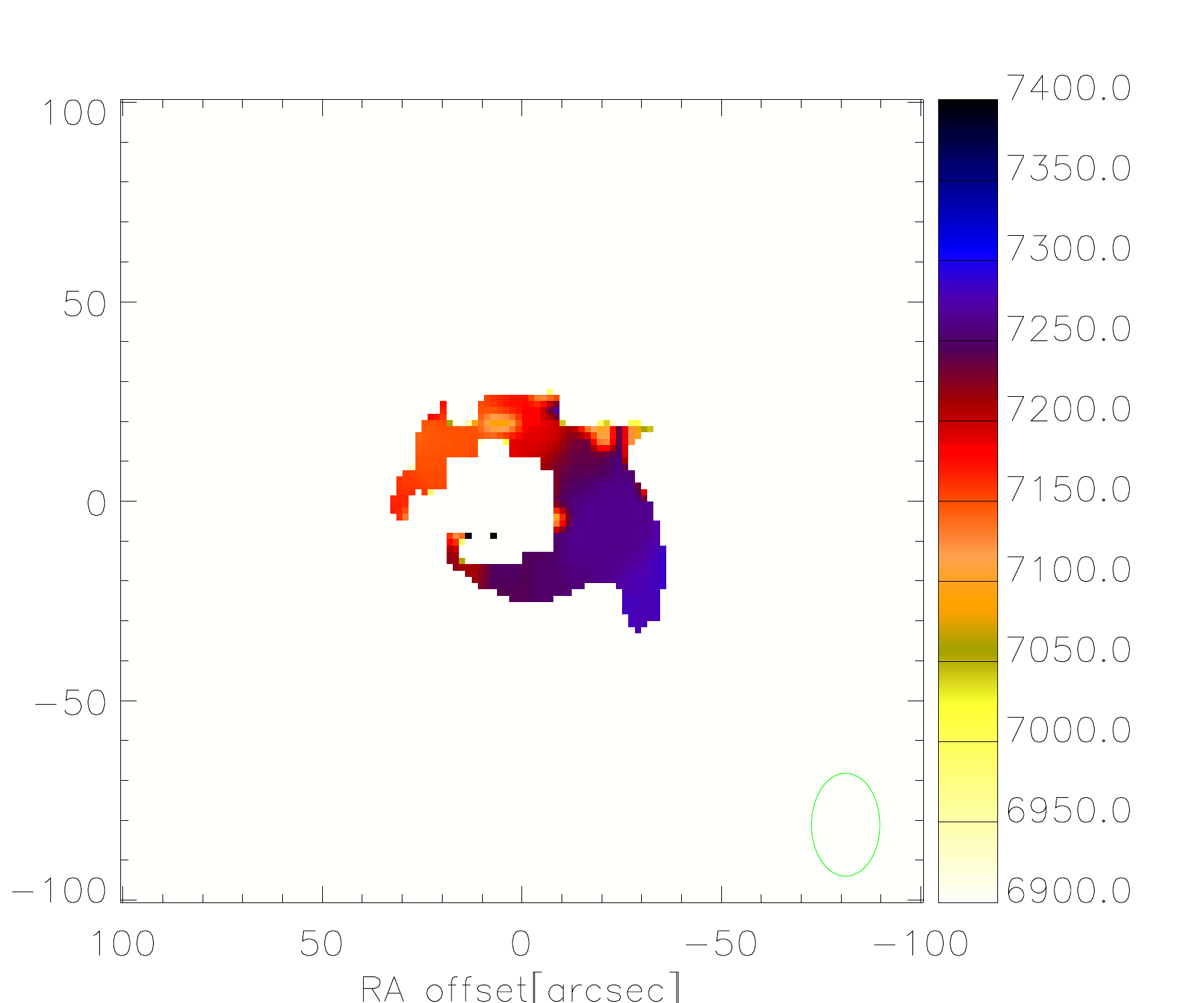}
\includegraphics[width=0.30\textwidth]{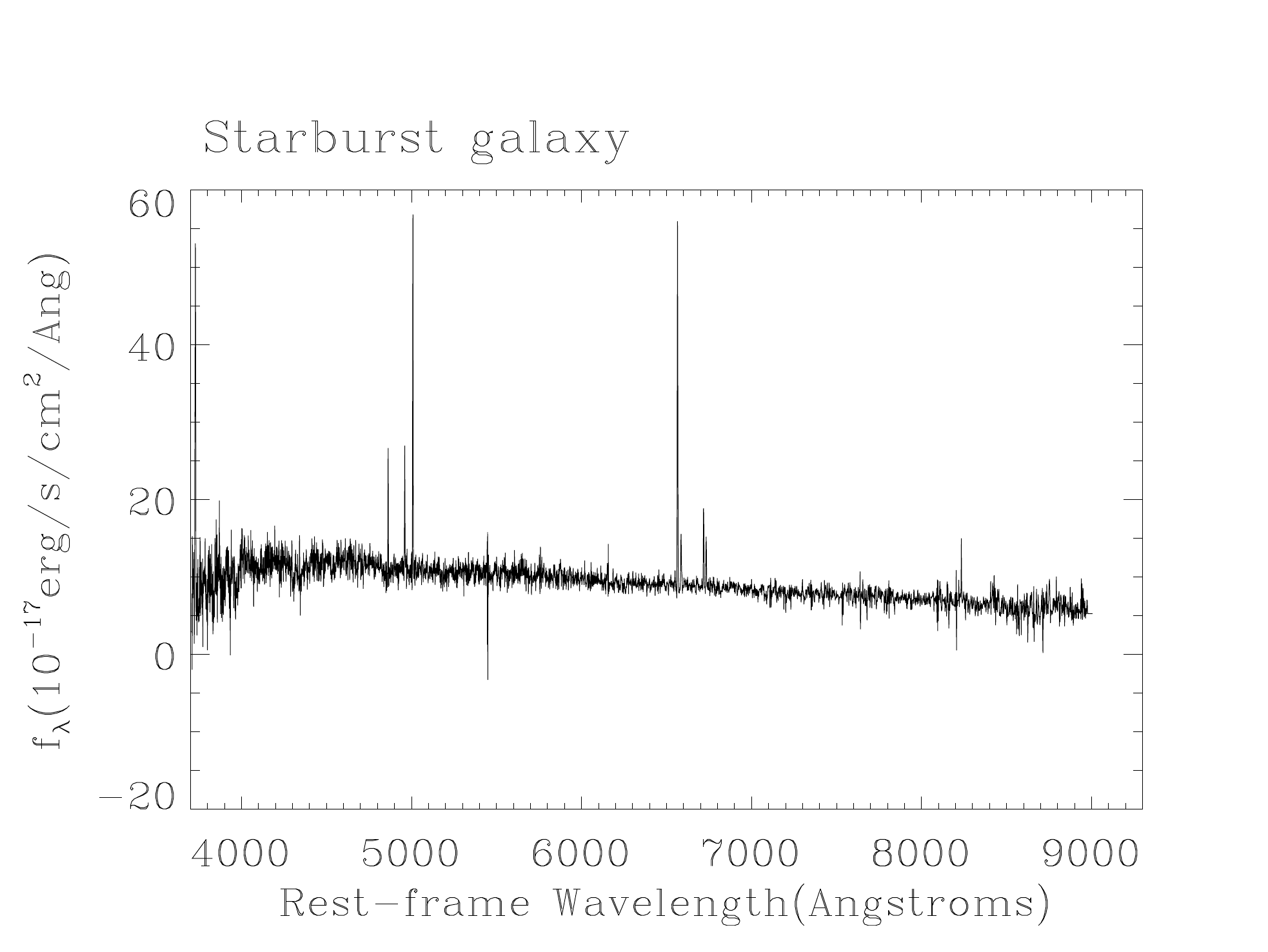}
}\par
 \caption{H{\sc{I}} column density contours overlaid on optical images; H{\sc{I}} velocity maps and SDSS spectra for outliers.
 The galaxy ID, stellar mass and the predicted H{\sc{I}} mass fraction
 are labeled in the top-left-hand corner of each left panel.
 The white, green, red and yellow H{\sc{I}} contours represent 2.0, 4.0, 14.0 and 20.0 times 
 the median SNR of the outermost contour, respectively.
 The shape of the beam is plotted in green at the bottom-right corner of each map. All the maps are displayed as
 north up and east left. }
 \label{fig10}
\end{figure*}

\addtocounter{figure}{-1}
\begin{figure*}
%\mbox{
%  \epsfig{figure=./plots/outliers/1323_m.ps,width=0.25\textwidth}
%  \epsfig{figure=./plots/outliers/1323_v.ps,width=0.3\textwidth}
%  \epsfig{figure=./plots/outliers/SDSS/1323_s.eps,width=0.3\textwidth}
%}\par
\mbox{
\includegraphics[width=0.25\textwidth]{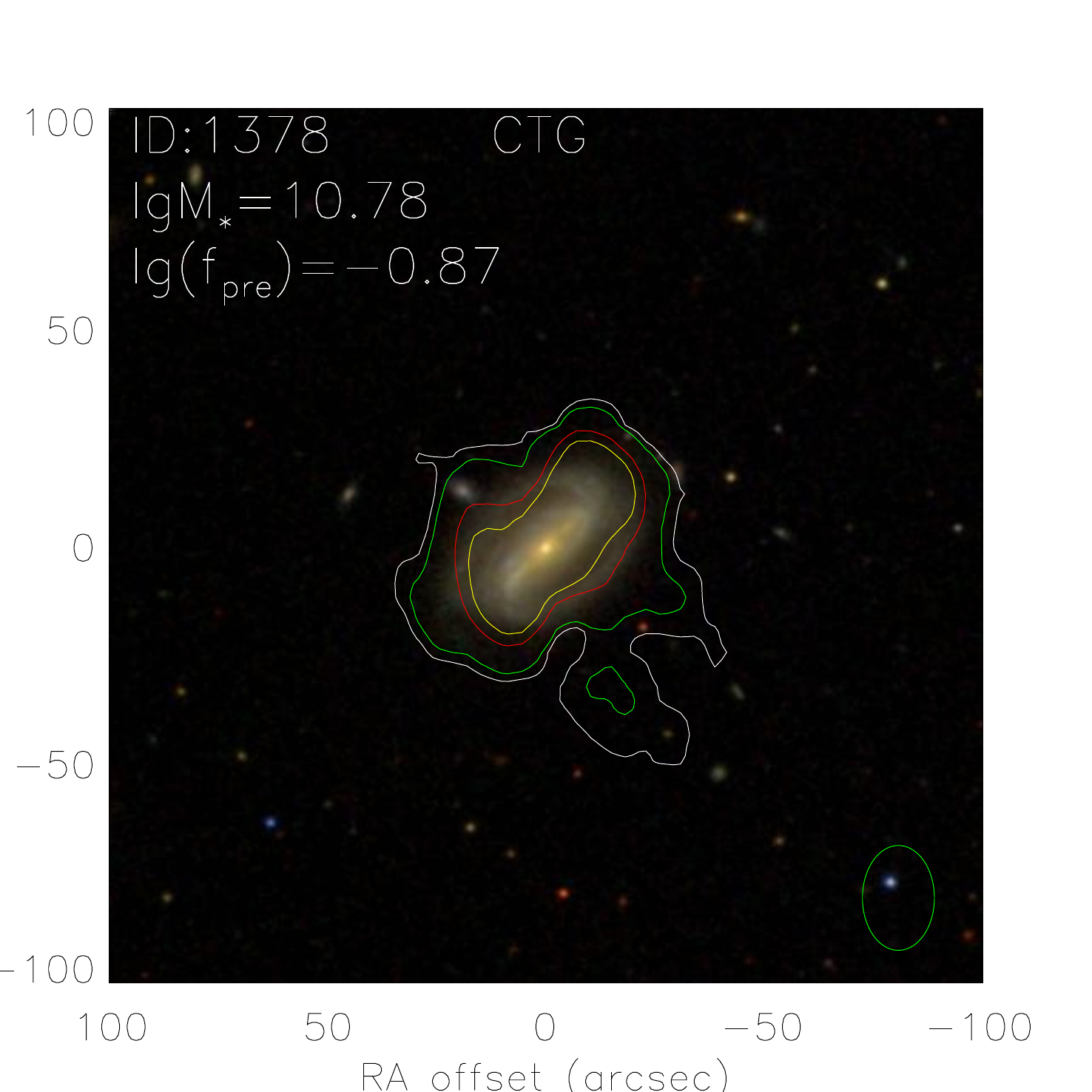}
\includegraphics[width=0.30\textwidth]{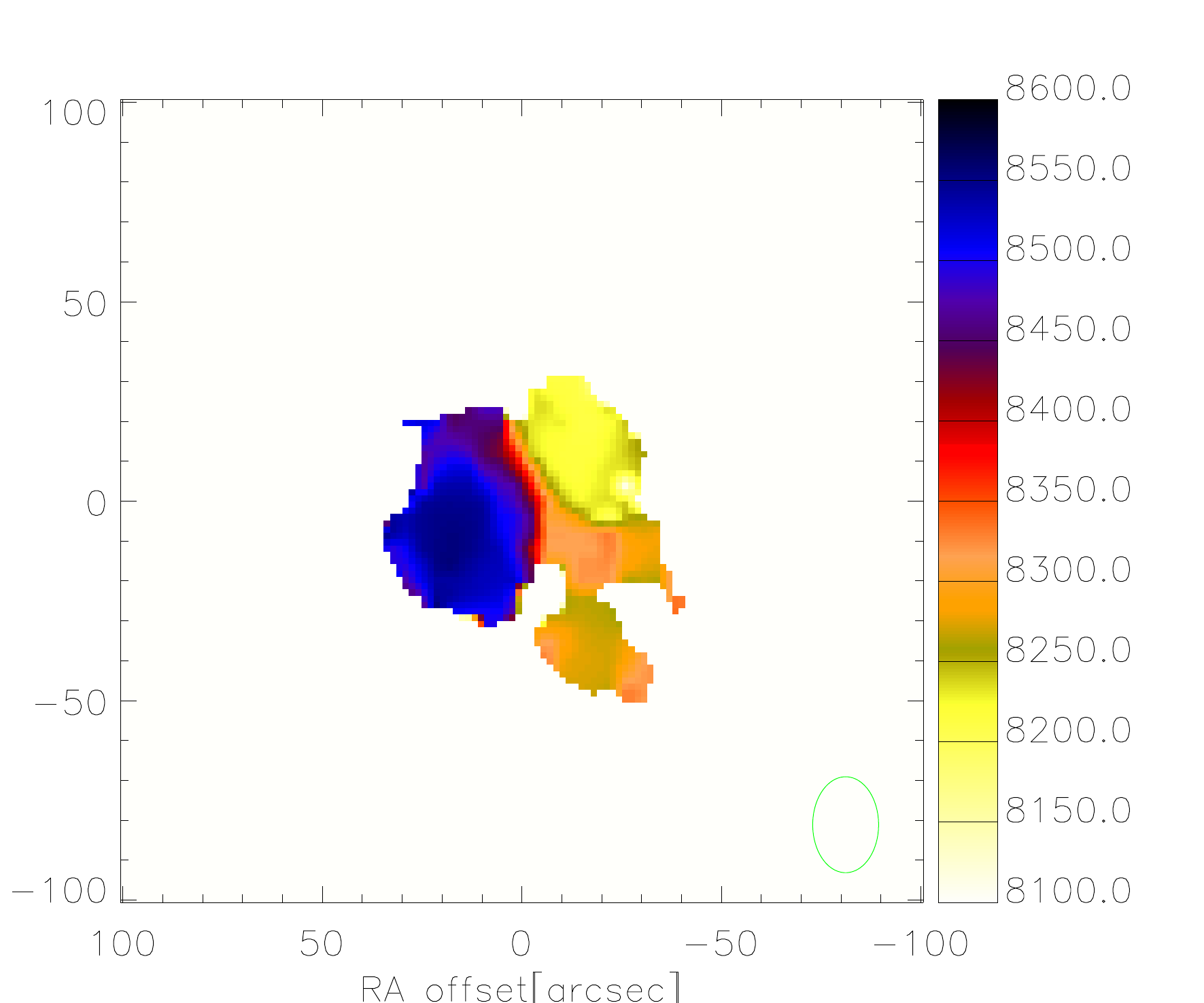}
\includegraphics[width=0.30\textwidth]{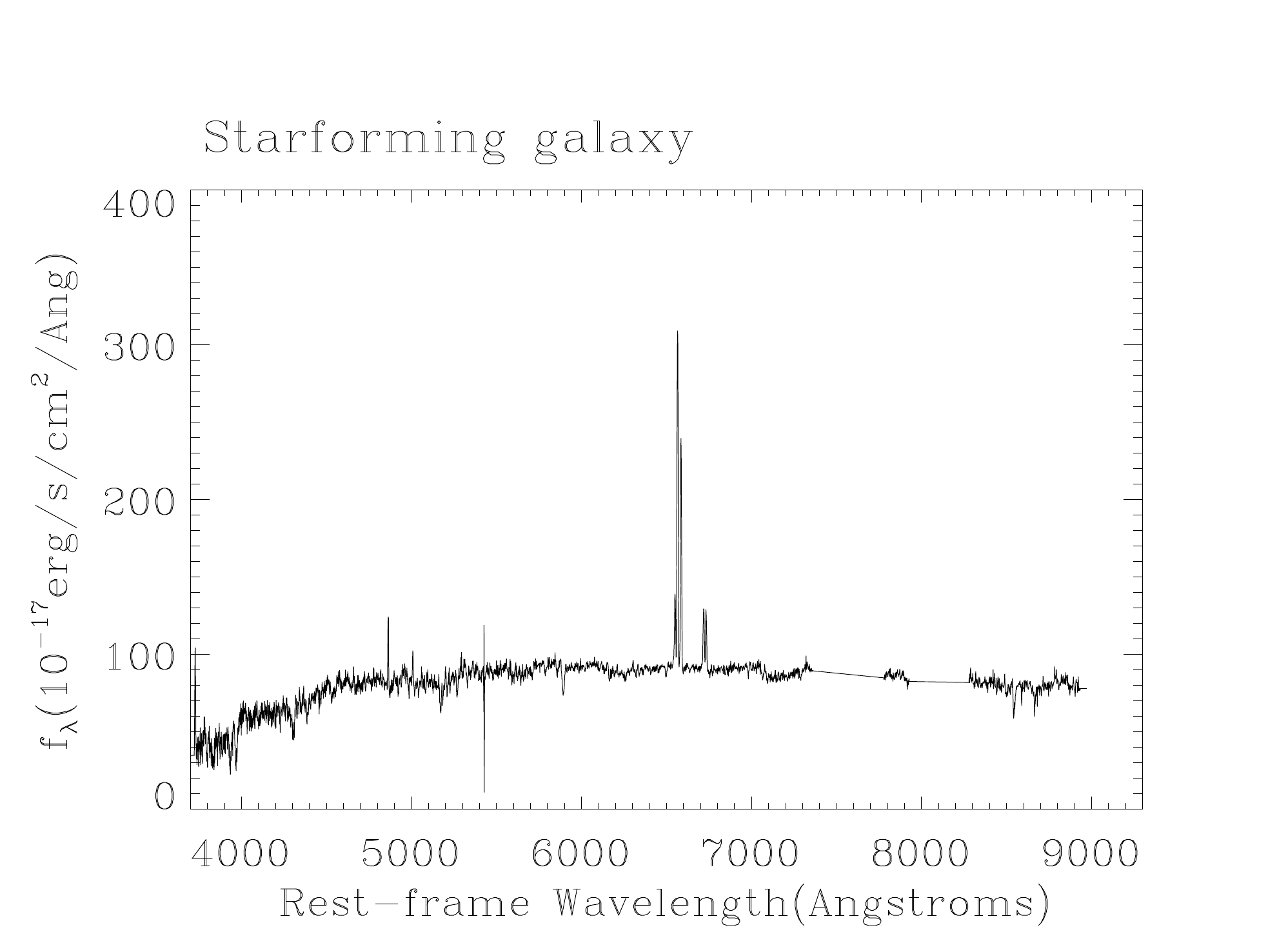}
}\par
\mbox{
\includegraphics[width=0.25\textwidth]{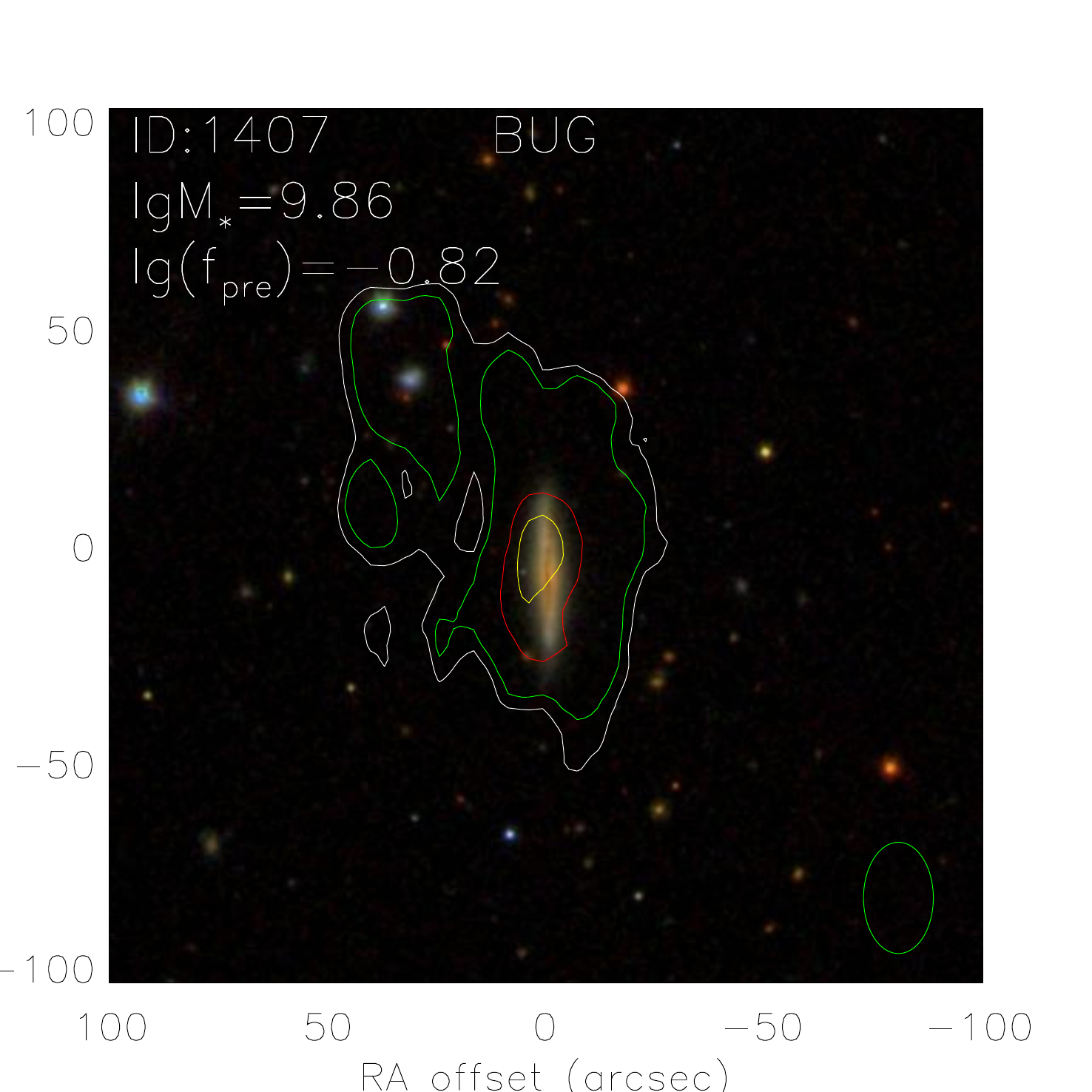}
\includegraphics[width=0.30\textwidth]{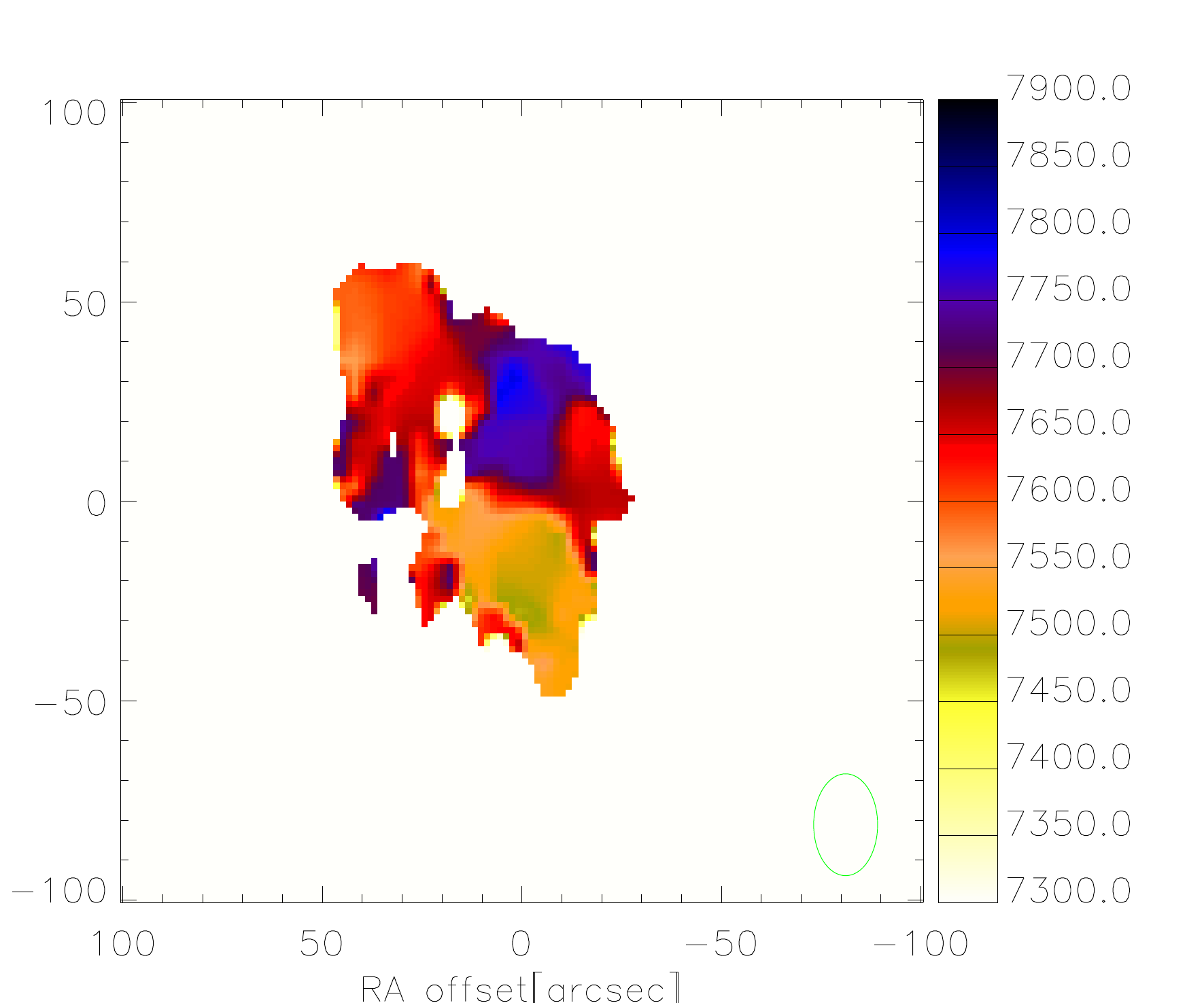}
\includegraphics[width=0.30\textwidth]{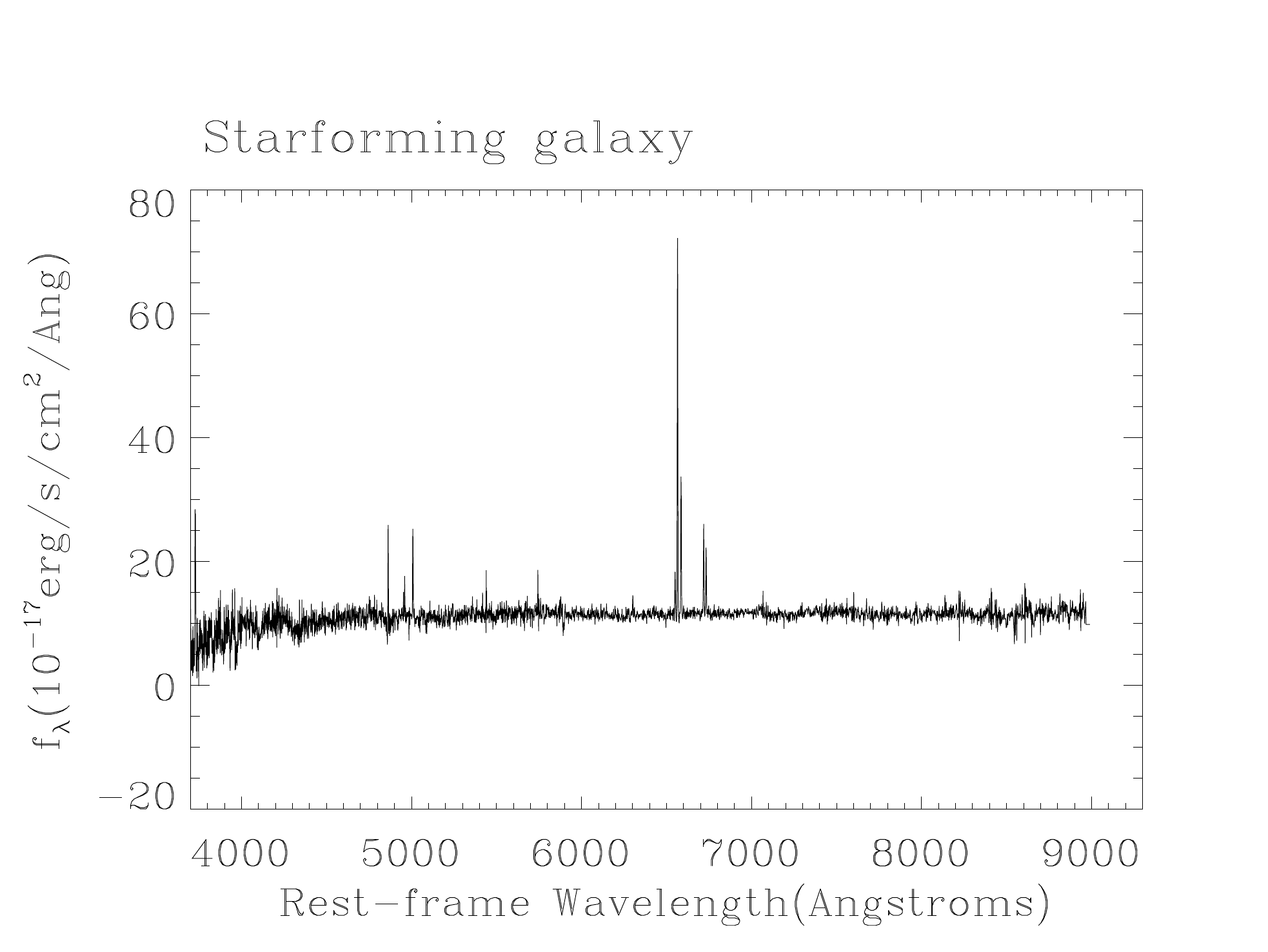}
}\par
\mbox{
\includegraphics[width=0.25\textwidth]{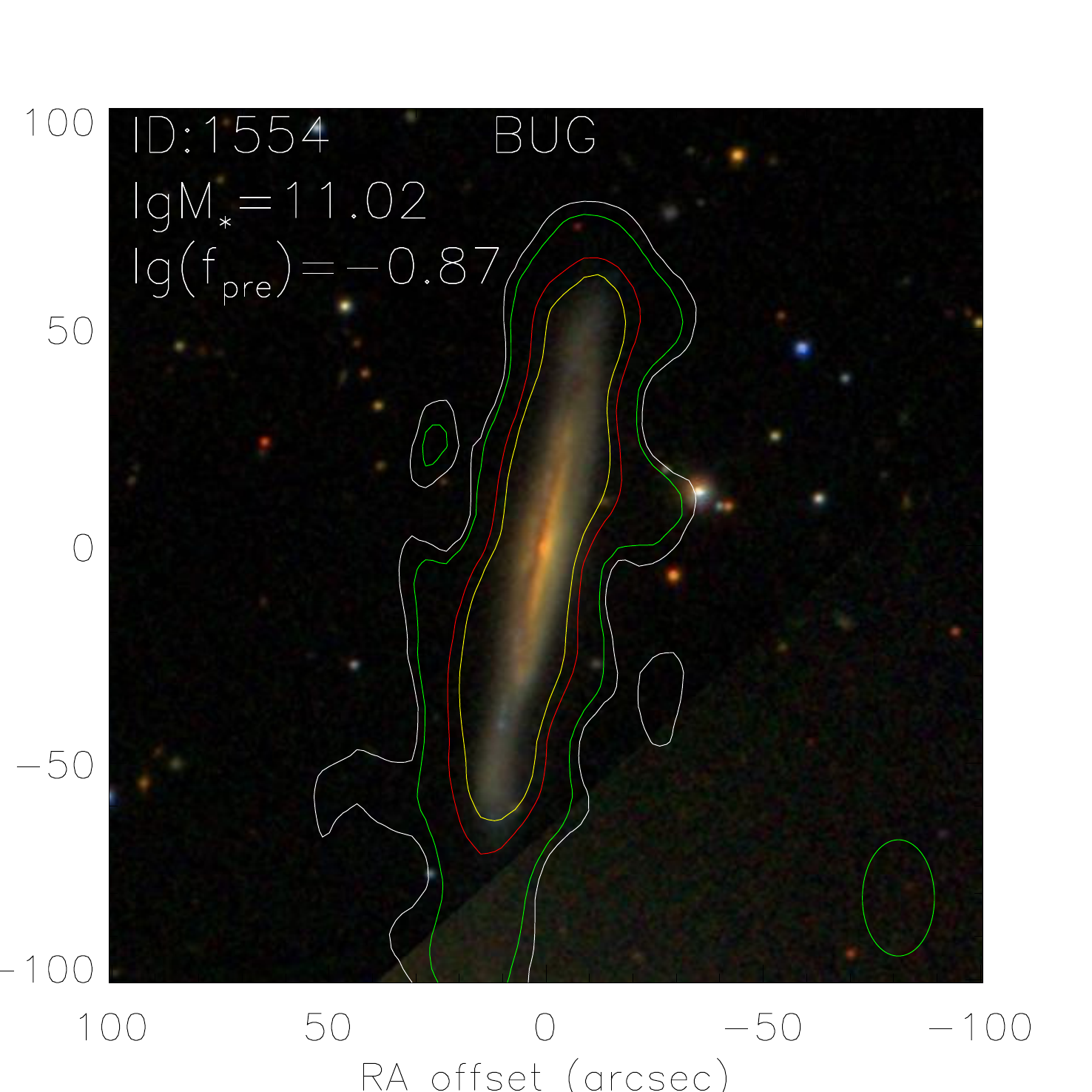}
\includegraphics[width=0.30\textwidth]{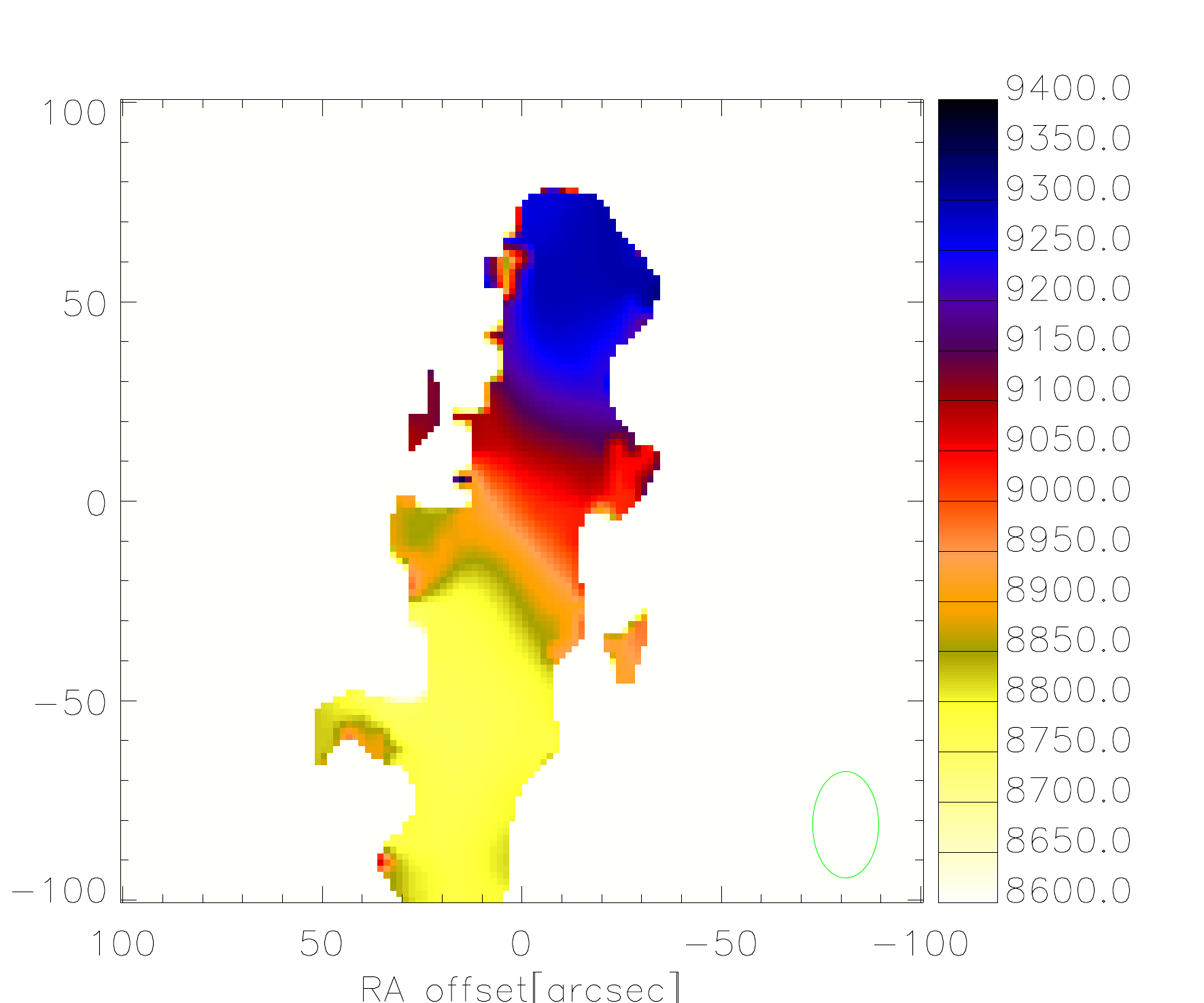}
\includegraphics[width=0.30\textwidth]{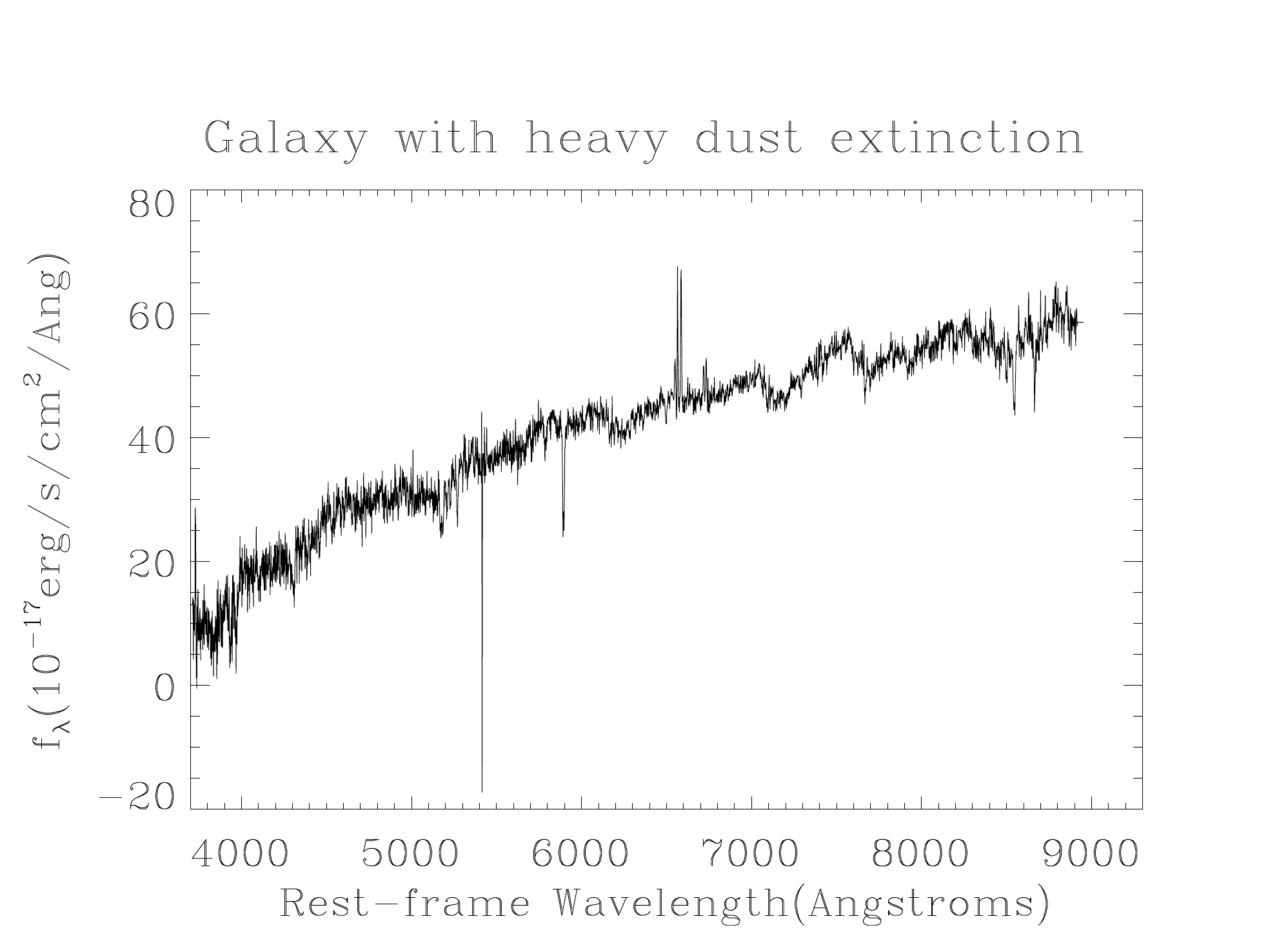}
}\par
\mbox{
\includegraphics[width=0.25\textwidth]{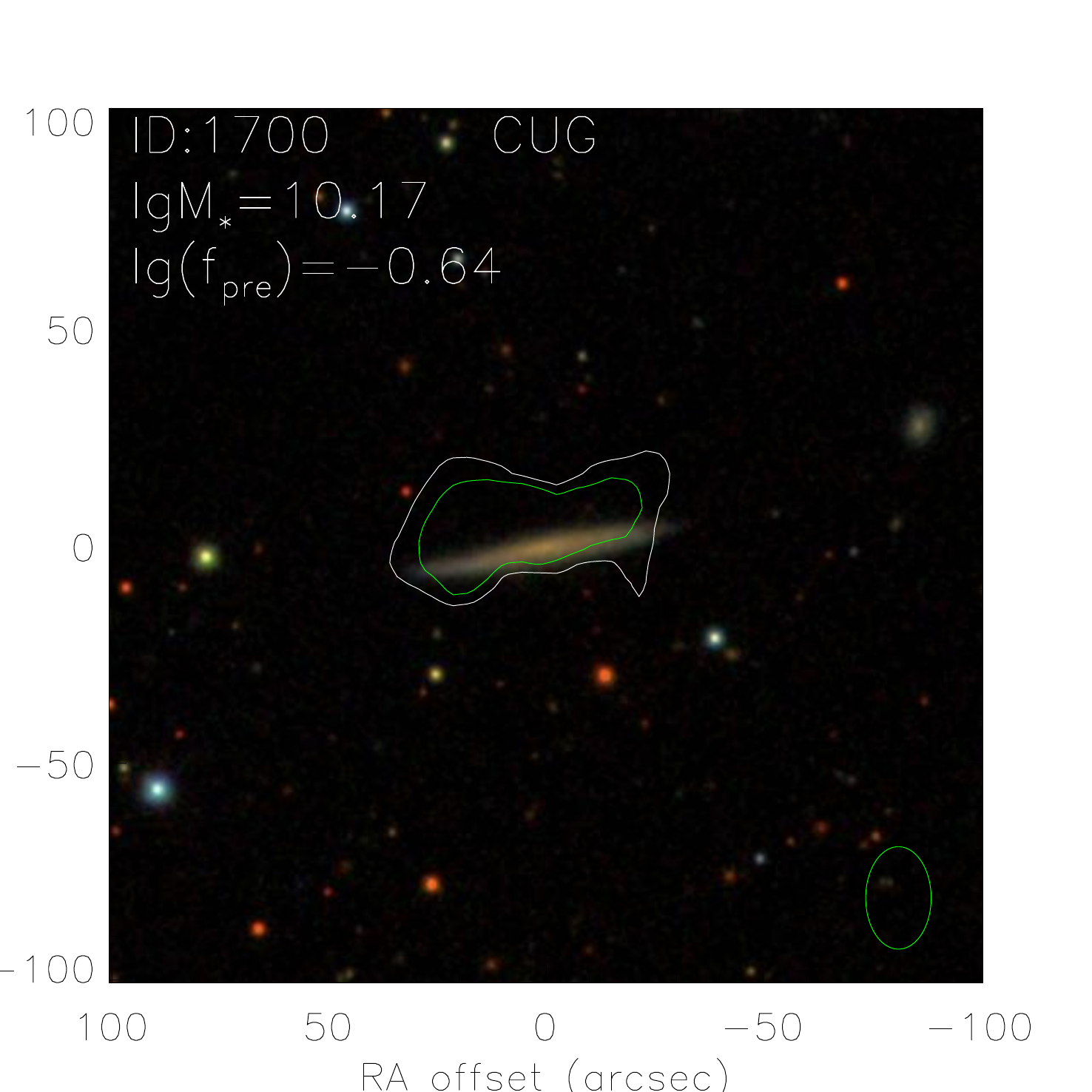}
\includegraphics[width=0.30\textwidth]{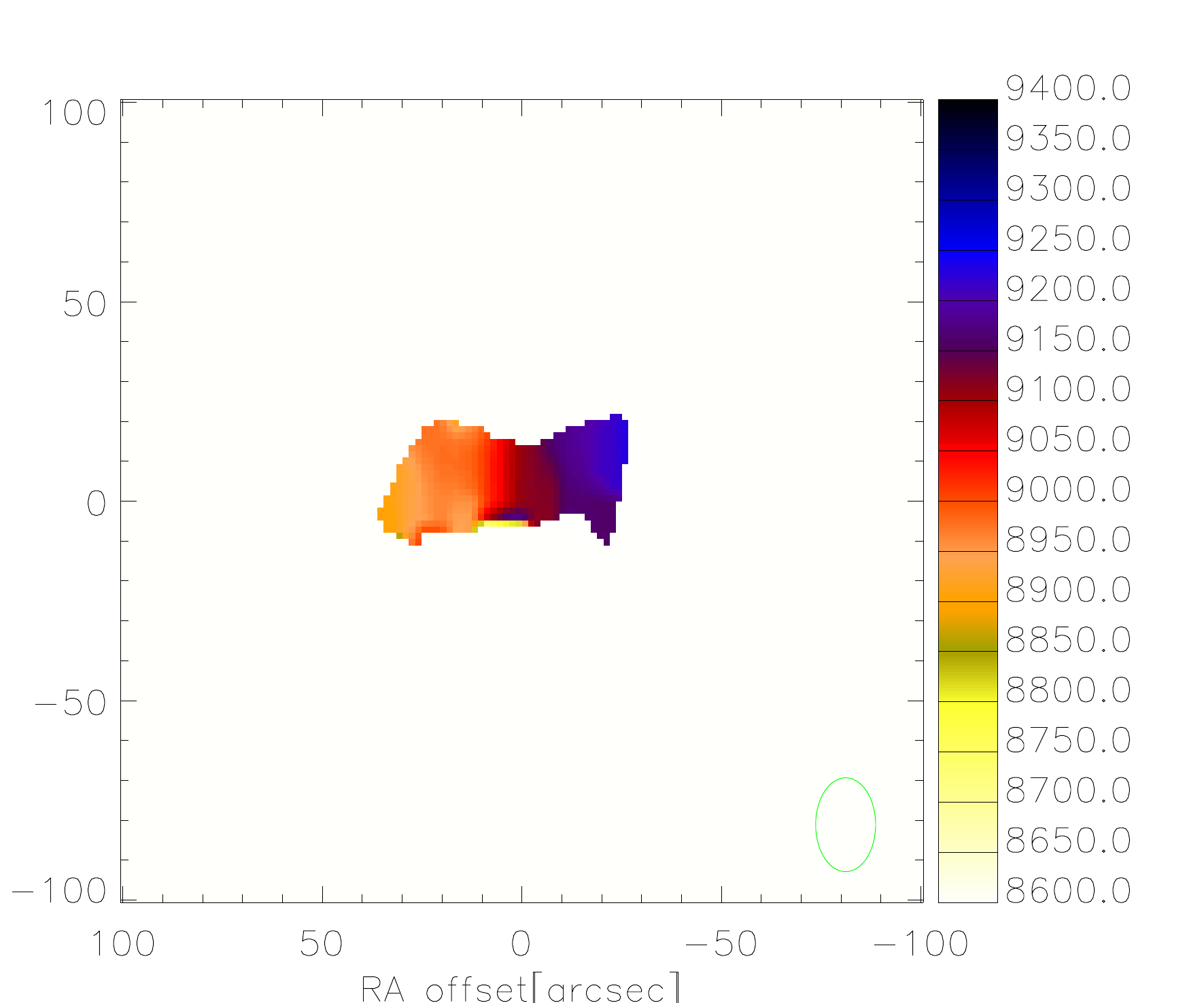}
\includegraphics[width=0.30\textwidth]{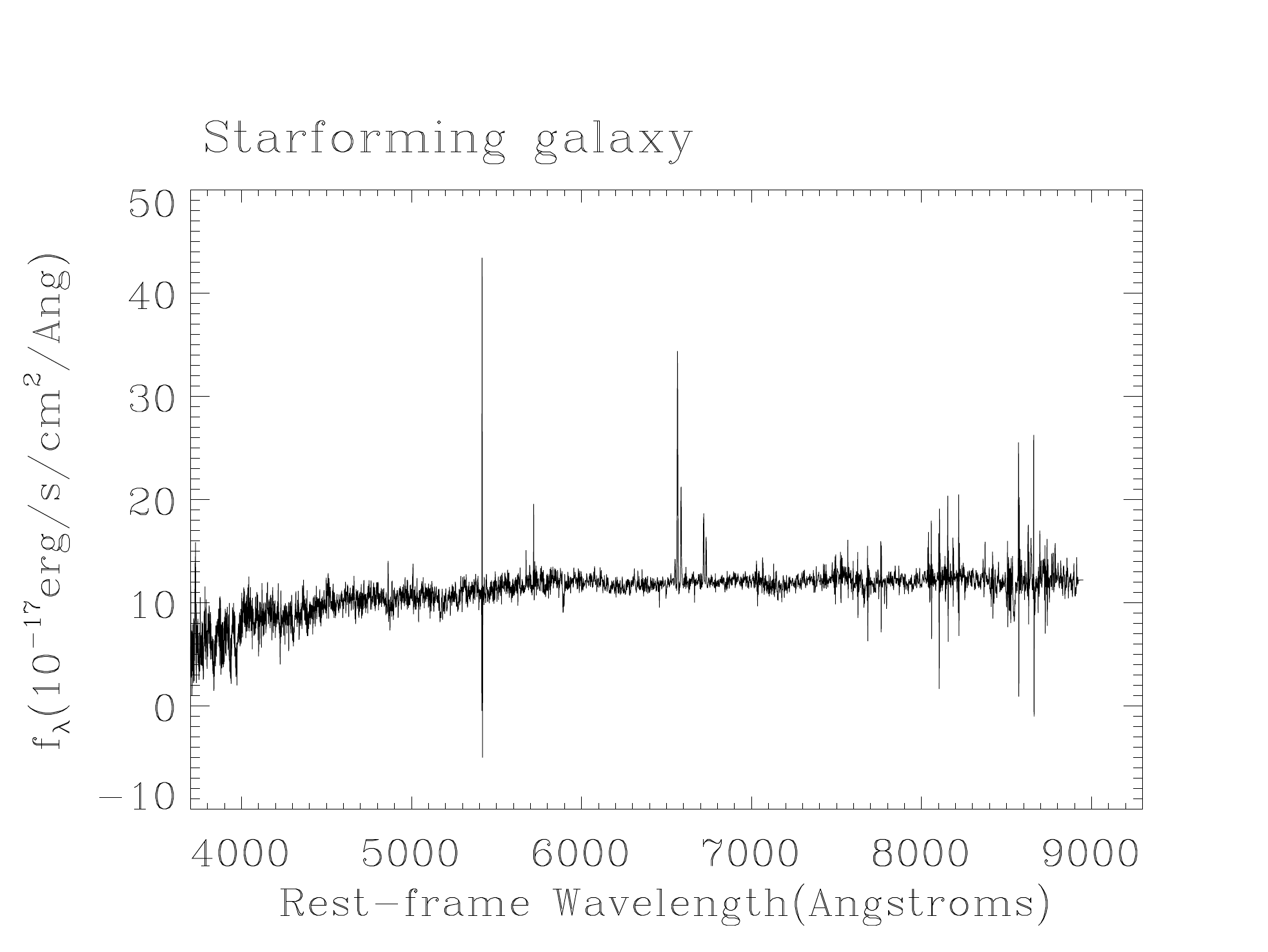}
}
 \caption{Continued}
\end{figure*}

\begin{figure*}
\mbox{
\includegraphics[width=0.25\textwidth]{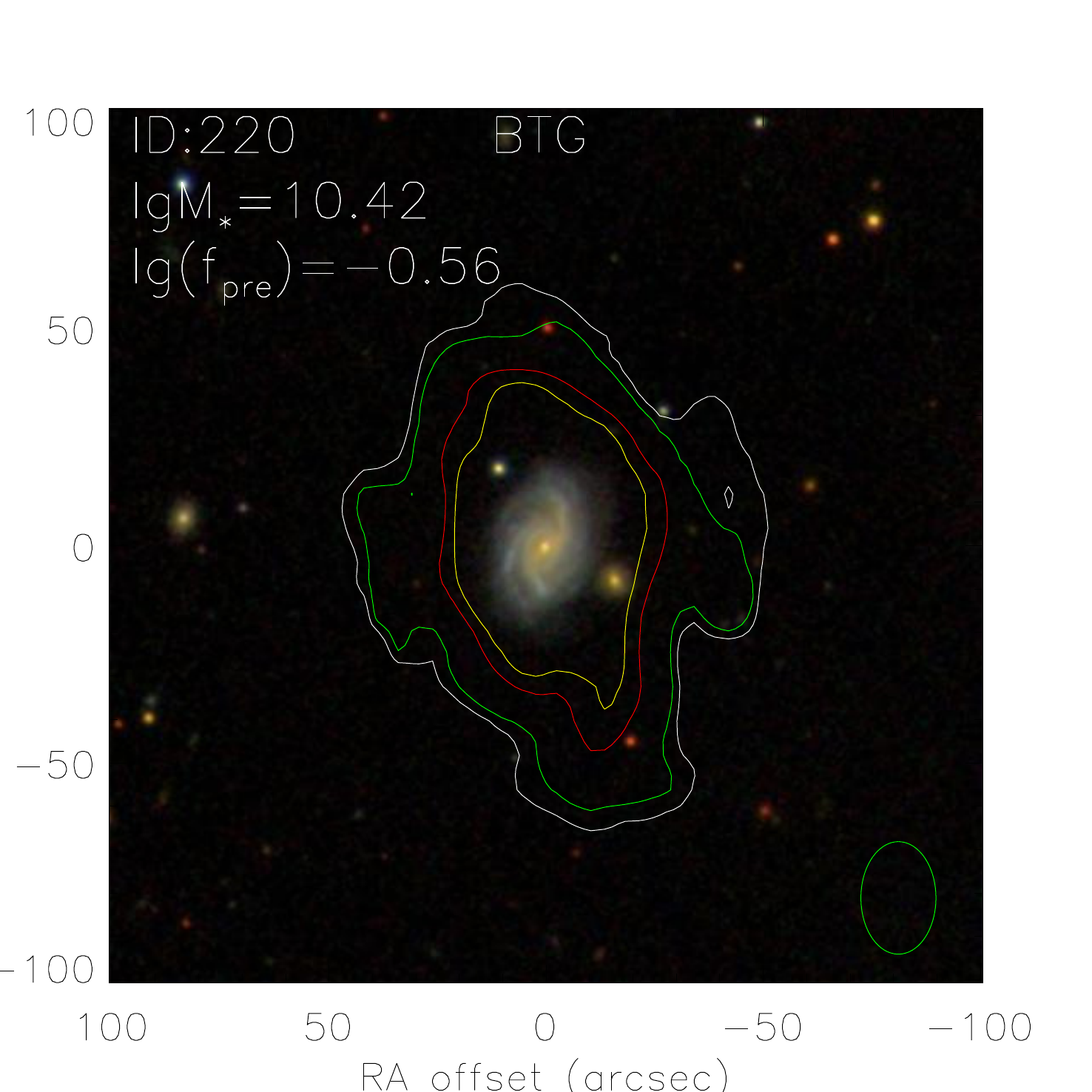}
\includegraphics[width=0.30\textwidth]{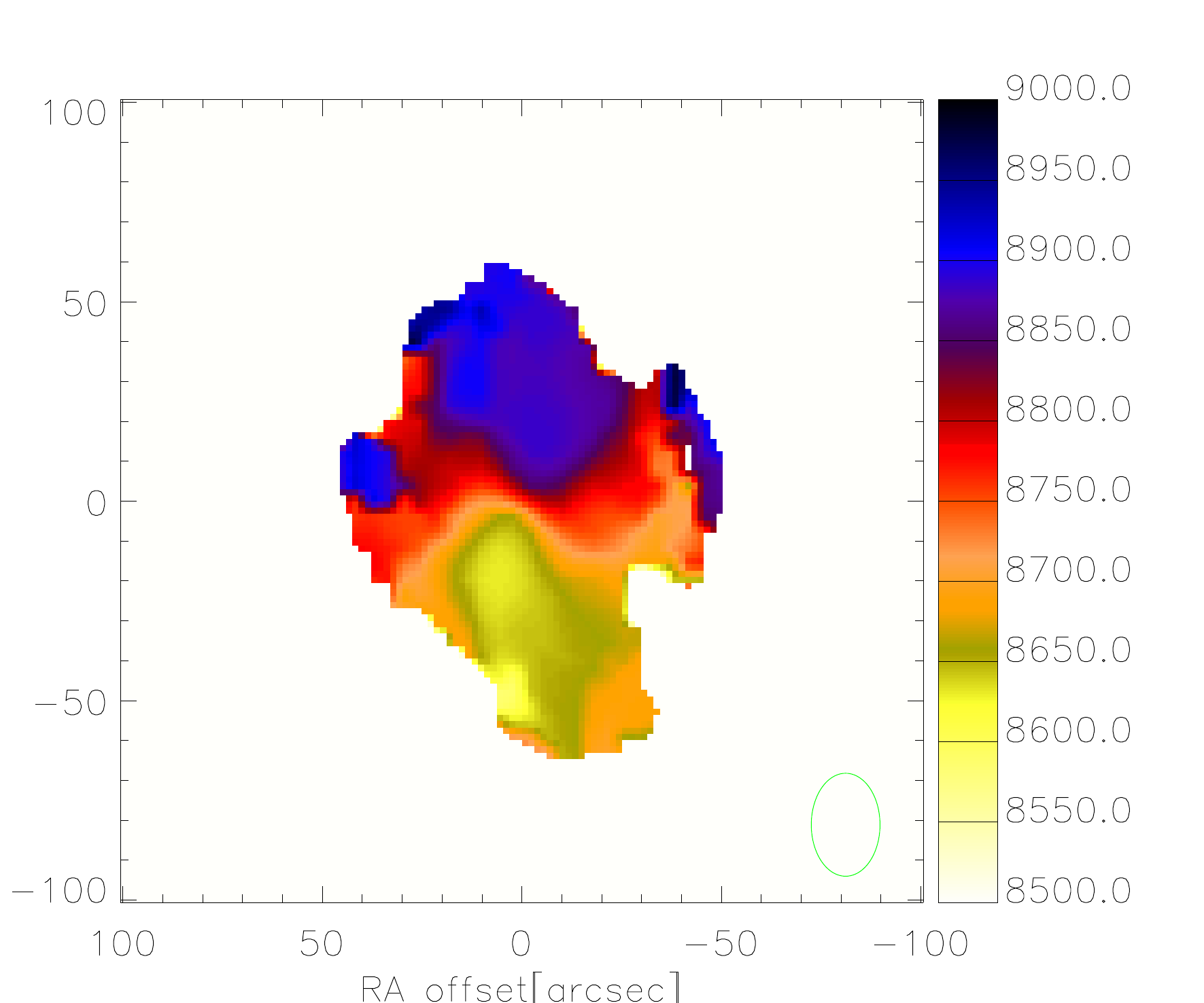}
\includegraphics[width=0.30\textwidth]{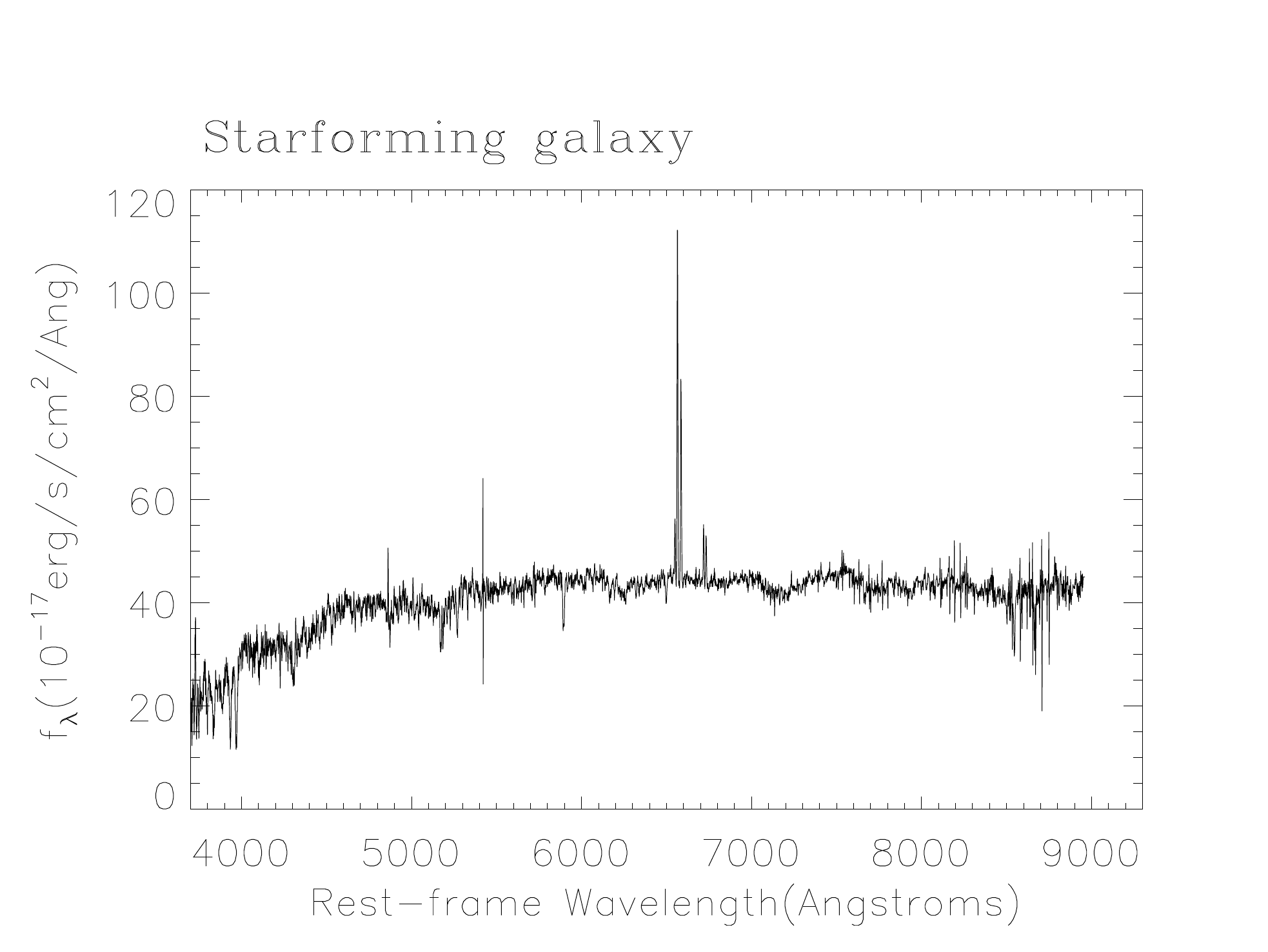}
}\par
\mbox{
\includegraphics[width=0.25\textwidth]{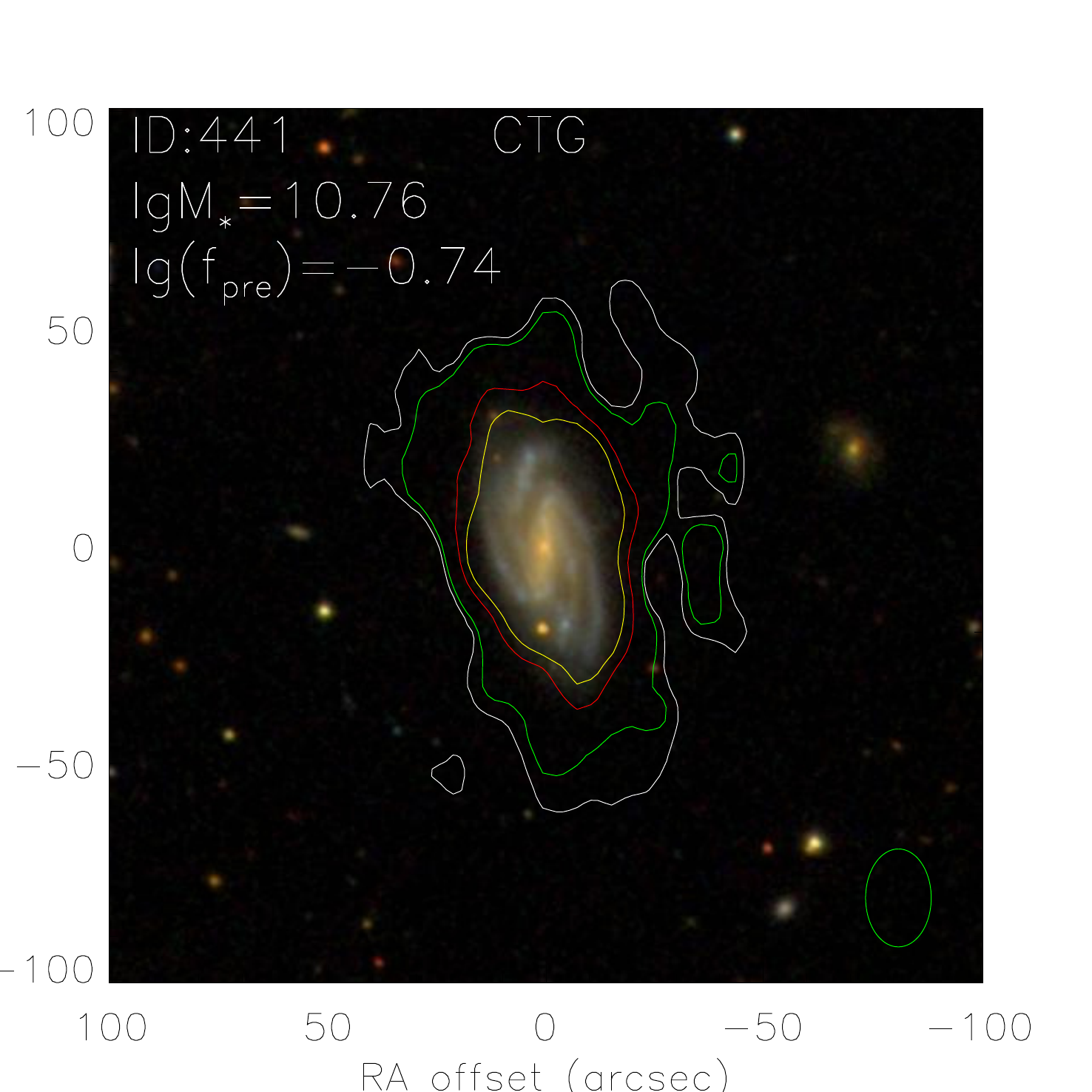}
\includegraphics[width=0.30\textwidth]{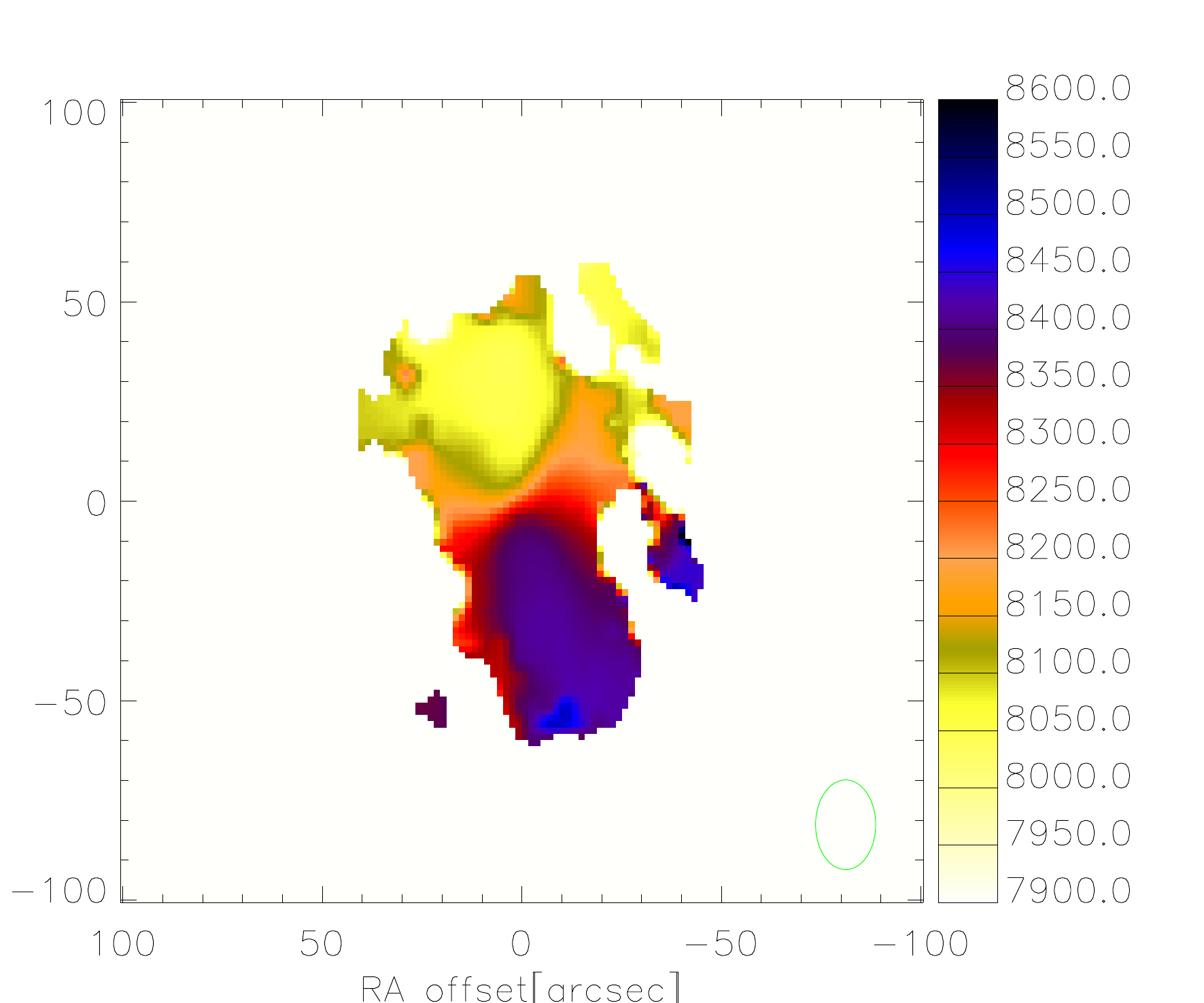}
\includegraphics[width=0.30\textwidth]{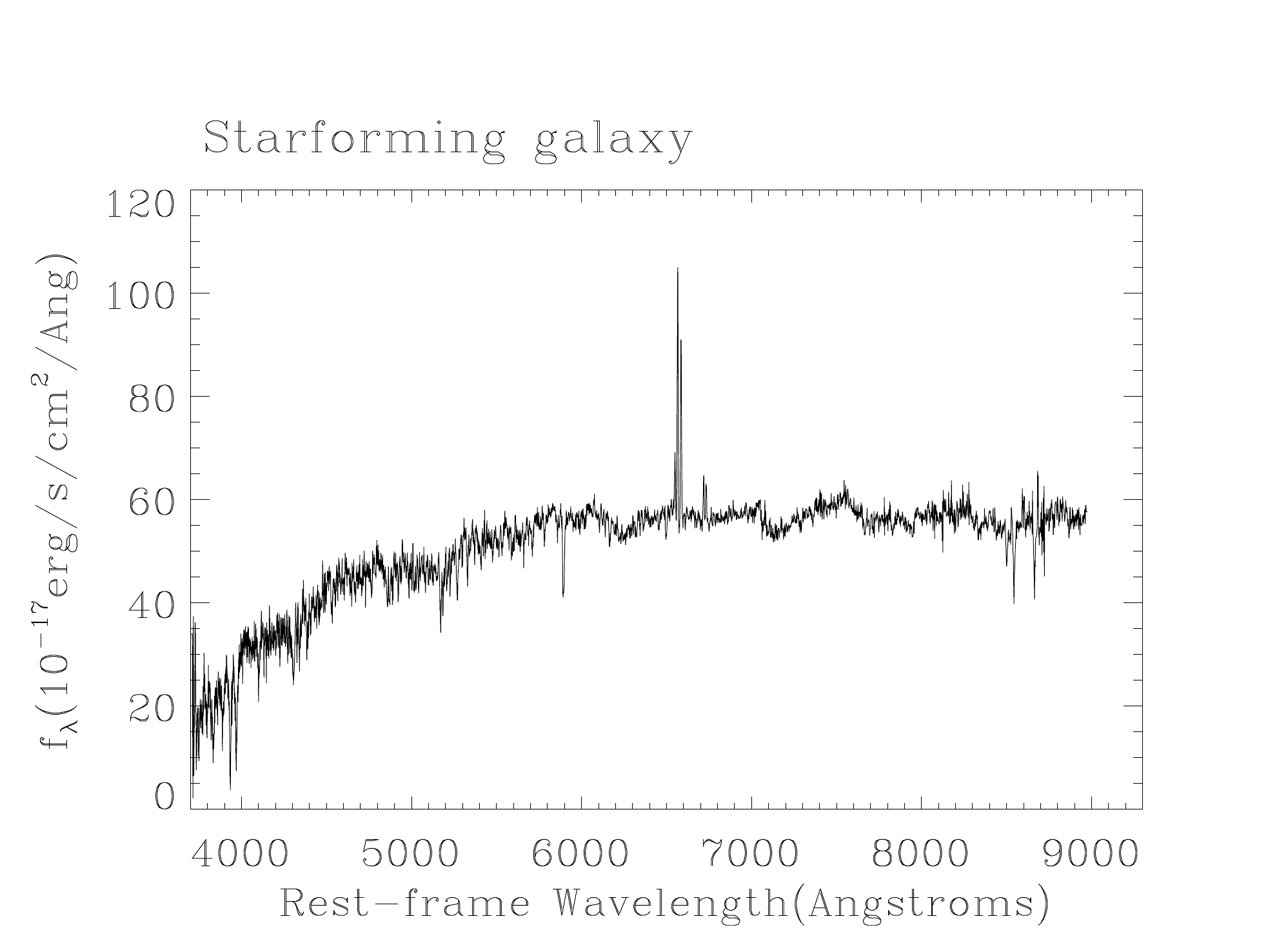}
}\par
\mbox{
\includegraphics[width=0.25\textwidth]{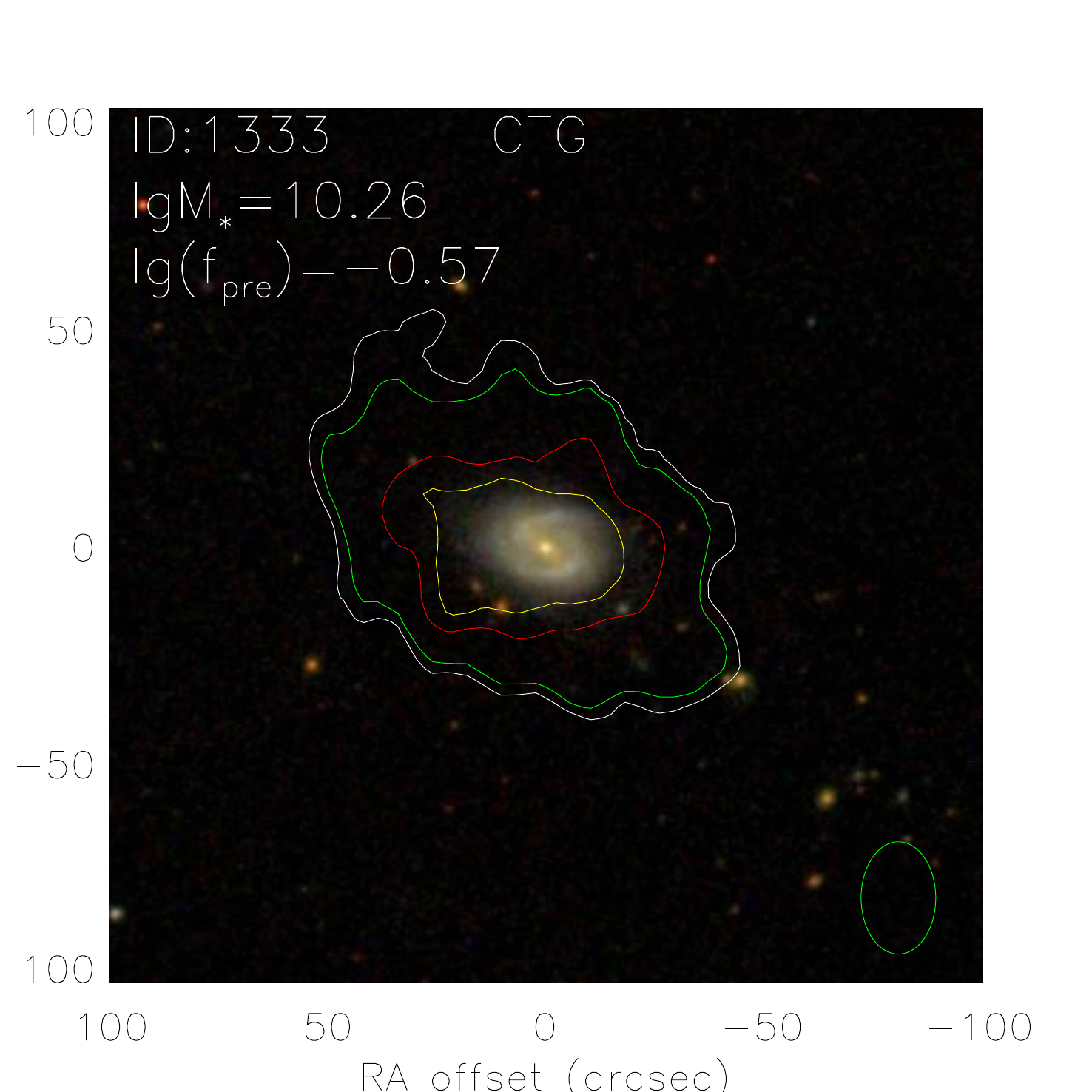}
\includegraphics[width=0.30\textwidth]{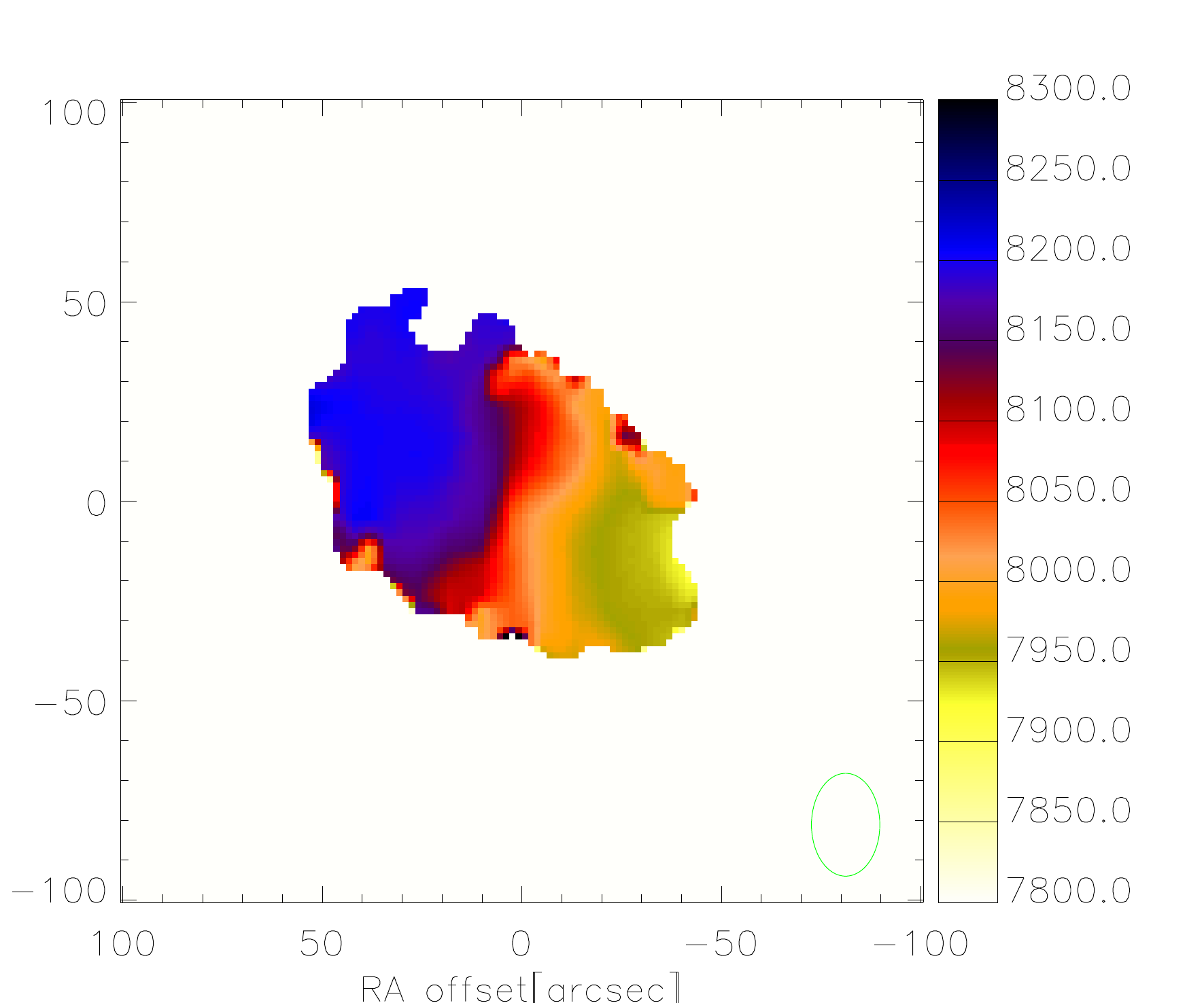}
\includegraphics[width=0.30\textwidth]{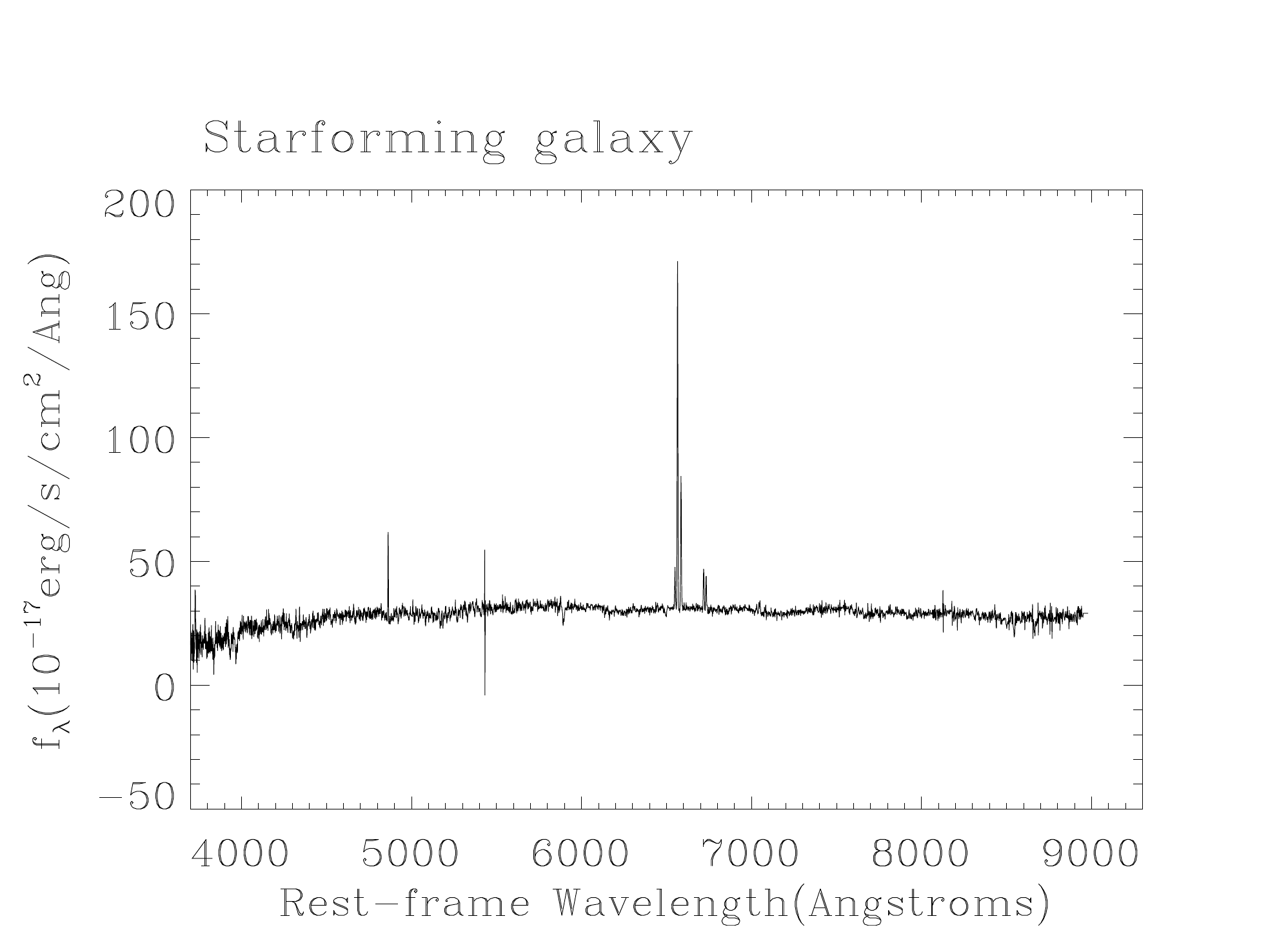}
}\par
\mbox{
\includegraphics[width=0.25\textwidth]{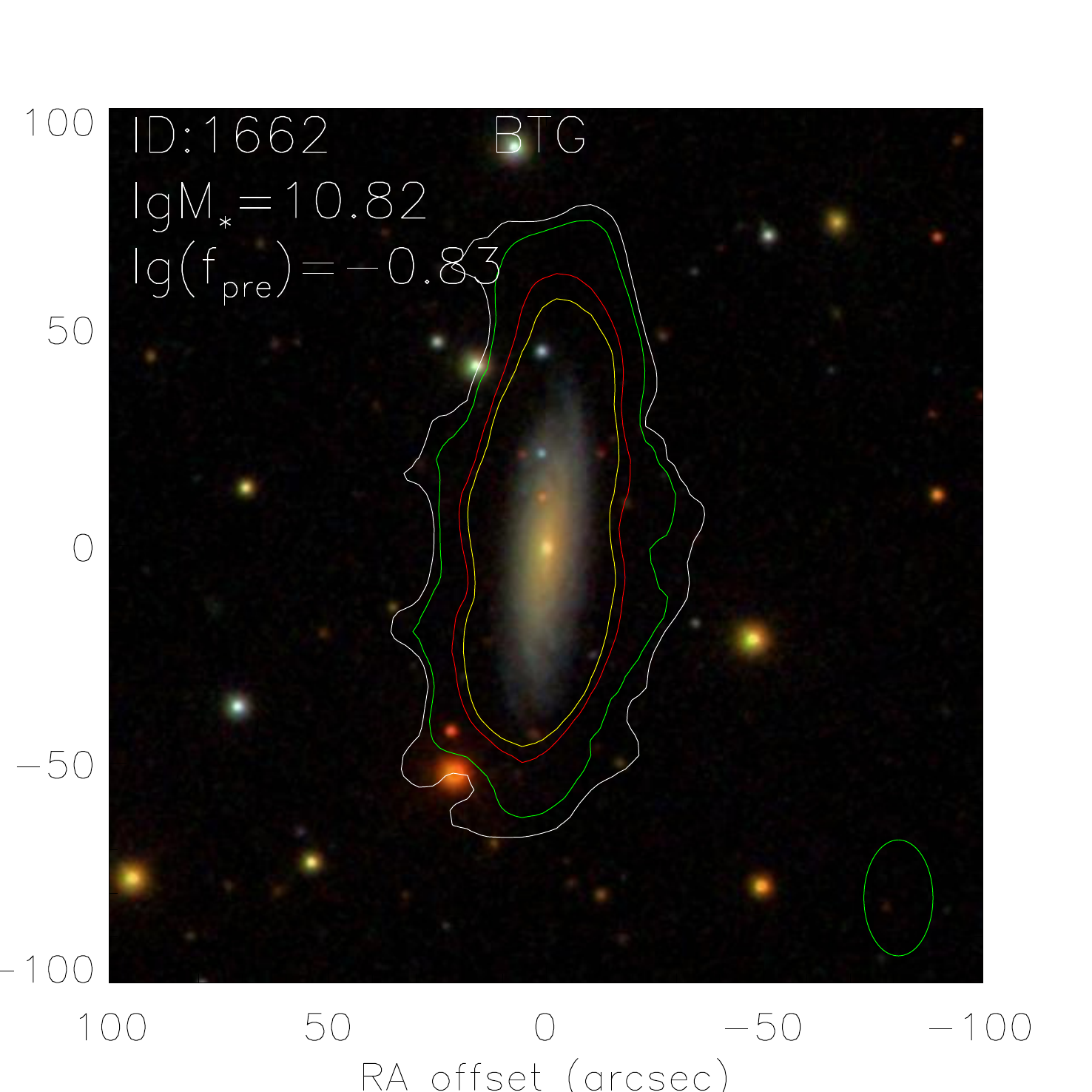}
\includegraphics[width=0.30\textwidth]{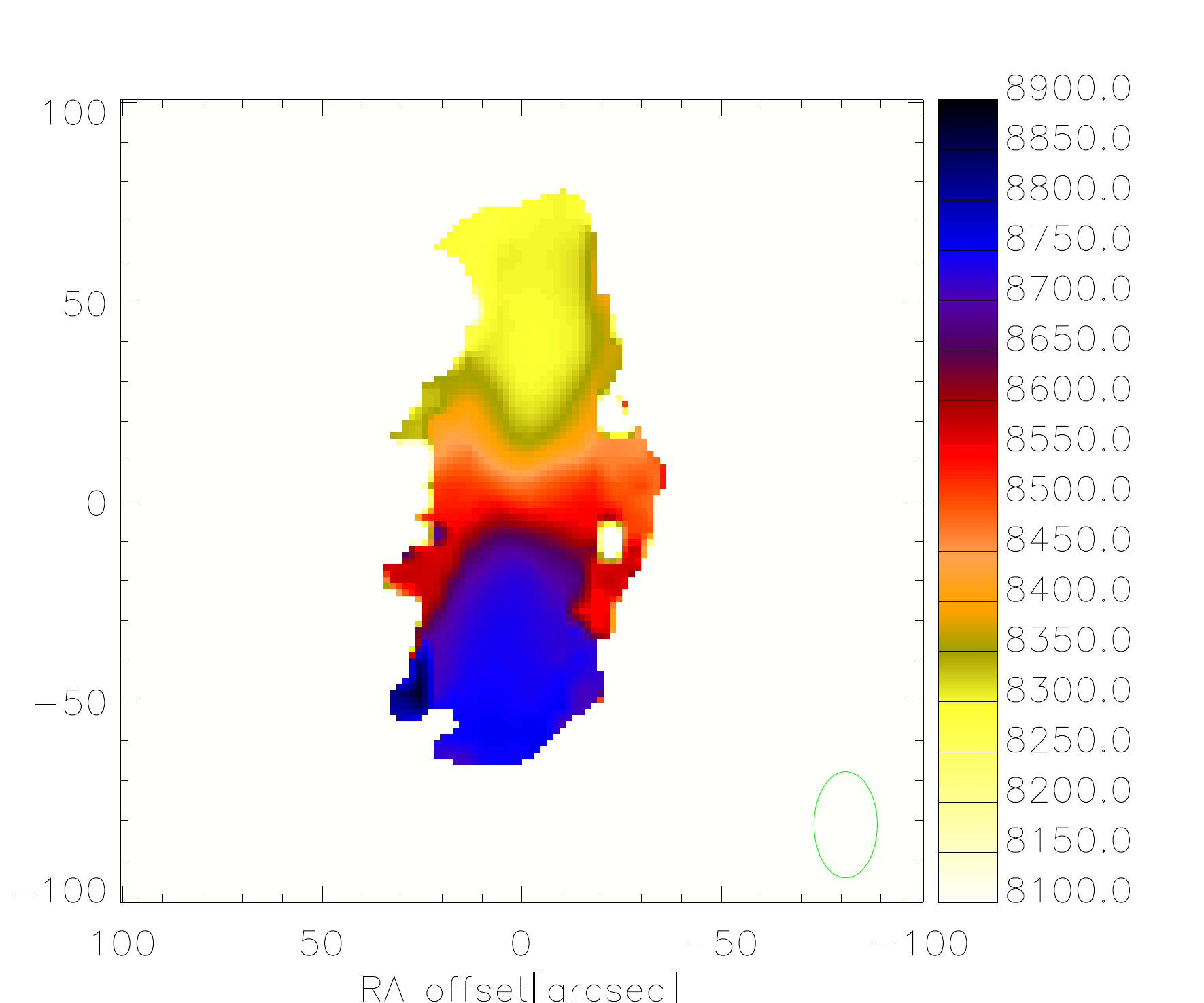}
\includegraphics[width=0.30\textwidth]{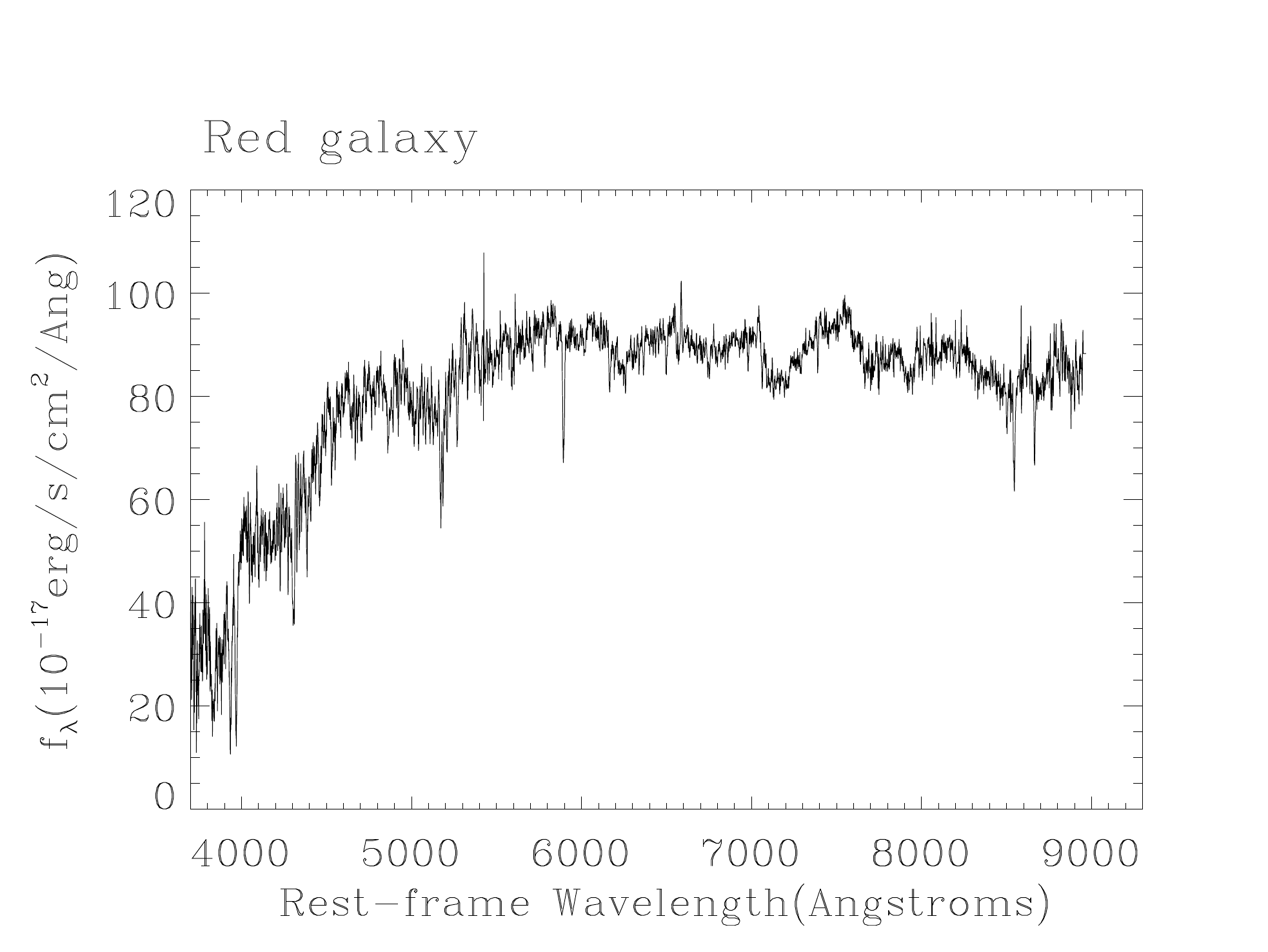}
}
\mbox{
\includegraphics[width=0.25\textwidth]{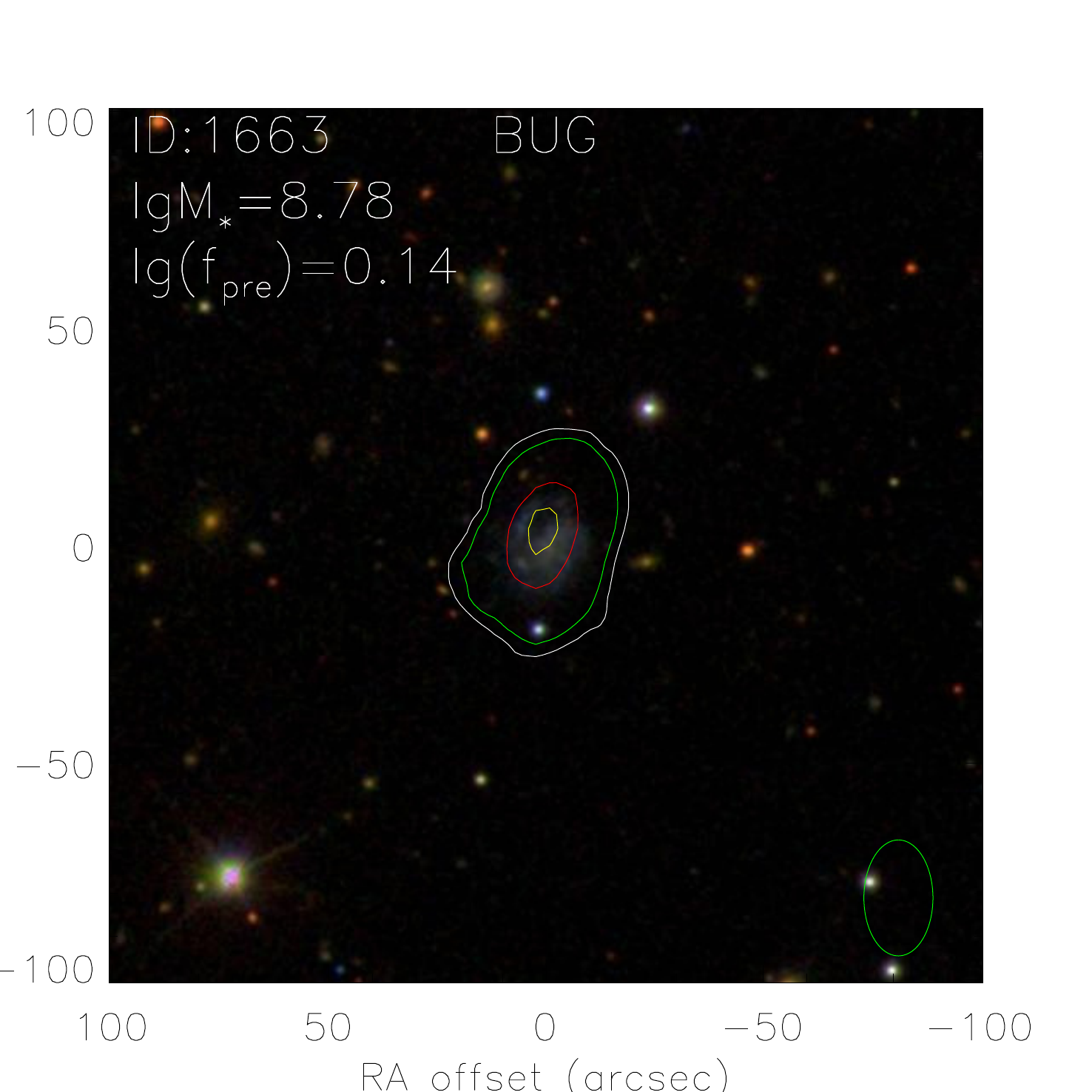}
\includegraphics[width=0.30\textwidth]{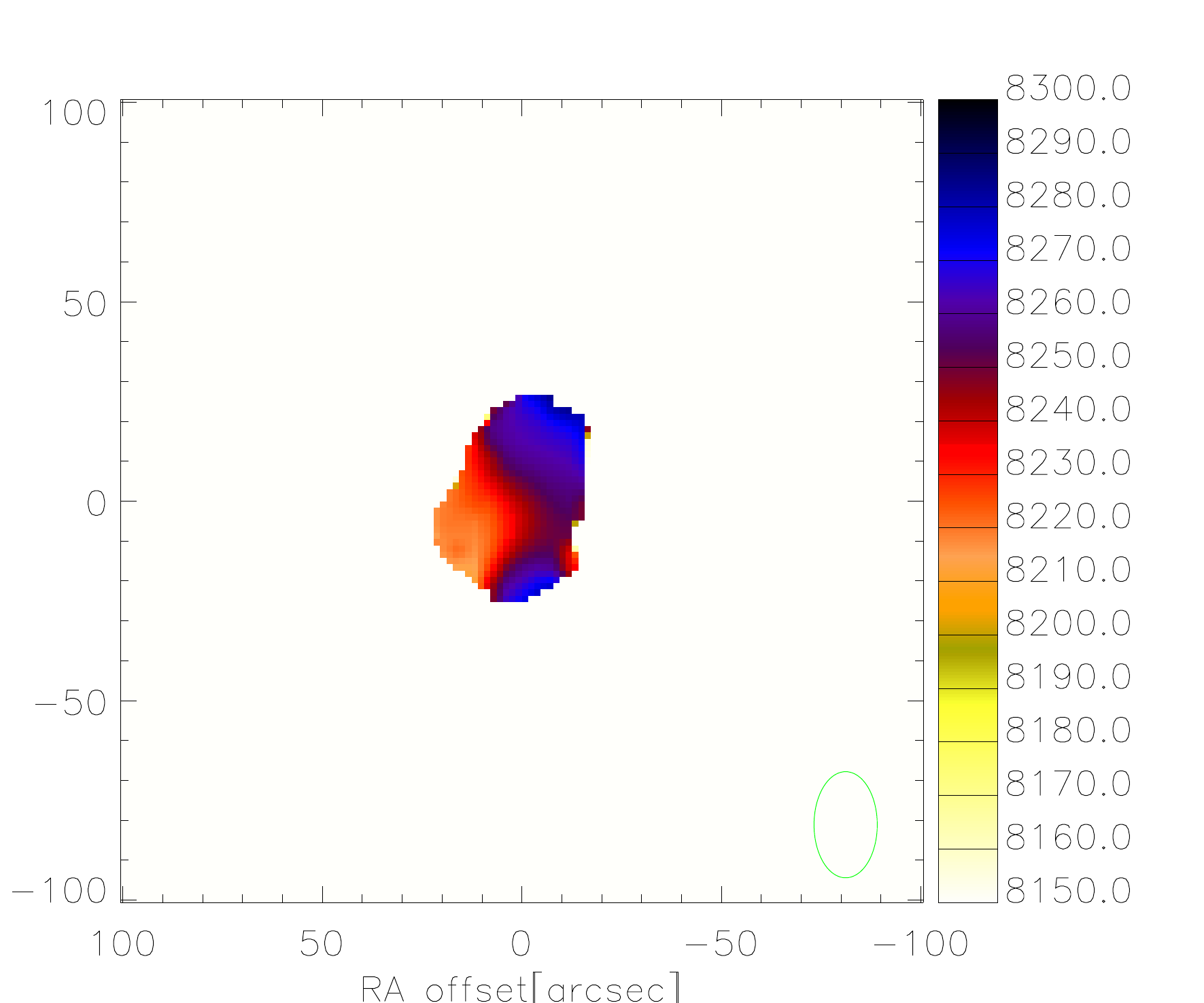}
\includegraphics[width=0.30\textwidth]{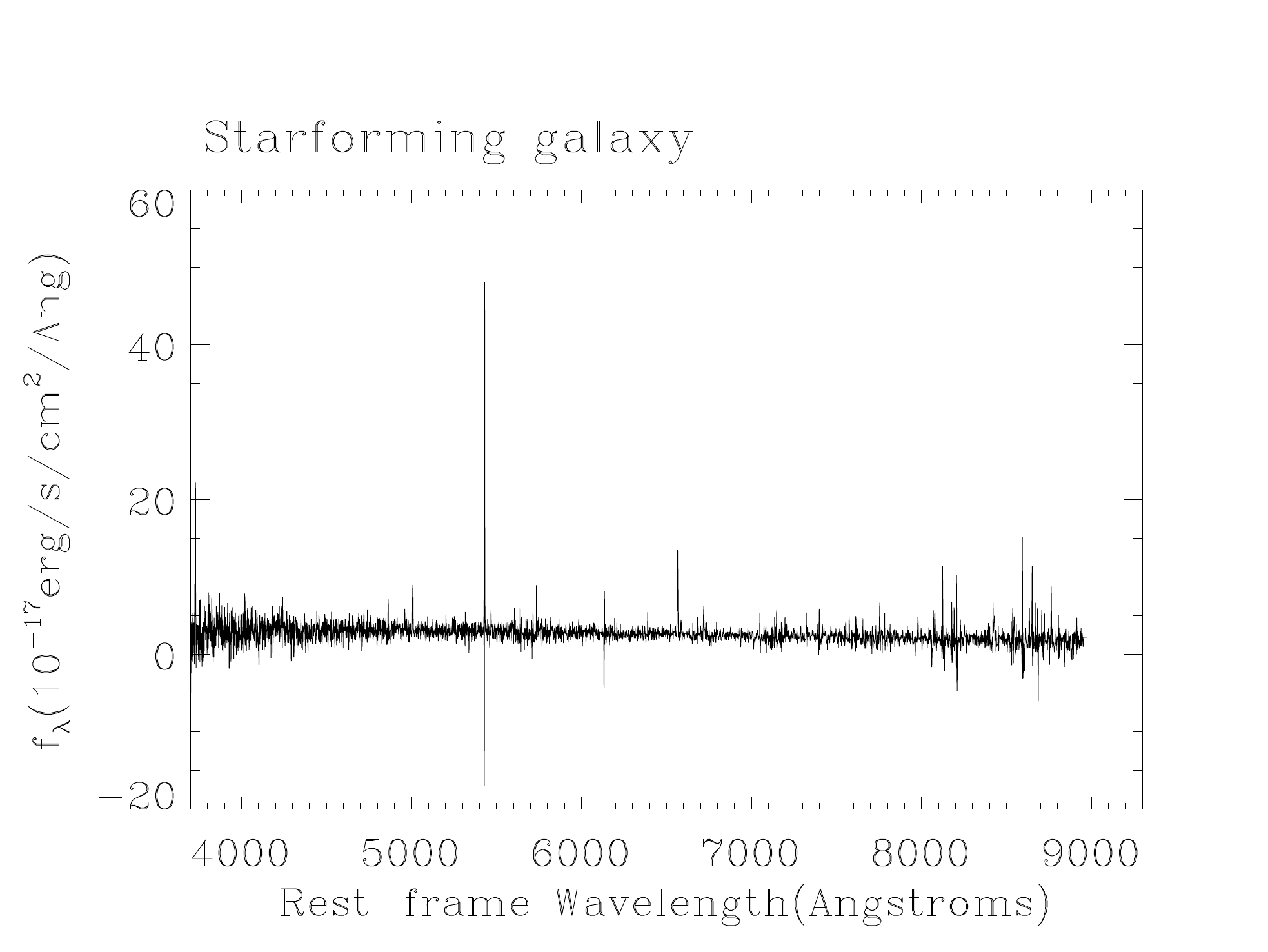}
}\par
\caption{The same as Fig. 10 but for some of the non-outlier galaxies, selected 
randomly from our sample galaxies with the offset to H{\sc I} mass-size relation less than 0.2 dex and the offset
 to H{\sc I}-plane less than 0.2 dex.}
 \label{fig11}
\end{figure*}

\begin{table*}
\begin{center}
\renewcommand{\arraystretch}{0.9}
\begin{tabular}{c c c c c c c c}
\hline \hline
\multicolumn{1}{c}{ID} & \multicolumn{1}{c}{Environment} & \multicolumn{1}{c}{$\log_{10} M_{\rm halo}$} & \multicolumn{1}{c}{H{\sc{I}} MS}  & \multicolumn{1}{c}{H{\sc{I}}-plane} & \multicolumn{1}{c}{$\Delta$Center} & \multicolumn{1}{c}{$R_{\rm 90,H{\sc I}}/R_{\rm 50,H{\sc I}}$} \\
\multicolumn{1}{c}{ } & \multicolumn{1}{c}{ } &\multicolumn{1}{c}{$\rm M_{\odot}$}  & \multicolumn{1}{c}{dex} & \multicolumn{1}{c}{dex} & \multicolumn{1}{c}{ } & \multicolumn{1}{c}{}\\
\hline
%179 & isolated & - &       0.72&  0.07 \\
941 & isolated & - &        1.15&     0.65  & 0.252 & 1.68 \\
1323 & -        & - &      0.51&   0.08  & 0.3981 & 1.51 \\
1554 & isolated & 12.59 &       0.97&     0.27  &0.350 & 2.09\\
1700 & isolated  & -     &       1.63&    0.15  &0.960 & 1.73\\
     &       &        &           &      &    &   \\
890 & isolated & 12.37 &      -0.14&    -0.87  & 0.447  & 1.85\\
983 & isolated & -     &       0.55&    -0.70  & 0.517 & 1.52\\
1104 & isolated & -    &     0.09&     0.66    & 0.129  &2.16\\
1378 & isolated & 12.37 &    -0.02&    -0.50   & 0.320 &1.76\\
1407 & -        & -     &     0.12&     0.62   & 0.517 & 1.87\\
\hline \hline
\end{tabular}
\caption{The list of outliers of H{\sc{I}} mass-size relation and H{\sc{I}}-plane.
 From left to right, columns represent galaxy ID, environment, halo mass,
 offset from normal H{\sc{I}} mass-size
 relation, offset from normal H{\sc{I}}-plane, $\Delta$Center and $R_{\rm 90,H{\sc I}}/R_{\rm 50,H{\sc I}}$, respectively.}
\label{table}
\end{center}
\end{table*}

\begin{figure*}
\includegraphics[width=0.9\textwidth]{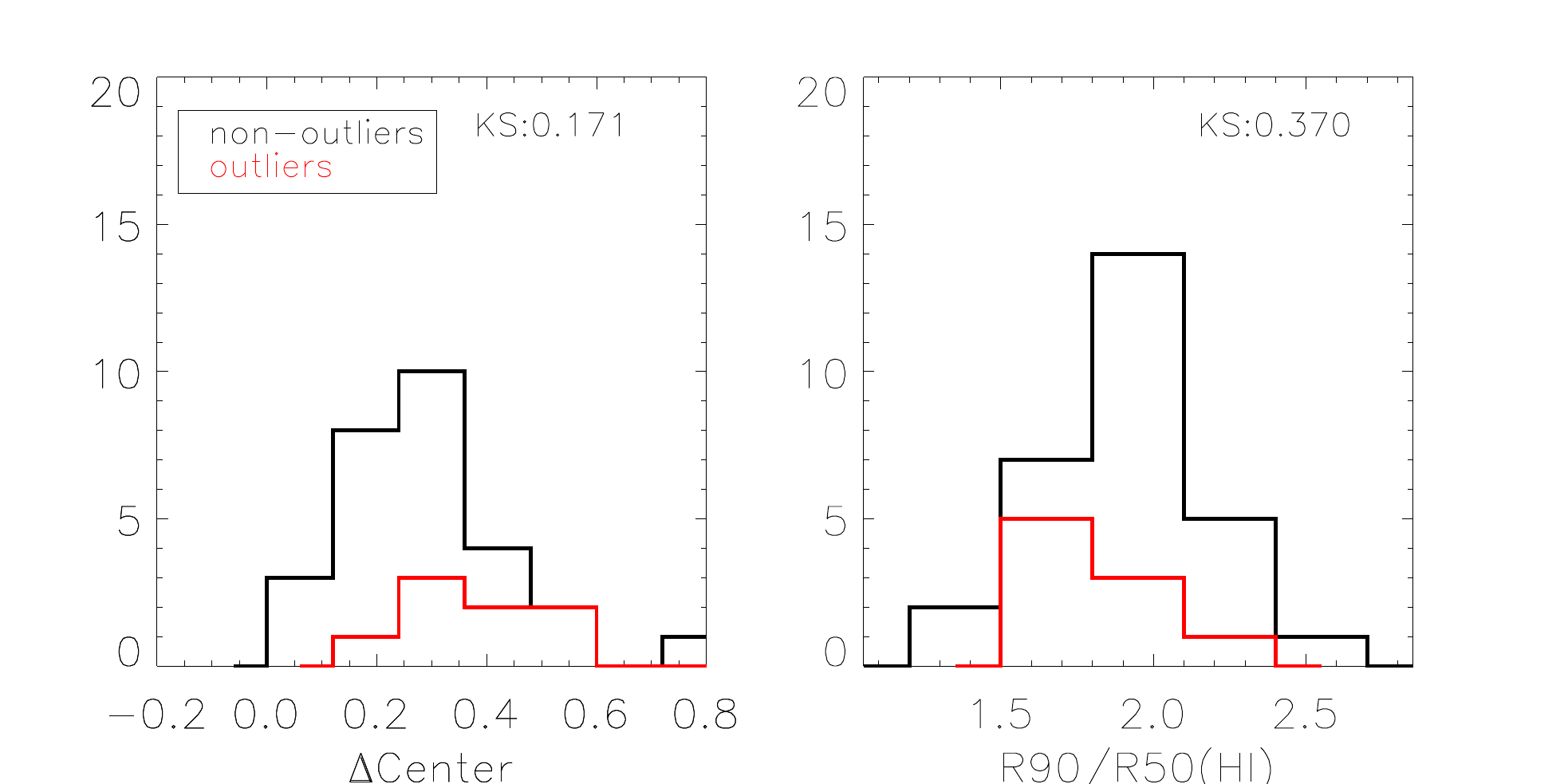}
\caption{The distributions of $\Delta$Center and $R_{\rm 90,H{\sc I}}/R_{\rm 50,H{\sc I}}$ for outliers (red) and non-outliers (black).
The K-S test probability is given in the top right corner of each panel. The 
non-outliers are selected from all the galaxies with both the offset to H{\sc I} 
mass-size relation and the offset to H{\sc I}-plane less than 0.2 dex.}
\label{fig12}
\end{figure*}

Galaxies accrete cold gas directly from the intergalactic medium or through interaction with companions, and
lose their gas through tidal or ram pressure stripping. 
Some authors have tried to connect specific H{\sc{I}} structures in galaxies with ongoing cold gas accretion, such
as the existence of extra-planar gas and warped structures of H{\sc{I}} distribution in
galaxies \citep[]{Wakker-07, Thilker-07, Oosterloo-10}. 
Ram pressure striping, especially
in clusters also affect the morphology of H{\sc{I}} disks \citep[]{McConnachie-07,Bernard-12,Serra-13,Zhang-13}.
In this section, we investigate the structures of those galaxies which are outliers in the
H{\sc{I}} mass-size relation (see Fig. 5) or the H{\sc{I}}-plane (see Fig. 7). 
We argue that (after close inspection) either offset is an indication for ongoing interaction
with the environment.

Outliers are defined to include galaxies that are offset in their H{\sc{I}} mass by more than 0.4 dex from the 
normal H{\sc{I}} mass-size relation or offset more than 0.5 dex from the H{\sc{I}}-plane. 
Here we just discuss the outliers with $M_{*}>M_{\rm H{\sc I}}$, since the H{\sc{I}} plane would underestimate
the H{\sc{I}} fraction for galaxies with $\log_{10}(M_{*}/M_{\rm H{\sc I}})>0$.
We select the outliers from all cubes with ``peculiar cubes'' (Section 3) excluded.
Note that there is no indication that outliers tend to fall preferentially into
either group of galaxies around H{\sc I}-rich galaxies (blue sample) or control galaxies. 
In table 2, we list the environment of galaxies, host system Dark Matter (DM) mass, offset from the H{\sc{I}} mass-size
relation, offset from the H{\sc{I}}-plane, $\Delta$Center and $R_{\rm 90,H{\sc I}}/R_{\rm 50,H{\sc I}}$. 
The environment of galaxies (isolated or in group) and 
the host halo mass estimates are from \citet{Yang-07}. 
Fig. 10 shows H{\sc{I}} total intensity contours overlaid on an SDSS colour image, an H{\sc{I}} velocity field and an SDSS
spectrum for each of these outliers. The white, green, red and yellow H{\sc{I}} contours 
represent 2.0, 4.0, 14.0 and 20.0 times the median SNR of the outermost H{\sc I} contour, respectively.
The velocity fields are derived from a Gauss-Hermite fitting procedure (den Heijer et al., in preparation). 

\subsection{Outliers in the  H{\sc{I}} mass-size relation}
 Galaxies marked 941, 1323, 1554 and 1700 are outliers from the H{\sc{I}} mass-size relation. 
 Galaxy 1554 and 1700 are edge-on galaxies, while Galaxy 941, 1323 and 1622 are not. 
 We will discuss Galaxy 941, 1323, 1554 and 1700 in detail.
 Galaxy 983 is investigated in next subsection because 
 it is also offset from the H{\sc{I}} fundamental plane.
 
 \textbf{Galaxy 941 and 1323} are both starburst dwarf galaxies, but most of their H{\sc{I}} is distributed asymmetrically,
 and is also offset from the optical disks. The H{\sc{I}} of Galaxy 941 is distributed in patches 
 that extend 10 times further than the optical disks. The nearby object seen in the SDSS image,
 is found to be a foreground star. 
 The H{\sc{I}} disk of Galaxy 1323 shows a big void in the region of the optical disk.
 There are two possible explanations for these disturbed
 morphologies. One is feedback processes (by AGN or by superwinds powered by SNe and
 stellar winds in the starburst) that push gas from inside the galaxy to the outside.
 The other explanation is that the H{\sc{I}} gas in the stellar disk regions have been converted to  molecular gas
 to sustain the fast star formation, while the H{\sc{I}} gas in the outer regions has not had enough time to flow into the 
 stellar disk.
 The stellar masses of the galaxies and their position on the ``Baldwin, Phillips \& Terlevich'' (BPT) 
 diagram suggests that they do not  have AGNs  in their cores.
 We check the star formation time scale for these two galaxies. 
 It would take more than 0.4 Gyr
 to consume ten percent of their total H{\sc{I}} gas with the current star formation rate. 
 We conclude that SNe explosions and stellar winds are the most likely cause for
 the holes in the H{\sc{I}} intensity maps.
 In similar case, \cite{Muhle-05} argued that a huge hole in 
 H{\sc{I}} distribution of NGC 1569 is probably driven from SNe feedback in the center of the galaxy over the past 20Myr.
 
 \textbf{Galaxy 1554 and 1700} are edge-on galaxies. As can be seen from the SDSS spectrum,
  the dust extinction is very high in Galaxy 1554. 
  The H{\sc{I}} gas is concentrated and distributed in the stellar disk. Galaxy 1700 is a 
  star-forming galaxy in which the  H{\sc{I}} seems to be vertically offset with respect to the stellar disk. 
  The inclination correction creates significant  uncertainty when calculating H{\sc{I}} sizes 
  for edge-on galaxies, since the H{\sc{I}} disks
  are usually thicker than stellar disks. This may result in the offset in the H{\sc{I}} mass-size relation.

\subsection{Outliers from the H{\sc{I}}-plane}

 \textbf{Galaxy 983} exhibits  a large offset from both the H{\sc{I}} mass-size relation and the H{\sc{I}}-plane. 
 While it  appears to be H{\sc{I}} deficient with respect to the plane, its H{\sc{I}} is more concentrated
 than predicted by the H{\sc{I}} mass-size relation. In addition, a significant part of its H{\sc{I}} appears
 to show irregular kinematics and is offset from the galaxy disk. Hence, morphology and kinematics
 suggest recent tidal or ram-pressure stripping to have removed gas in the outskirts of the galaxy.

 \textbf{Galaxy 890} has its H{\sc{I}} gas compressed against the stellar disk on one side and
 extended to larger radius on the opposite side.
 Its morphological H{\sc{I}} and optical centers are distinctly offset from each other. This 
 is very similar to the ram-pressure stripped 
 galaxies observed in clusters \citep[]{Vollmer-04,Koopmann-Kenney-04,Crowl-05, Chung-09}.
 The observed  H{\sc{I}} gas mass is just 12\% of the predicted value. 
 The velocity map shows a nice spider diagram in the optical region, marked in black ellipse (see Fig. 10).
 For the west extended part, which also contains most of the H{\sc{I}}, it shows a clear decrease in 
 velocity from the inner region to the outer region. 
 So it is likely that the direction of ram-pressure stripping 
 is from east to west and from far to near.

 \textbf{Galaxy 1104 and Galaxy 1407} have at least
 3 times more H{\sc{I}} than predicted (see Fig. 7). 
 Galaxy 1104 is a warped edge-on galaxy with a high warp amplitude. The 
 H{\sc{I}} distribution is also rather asymmetric, suggesting a recent interaction.
 Galaxy 1407 is a star-forming galaxy with a large amount of H{\sc{I}} offset from the regularly rotating disk. 
 It seems to be interacting with two close companions.   
 
 \textbf{Galaxy 1378} has  less H{\sc{I}} gas than predicted and its H{\sc{I}} gas is highly 
 lopsided.
 It is an isolated galaxy with strong on-going star formation at the center.
 This galaxy possesses a large amount of gas outside the regularly
 rotating disk, suggesting ram-pressure stripping or tidal stripping of H{\sc{I}}
 from the galaxy. 

%\subsection{Comparison with non-outliers}
For comparison, we also present a set of ``non-outlier'' galaxies in Fig. 11. These are selected
randomly from our sample galaxies with both the offset to H{\sc I} mass-size relation and the offset
 to H{\sc I}-plane less than 0.2 dex. Generally speaking, the H{\sc I} shapes of these non-outliers are
more regular compared to the H{\sc I} shapes of outliers. Most of them are less asymmetric and have less 
H{\sc I}-gas clouds in their outer region, which suggests no violent interaction with IGM.
The comparison between Fig. 10 and Fig. 11 confirms that most of the outliers really have very 
irregular H{\sc I} morphologies.

Fig. 12 compares $\Delta$Center (left panel) and $R_{\rm 90,H{\sc I}}/R_{\rm 50,H{\sc I}}$ 
(right panel) between outliers and non-outliers.
Note that the non-outlier sample includes all the non-outliers defined in the
same way as above, not only those shown in Fig. 11.
The outliers appear to have larger $\Delta$Center and smaller $R_{\rm 90,H{\sc I}}/R_{\rm 50,H{\sc I}}$
 when compared to the non-outliers. This result is 
consistent with the conclusion above, which are drawn from the example H{\sc I} maps
in Fig. 11. However, given the small sample size and the resulting poor statistics indicated by
 the K-S tests (see the K-S probabilities quoted in the figure),
this result should not be overemphasized.
 We list $\Delta$Center and $R_{\rm 90,H{\sc I}}/R_{\rm 50,H{\sc I}}$ for the outliers
 in table 2. Galaxy 1700 has very large 
$\Delta$Center ($\sim$0.9) that is consistent with its clearly biased H{\sc I} distribution. 
Galaxy 1407 and Galaxy 983 have relatively large $\Delta$Center ($\sim$0.5), 
which are attributed to their extra-planar gas. 
Galaxy 1104, Galaxy 1554 and Galaxy 1407 have the largest $R_{\rm 90,H{\sc I}}/R_{\rm 50,H{\sc I}}$ among others. 
Their location on H{\sc I}-plane are all above the mean relation. This is consistent with what we find in Fig. 9. 
However, we don't compare $rs$ and $rs/R1$ for the outliers, as their $rs$ 
cannot be well determined.

\section{Summary and discussion}
In this paper, we present a catalogue of galaxies from the Bluedisk H{\sc{I}} galaxy survey that
includes sources within the cubes that were not specifically targeted for observation.
These galaxies are nevertheless very interesting, because they probe the environments
of unusually H{\sc I}-rich galaxies, as well as a control sample of galaxies with similar
masses and structural properties, but with more normal H{\sc I} content.

We present the distribution of H{\sc{I}} morphological parameters, the H{\sc I} mass-size relation and 
scaling relations between H{\sc I} gas mass fraction and galaxy mass, structure and colour.   
The main results in this work are listed below.
\begin{itemize}
\item Our sample follows established  H{\sc{I}} scaling relations as
function of stellar mass, stellar surface density and colour very well, and fits  the
 H{\sc{I}} mass-size relation, except for a few outliers. 
\item  Galaxies in the H{\sc I}-rich cubes are displaced to higher H{\sc I} gas mass fractions than predicted
by the optical properties, compared to galaxies in the control cubes. 
\item We inspect the H{\sc{I}} intensity maps and velocity fields of the outliers from the H{\sc{I}} mass-size
relation and the  plane.
 We find that all these galaxies are likely to have undergone recent
interaction  with their environment.
\end{itemize}

The phenomenon of  galactic conformity was first discovered  by \cite{Weinmann-06}, who argued that
the properties of satellite galaxies are strongly correlated with those of  their central
galaxies. In particular,  early-type central galaxies have  a larger fraction of 
early-type satellites than late-type central galaxies with the same stellar mass.
Subsequently, \cite{Kauffmann-Li-Heckman-10} found that the total mass of gas in satellites
has a strong correlation with the colours and specific star formation 
rates of central galaxies of 
all stellar masses, and that this correlation extends out to
radii of 1  Mpc or more. This suggests that more gas-rich galaxies
should have more gas in satellites in their immediate surroundings.
This work was, however, based on optical proxies for H{\sc I} content and not on real H{\sc I} data.
In this paper, we find that galaxies in the large-scale environment of H{\sc{I}}-rich targeted galaxies
tend to be H{\sc{I}}-rich and to have a larger $R_{\rm 90,H{\sc I}}/R_{\rm 50,H{\sc I}}$.
Our findings thus support the conjectures presented in \citep{Kauffmann-Li-Heckman-10}.

\citet{Weinmann-06} and \citet{Ann-Park-Choi-08} argued that the X-ray-emitting hot gas of 
host early-type central galaxies can deprive their satellites of their gas reservoirs 
through hydrodynamic interactions. However, this cannot explain the conformity 
effect in low mass halos. \citet{Kauffmann-Li-Heckman-10} argued that satellite galaxies 
trace the high density peaks of underlying reservoir of ionized gas, which provides 
fuel for star formation in central galaxies.

In this work  we have been able to study a few  galaxies which have irregular H{\sc{I}} shapes and  
anomalous  H{\sc{I}} gas content. In most cases, we find signatures of interaction with the environment  
that is suggestive of tidal or ram pressure stripping, 
though two galaxies are found (1407 and 1104) that
may be  accreting  H{\sc{I}} clouds.

\section*{Acknowledgments}

EW and CL would like to thank the hospitality of the Max Planck 
Institute for Astrophysics while this work was being initiated.
EW is grateful to Paolo Serra and Zhixiong Liang for helpful discussion on 
data analysis progress, to Milan den Heijer for providing 
the Gauss-Hermite velocity maps, and to Attila Popping
for readily providing his primary beam attenuation model.
This work is supported by National Key Basic Research Program 
of China (No. 2015CB857004), NSFC (Grant No. 11173045, 11233005, 
11325314, 11320101002), the Strategic Priority Research 
Program ``The Emergence of Cosmological Structures'' of CAS 
(Grant No. XDB09000000), and the exchange program between CAS
and the Max Panck Society.

Funding for  the SDSS and SDSS-II  has been provided by  the Alfred P.
Sloan Foundation, the Participating Institutions, the National Science
Foundation, the  U.S.  Department of Energy,  the National Aeronautics
and Space Administration, the  Japanese Monbukagakusho, the Max Planck
Society,  and the Higher  Education Funding  Council for  England. The
SDSS Web  Site is  http://www.sdss.org/.  The SDSS  is managed  by the
Astrophysical    Research    Consortium    for    the    Participating
Institutions. The  Participating Institutions are  the American Museum
of  Natural History,  Astrophysical Institute  Potsdam,  University of
Basel,  University  of  Cambridge,  Case Western  Reserve  University,
University of Chicago, Drexel  University, Fermilab, the Institute for
Advanced   Study,  the  Japan   Participation  Group,   Johns  Hopkins
University, the  Joint Institute  for Nuclear Astrophysics,  the Kavli
Institute  for   Particle  Astrophysics  and   Cosmology,  the  Korean
Scientist Group, the Chinese  Academy of Sciences (LAMOST), Los Alamos
National  Laboratory, the  Max-Planck-Institute for  Astronomy (MPIA),
the  Max-Planck-Institute  for Astrophysics  (MPA),  New Mexico  State
University,   Ohio  State   University,   University  of   Pittsburgh,
University  of  Portsmouth, Princeton  University,  the United  States
Naval Observatory, and the University of Washington.

%\bibliography{ref}

\begin{thebibliography}{50}
\expandafter\ifx\csname natexlab\endcsname\relax\def\natexlab#1{#1}\fi

\bibitem[{{Ann}, {Park} \& {Choi}(2008){Ann}, {Park}, \&
  {Choi}}]{Ann-Park-Choi-08}
{Ann} H.~B., {Park} C., {Choi} Y.-Y., 2008, \mnras, 389, 86

\bibitem[{{Becker} {et~al}\mbox{.}(2012){Becker}, {Helfand}, {White}, {Gregg},
  \& {Laurent-Muehlheisen}}]{Becker-12}
{Becker} R.~H., {Helfand} D.~J., {White} R.~L., {Gregg} M.~D.,
  {Laurent-Muehlheisen} S.~A., 2012, VizieR Online Data Catalog, 8090, 0

\bibitem[{{Becker}, {White} \& {Helfand}(1995){Becker}, {White}, \&
  {Helfand}}]{Becker-White-Helfand-95}
{Becker} R.~H., {White} R.~L., {Helfand} D.~J., 1995, \apj, 450, 559

\bibitem[{{Bernard} {et~al}\mbox{.}(2012){Bernard}, {Ferguson}, {Barker},
  {Irwin}, {Jablonka}, \& {Arimoto}}]{Bernard-12}
{Bernard} E.~J., {Ferguson} A.~M.~N., {Barker} M.~K., {Irwin} M.~J., {Jablonka}
  P., {Arimoto} N., 2012, \mnras, 426, 3490

\bibitem[{{Binney}, {Dehnen} \& {Bertelli}(2000){Binney}, {Dehnen}, \&
  {Bertelli}}]{Binney-Dehnen-Bertelli-00}
{Binney} J., {Dehnen} W., {Bertelli} G., 2000, \mnras, 318, 658

\bibitem[{{Boomsma} {et~al}\mbox{.}(2005){Boomsma}, {Oosterloo}, {Fraternali},
  {van der Hulst}, \& {Sancisi}}]{Boomsma-05}
{Boomsma} R., {Oosterloo} T.~A., {Fraternali} F., {van der Hulst} J.~M.,
  {Sancisi} R., 2005, \aap, 431, 65

\bibitem[{{Broeils} \& {Rhee}(1997)}]{Broeils-Rhee-97}
{Broeils} A.~H., {Rhee} M.-H., 1997, \aap, 324, 877

\bibitem[{{Calc{\'a}neo-Rold{\'a}n}
  {et~al}\mbox{.}(2000){Calc{\'a}neo-Rold{\'a}n}, {Moore}, {Bland-Hawthorn},
  {Malin}, \& {Sadler}}]{Calcaneo-Roldan-00}
{Calc{\'a}neo-Rold{\'a}n} C., {Moore} B., {Bland-Hawthorn} J., {Malin} D.,
  {Sadler} E.~M., 2000, \mnras, 314, 324

\bibitem[{{Catinella} {et~al}\mbox{.}(2010){Catinella}, {Schiminovich},
  {Kauffmann}, {Fabello}, {Wang}, {Hummels}, {Lemonias}, {Moran}, \& {et
  al.}}]{Catinella-10}
{Catinella} B. {et~al.}, 2010, \mnras, 403, 683

\bibitem[{{Chang}, {Macci{\`o}} \& {Kang}(2013){Chang}, {Macci{\`o}}, \&
  {Kang}}]{Chang-Maccio-Kang-13}
{Chang} J., {Macci{\`o}} A.~V., {Kang} X., 2013, \mnras, 431, 3533

\bibitem[{{Chaves} \& {Irwin}(2001)}]{Chaves-Irwin-01}
{Chaves} T.~A., {Irwin} J.~A., 2001, \apj, 557, 646

\bibitem[{{Chung} {et~al}\mbox{.}(2009){Chung}, {van Gorkom}, {Kenney},
  {Crowl}, \& {Vollmer}}]{Chung-09}
{Chung} A., {van Gorkom} J.~H., {Kenney} J.~D.~P., {Crowl} H., {Vollmer} B.,
  2009, \aj, 138, 1741

\bibitem[{{Condon} {et~al}\mbox{.}(1998){Condon}, {Cotton}, {Greisen}, {Yin},
  {Perley}, {Taylor}, \& {Broderick}}]{Condon-98}
{Condon} J.~J., {Cotton} W.~D., {Greisen} E.~W., {Yin} Q.~F., {Perley} R.~A.,
  {Taylor} G.~B., {Broderick} J.~J., 1998, \aj, 115, 1693

\bibitem[{{Conselice} {et~al}\mbox{.}(2013){Conselice}, {Mortlock}, {Bluck},
  {Gr{\"u}tzbauch}, \& {Duncan}}]{Conselice-13}
{Conselice} C.~J., {Mortlock} A., {Bluck} A.~F.~L., {Gr{\"u}tzbauch} R.,
  {Duncan} K., 2013, \mnras, 430, 1051

\bibitem[{{Cortese} {et~al}\mbox{.}(2011){Cortese}, {Catinella}, {Boissier},
  {Boselli}, \& {Heinis}}]{Cortese-11}
{Cortese} L., {Catinella} B., {Boissier} S., {Boselli} A., {Heinis} S., 2011,
  \mnras, 415, 1797

\bibitem[{{Crowl} {et~al}\mbox{.}(2005){Crowl}, {Kenney}, {van Gorkom}, \&
  {Vollmer}}]{Crowl-05}
{Crowl} H.~H., {Kenney} J.~D.~P., {van Gorkom} J.~H., {Vollmer} B., 2005, \aj,
  130, 65

\bibitem[{{Evoli} {et~al}\mbox{.}(2011){Evoli}, {Salucci}, {Lapi}, \&
  {Danese}}]{Evoli-11}
{Evoli} C., {Salucci} P., {Lapi} A., {Danese} L., 2011, \apj, 743, 45

\bibitem[{{Huang} {et~al}\mbox{.}(2012){Huang}, {Haynes}, {Giovanelli}, \&
  {Brinchmann}}]{Huang-12}
{Huang} S., {Haynes} M.~P., {Giovanelli} R., {Brinchmann} J., 2012, \apj, 756,
  113

\bibitem[{{Jasche} {et~al}\mbox{.}(2010){Jasche}, {Kitaura}, {Li}, \&
  {En{\ss}lin}}]{Jasche-10}
{Jasche} J., {Kitaura} F.~S., {Li} C., {En{\ss}lin} T.~A., 2010, \mnras, 409,
  355

\bibitem[{{Kapferer} {et~al}\mbox{.}(2008){Kapferer}, {Kronberger}, {Ferrari},
  {Riser}, \& {Schindler}}]{Kapferer-08}
{Kapferer} W., {Kronberger} T., {Ferrari} C., {Riser} T., {Schindler} S., 2008,
  \mnras, 389, 1405

\bibitem[{{Kauffmann}, {Li} \& {Heckman}(2010){Kauffmann}, {Li}, \&
  {Heckman}}]{Kauffmann-Li-Heckman-10}
{Kauffmann} G., {Li} C., {Heckman} T.~M., 2010, \mnras, 409, 491

\bibitem[{{Koopmann} \& {Kenney}(2004)}]{Koopmann-Kenney-04}
{Koopmann} R.~A., {Kenney} J.~D.~P., 2004, \apj, 613, 866

\bibitem[{{Kregel} \& {Sancisi}(2001)}]{Kregel-Sancisi-01}
{Kregel} M., {Sancisi} R., 2001, \aap, 376, 59

\bibitem[{{Li} {et~al}\mbox{.}(2012){Li}, {Kauffmann}, {Fu}, {Wang},
  {Catinella}, {Fabello}, {Schiminovich}, \& {Zhang}}]{Li-12}
{Li} C., {Kauffmann} G., {Fu} J., {Wang} J., {Catinella} B., {Fabello} S.,
  {Schiminovich} D., {Zhang} W., 2012, \mnras, 424, 1471

\bibitem[{{Mayer} {et~al}\mbox{.}(2006){Mayer}, {Mastropietro}, {Wadsley},
  {Stadel}, \& {Moore}}]{Mayer-06}
{Mayer} L., {Mastropietro} C., {Wadsley} J., {Stadel} J., {Moore} B., 2006,
  \mnras, 369, 1021

\bibitem[{{McCarthy} {et~al}\mbox{.}(2008){McCarthy}, {Frenk}, {Font}, {Lacey},
  {Bower}, {Mitchell}, {Balogh}, \& {Theuns}}]{McCarthy-08}
{McCarthy} I.~G., {Frenk} C.~S., {Font} A.~S., {Lacey} C.~G., {Bower} R.~G.,
  {Mitchell} N.~L., {Balogh} M.~L., {Theuns} T., 2008, \mnras, 383, 593

\bibitem[{{McConnachie} {et~al}\mbox{.}(2007){McConnachie}, {Venn}, {Irwin},
  {Young}, \& {Geehan}}]{McConnachie-07}
{McConnachie} A.~W., {Venn} K.~A., {Irwin} M.~J., {Young} L.~M., {Geehan}
  J.~J., 2007, \apjl, 671, L33

\bibitem[{{Moore} {et~al}\mbox{.}(1996){Moore}, {Katz}, {Lake}, {Dressler}, \&
  {Oemler}}]{Moore-96}
{Moore} B., {Katz} N., {Lake} G., {Dressler} A., {Oemler} A., 1996, \nat, 379,
  613

\bibitem[{{Moran} {et~al}\mbox{.}(2012){Moran}, {Heckman}, {Kauffmann},
  {Dav{\'e}}, {Catinella}, {Brinchmann}, {Wang}, {Schiminovich}, \& {et
  al.}}]{Moran-12}
{Moran} S.~M. {et~al.}, 2012, \apj, 745, 66

\bibitem[{{M{\"u}hle} {et~al}\mbox{.}(2005){M{\"u}hle}, {Klein}, {Wilcots}, \&
  {H{\"u}ttemeister}}]{Muhle-05}
{M{\"u}hle} S., {Klein} U., {Wilcots} E.~M., {H{\"u}ttemeister} S., 2005, \aj,
  130, 524

\bibitem[{{Noordermeer} {et~al}\mbox{.}(2005){Noordermeer}, {van der Hulst},
  {Sancisi}, {Swaters}, \& {van Albada}}]{Noordermeer-05}
{Noordermeer} E., {van der Hulst} J.~M., {Sancisi} R., {Swaters} R.~A., {van
  Albada} T.~S., 2005, \aap, 442, 137

\bibitem[{{Oosterloo} {et~al}\mbox{.}(2010){Oosterloo}, {Morganti}, {Crocker},
  {J{\"u}tte}, {Cappellari}, {de Zeeuw}, {Krajnovi{\'c}}, {McDermid}, \& {et
  al.}}]{Oosterloo-10}
{Oosterloo} T. {et~al.}, 2010, \mnras, 409, 500

\bibitem[{{Popping} \& {Braun}(2008)}]{Popping-Braun-08}
{Popping} A., {Braun} R., 2008, \aap, 479, 903

\bibitem[{{Sancisi}(1976)}]{Sancisi-76}
{Sancisi} R., 1976, \aap, 53, 159

\bibitem[{{Sancisi} {et~al}\mbox{.}(2008){Sancisi}, {Fraternali}, {Oosterloo},
  \& {van der Hulst}}]{Sancisi-08}
{Sancisi} R., {Fraternali} F., {Oosterloo} T., {van der Hulst} T., 2008, \aapr,
  15, 189

\bibitem[{{Sault}, {Teuben} \& {Wright}(1995){Sault}, {Teuben}, \&
  {Wright}}]{Sault-Teuben-Wright-95}
{Sault} R.~J., {Teuben} P.~J., {Wright} M.~C.~H., 1995, in Astronomical Data
  Analysis Software and Systems IV, {Shaw} R.~A., {Payne} H.~E., {Hayes}
  J.~J.~E., eds., Vol.~77, p. 433

\bibitem[{{Serra} {et~al}\mbox{.}(2013){Serra}, {Koribalski}, {Duc},
  {Oosterloo}, {McDermid}, {Michel-Dansac}, {Emsellem}, {Cuillandre}, \& {et
  al.}}]{Serra-13}
{Serra} P. {et~al.}, 2013, \mnras, 428, 370

\bibitem[{{Serra} {et~al}\mbox{.}(2012){Serra}, {Oosterloo}, {Morganti},
  {Alatalo}, {Blitz}, {Bois}, {Bournaud}, {Bureau}, \& {et al.}}]{Serra-12}
{Serra} P. {et~al.}, 2012, \mnras, 422, 1835

\bibitem[{{Shang} {et~al}\mbox{.}(1998){Shang}, {Zheng}, {Brinks}, {Chen},
  {Burstein}, {Su}, {Byun}, {Deng}, \& {et al.}}]{Shang-98}
{Shang} Z. {et~al.}, 1998, \apjl, 504, L23

\bibitem[{{Silk} \& {Mamon}(2012)}]{Silk-Mamon-12}
{Silk} J., {Mamon} G.~A., 2012, Research in Astronomy and Astrophysics, 12, 917

\bibitem[{{Swaters} {et~al}\mbox{.}(2002){Swaters}, {van Albada}, {van der
  Hulst}, \& {Sancisi}}]{Swaters-02}
{Swaters} R.~A., {van Albada} T.~S., {van der Hulst} J.~M., {Sancisi} R., 2002,
  \aap, 390, 829

\bibitem[{{Thilker} {et~al}\mbox{.}(2007){Thilker}, {Bianchi}, {Meurer}, {Gil
  de Paz}, {Boissier}, {Madore}, {Boselli}, {Ferguson}, \& {et
  al.}}]{Thilker-07}
{Thilker} D.~A. {et~al.}, 2007, \apjs, 173, 538

\bibitem[{{Vollmer} {et~al}\mbox{.}(2004){Vollmer}, {Balkowski}, {Cayatte},
  {van Driel}, \& {Huchtmeier}}]{Vollmer-04}
{Vollmer} B., {Balkowski} C., {Cayatte} V., {van Driel} W., {Huchtmeier} W.,
  2004, \aap, 419, 35

\bibitem[{{Wakker} {et~al}\mbox{.}(2007){Wakker}, {York}, {Howk}, {Barentine},
  {Wilhelm}, {Peletier}, {van Woerden}, {Beers}, \& {et al.}}]{Wakker-07}
{Wakker} B.~P. {et~al.}, 2007, \apjl, 670, L113

\bibitem[{{Wang} {et~al}\mbox{.}(2014){Wang}, {Fu}, {Aumer}, {Kauffmann},
  {J{\'o}zsa}, {Serra}, {Huang}, {Brinchmann}, \& {et al.}}]{Wang-14}
{Wang} J. {et~al.}, 2014, \mnras, 441, 2159

\bibitem[{{Wang} {et~al}\mbox{.}(2013){Wang}, {Kauffmann}, {J{\'o}zsa},
  {Serra}, {van der Hulst}, {Bigiel}, {Brinchmann}, {Verheijen}, \& {et
  al.}}]{Wang-13}
{Wang} J. {et~al.}, 2013, \mnras, 433, 270

\bibitem[{{Wang} {et~al}\mbox{.}(2011){Wang}, {Kauffmann}, {Overzier},
  {Catinella}, {Schiminovich}, {Heckman}, {Moran}, {Haynes}, \& {et
  al.}}]{Wang-11}
{Wang} J. {et~al.}, 2011, \mnras, 412, 1081

\bibitem[{{Weinmann} {et~al}\mbox{.}(2006){Weinmann}, {van den Bosch}, {Yang},
  \& {Mo}}]{Weinmann-06}
{Weinmann} S.~M., {van den Bosch} F.~C., {Yang} X., {Mo} H.~J., 2006, \mnras,
  366, 2

\bibitem[{{Yang} {et~al}\mbox{.}(2007){Yang}, {Mo}, {van den Bosch},
  {Pasquali}, {Li}, \& {Barden}}]{Yang-07}
{Yang} X., {Mo} H.~J., {van den Bosch} F.~C., {Pasquali} A., {Li} C., {Barden}
  M., 2007, \apj, 671, 153

\bibitem[{{Zhang} {et~al}\mbox{.}(2013){Zhang}, {Li}, {Kauffmann}, \&
  {Xiao}}]{Zhang-13}
{Zhang} W., {Li} C., {Kauffmann} G., {Xiao} T., 2013, \mnras, 429, 2191

\end{thebibliography}

\bsp
\label{lastpage}
\fi
\end{document}